\documentclass{article}

\usepackage{PRIMEarxiv_modi}

\usepackage[utf8]{inputenc} 
\usepackage[T1]{fontenc}    
\usepackage{url}            
\usepackage{booktabs}       
\usepackage{amsfonts}       
\usepackage{nicefrac}       
\usepackage{microtype}      
\usepackage{lipsum}
\usepackage{amsmath}
\usepackage{amsthm}
\usepackage{amssymb}

\usepackage{algorithm}
\usepackage{algpseudocode}

\usepackage{siunitx}
\usepackage{bigints}
\usepackage{bm}
\usepackage{diagbox}
\usepackage{makecell}
\usepackage{multirow}
\usepackage{caption}
\usepackage{subcaption}
\usepackage{graphicx}

\usepackage{upgreek}
\usepackage{authblk}

\usepackage[numbers, sort&compress]{natbib}
\setcitestyle{super}
\setcitestyle{open={},close={}}
\usepackage{soul}  
\usepackage{setspace} 
\usepackage{tipa} 
\usepackage{listings}  

\usepackage{array}
\usepackage{makecell}
\usepackage{indentfirst}
\usepackage{color, xcolor}
\definecolor{lightgray}{gray}{0.9}
\usepackage{mathtools}
\usepackage{enumitem}

\usepackage{hyperref}
\hypersetup{
    colorlinks=true,
    citecolor=blue,
    linkcolor=blue,
    filecolor=blue,
    urlcolor=blue,
}

\usepackage{pifont}

\numberwithin{equation}{section}

\title{Physics-informed machine learning for combustion: A review}

\author[1]{Jiahao Wu}
\author[2]{Xutun Wang}
\author[1]{Guihua Zhang}
\author[1]{Jiayue Liu}
\author[1]{Xin Li}
\author[1]{Yang Zhang}
\author[1]{Hai Zhang}
\author[1]{Junfu Lyu}
\author[2]{Bing Wang}
\author[$\ \ $1]{Yuxin Wu\thanks{Corresponding author: wuyx09@tsinghua.edu.cn}}

\affil[1]{Department of Energy and Power Engineering, Key Laboratory for Thermal Science and Power Engineering of Ministry of Education, Tsinghua University, Beijing 100084, China}
\affil[2]{School of Aerospace Engineering, Key Laboratory of Advanced Power and Propulsion of Ministry of Education, Tsinghua University, Beijing 100084, China}

\date{} 

\begin{document}
\maketitle

\begin{abstract}

Physics-informed machine learning (PIML) represents an emerging paradigm that integrates various forms of physical knowledge into machine learning (ML) components, thereby enhancing the physical consistency of ML models compared to purely data-driven paradigms. The field of combustion, characterized by a rich foundation of physical laws and abundant data, is undergoing a transformation due to PIML. This paper aims to provide a comprehensive overview of PIML for combustion, systematically outlining fundamental principles, significant contributions, key advancements, and available resources. The application of PIML in combustion is categorized into three domains: combustion chemical kinetics, combustion reacting flows, and other combustion-related scenarios. Additionally, current challenges, potential solutions, and practical guidelines for researchers and engineers will be discussed. A primary focus of this review is to demonstrate how combustion laws can be integrated into ML, either through soft or hard constraints, via loss functions or representation models, and within coordinate-to-variable or field-to-field paradigms. This paper shows that PIML offers a unified framework linking physics, model, and data in combustion--integrating physical knowledge in model-to-data simulation and reconstruction tasks, as well as data-to-model modeling tasks--resulting in enhanced data, improved physical models, and more reliable ML models. PIML for combustion presents significant opportunities for both the combustion and ML communities, encouraging greater collaboration and cross-disciplinary engagement.

\end{abstract}


\begin{spacing}{0.5}
    \small
    \tableofcontents
\end{spacing}

\section{Introduction}
\label{sec: intro}

Combustion, the core process of energy conversion, plays a pivotal role in modern industrial systems, underpinning key sectors such as power generation, transportation, chemical manufacturing, and space exploration. The objective of combustion science is twofold: to harness the benefits of combustion while mitigating its hazards \cite{Reitz201408_reviewRCCI, Kohse202304_combusCN, Okafor201808_lowNOx-ammonia}, such as environmental pollution and fires. Achieving this goal requires a deep understanding of combustion, which is inherently complex. By definition, combustion is a high-temperature exothermic redox chemical reaction between a fuel and an oxidant. Therefore, it encompasses multiple coupled physical processes, including chemical reactions, heat transfer, mass transfer, and fluid flow, spanning various temporal and spatial scales \cite{Kee2005_react-flow, Poinsot2012_theo-nume-comb, Law2010_comb-phy, web_Chemkin-theory, web_Cantera-science}. Consequently, combustion research poses substantial scientific and engineering challenges.

\begin{figure}[ht]
    \centering
    \includegraphics[width=1.0\textwidth]{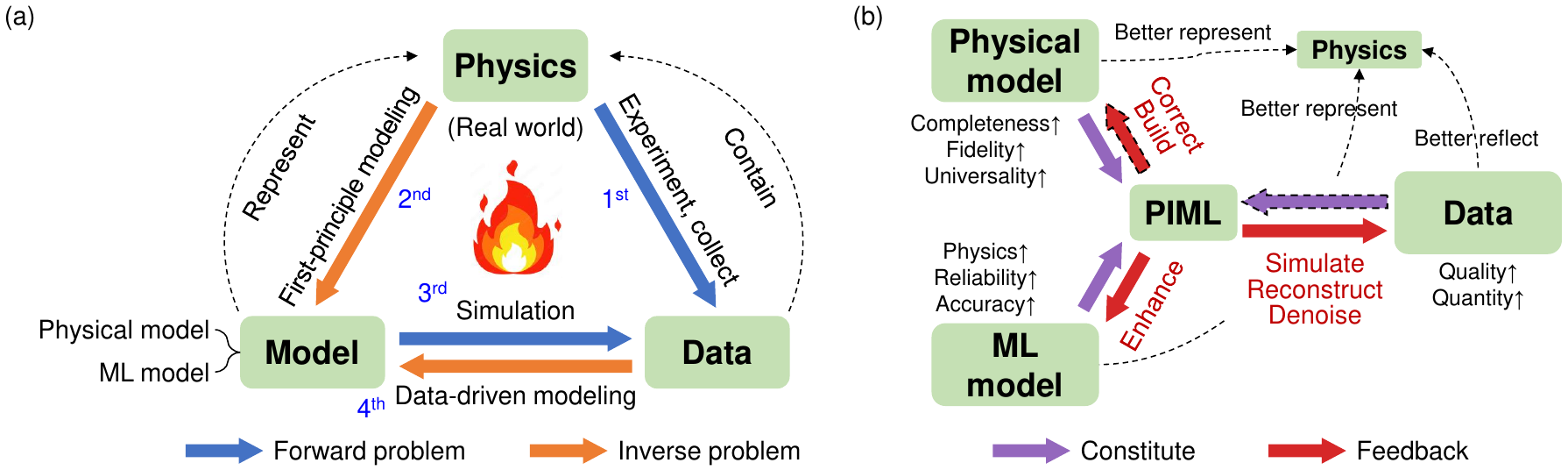}
    \caption{(a) Physics, model, and data are three elements in combustion. Physics represents the real world, physical models and ML models try to represent physics, while data are value records and usually contain physics. Their mutual transformations constitute the forward and inverse problems, which are defined as generating data and modeling, respectively. These transformations also correspond to four basic paradigms of science, as labeled in the figure. (b) PIML can act as a unified bridge connecting physics, model, and data. PIML is constituted by physical model, ML model, and possible data, then each of them can be enhanced by the other two. Models can better represent physics, while data can better reflect physics. The dashed colored arrows represent optional implementations.}
    \label{fig: physics-model-data}
\end{figure}

As shown in Figure \ref{fig: physics-model-data}(a), combustion research and applications involve three core elements: physics, model, and data. Physics represents the real world; models aim to represent physics either through theoretical/physical formulations or data-driven approaches; and data are recorded values that typically contain physics. Transformations among these elements constitute the forward and inverse problems in research. Forward problems include physics-to-data and model-to-data problems, which involve realizing combustion processes in the real and virtual world, respectively, to generate data. The former includes experimental research, such as designing methods for controllable combustion \cite{Liu202307_insta-TJF}, while the latter primarily refers to numerical simulations, including application to various combustion devices \cite{Wu201001_simu-gasifier, Smirnov201509_accumu-rocket, Urzay201801_super-combus-flight, Zhang202304_SOTA-flamelet} and the development of numerical algorithms \cite{Haworth200910_progress-PDF, vanOijen201610_SOTA-FGM}. In contrast, inverse problems include physics-to-model and data-to-model problems, which develop models or theories based on first principles \cite{Kee2005_react-flow, Poinsot2012_theo-nume-comb, Law2010_comb-phy, web_Chemkin-theory, web_Cantera-science} and data patterns, respectively. Model development can involve constructing models from scratch or determining parameters for existing models, which may take either interpretable or purely fitted forms. Forward problems and inverse problems are inherently interdependent. For example, the model-to-data forward problems require the complete form and parameters of the model to be known in advance, while data-to-model inverse problems often depend on physics-to-data experiments \cite{Liu202310_coflow-lift, Guo202304_turb-single-bio, Guo202412_turb-woody-bio, Guo202504_turb-4-fan, Li202508_H2-CH4-trans}. Consequently, theoretical, experimental, and numerical methods are often adopted simultaneously \cite{Poinsot201606_CI-engine}. However, these three types of methods differ substantially in implementation, resource requirements, cost, output form, and transferability, hindering their integration and slowing the research cycle. This highlights the need for a paradigm that can overcome these barriers, establish a unified framework linking physics, models, and data, and more effectively address a wide range of forward and inverse problems in combustion.

Artificial intelligence (AI), machine learning (ML) \cite{Jordan201507_ML}, and deep learning (DL) \cite{LeCun201505_DL} have made remarkable achievements in recent years, transforming fields such as computer vision (CV) \cite{He201606_ResNet}, natural language processing (NLP) \cite{OpenAI202304_GPT-4}, gaming \cite{Silver201601_AlphaGo}, and weather forecasting \cite{Wu202306_Corrformer, Zhang202307_NowcastNet, Bi202307_Pangu-Weather}. AI for science (AI4Sci) and scientific machine learning (SciML) \cite{Baker201902_need4SciML} have emerged as key areas of interest for both the AI and scientific communities. These approaches leverage AI/ML to analyze scientific data, uncover patterns \cite{Wang202308_SciDisc-AI}, and solve scientific problems, and are recognized as the fourth paradigm of science \cite{Hey2009_4th-parad, Agrawal201604_4th-parad-material, Karpatn201706_TGDS}. They have produced transformative results, such as the AlphaFold in life sciences \cite{Jumper202107_AlphaFold2, Abramson202405_AlphaFold3}. SciML has also made significant advances in fluid mechanics \cite{Kutz201701_DL-fluid, Brunton201909_ML4fluid, Zhang202507_AI4FM}, including both experimental fluid mechanics \cite{Vinuesa202308_ML4expFM} and computational fluid dynamics (CFD) \cite{Tompson201607_acceEuler, Kochkov202103_MLacceCFD, Vinuesa202206_enhanceCFD}. The combustion community has likewise embraced ML \cite{Zhou201111_ML4combus, Ihme202204_combusML, Ihme202408_AI-catalyst-combus, Deng202508_SciML-combus}, motivated not only by the power and versatility of ML methods, but also by the data-rich nature of combustion and the limitations of traditional approaches. ML has been applied to combustion model optimization \cite{Lapeyre201902_CNN-subgrid-rate, Seltz201908_LES2FGM, Chung202101_dyn-submodel, Shin202105_data-subgrid, Liu202112_ML-LES-PDF}, combustion chemistry \cite{D'Alessio201910_pre-part, Wan202007_CNN-reduce, Wan202007_NN-reduce, Nguyen202105_ML-inte-chem, Owoyele202109_ChemNODE}, combustion tabulation \cite{Ranade201911_ML-PDF-tab, Zhang202008_spray-FGM-ANN, Shadram202201_PANN-closure, Zhang202503_CPDF-RF, Zhang202507_FGM-5ML}, and surrogate modeling of combustion systems \cite{Wang202009_surr-supercritical, Lyu202106_LSTM-CNN-CI, Jung202302_surr-propellant, Deng202303_pred-bio-fire, Liu202401_pred-type-height}. The applied ML techniques include supervised, unsupervised, and semi-supervised learning, which respectively focus on learning mappings from labeled data, identifying patterns in unlabeled data, and combining both approaches. These methods are built on diverse ML models, including neural networks (NNs), support vector machines (SVMs), and random forests (RFs). For more applications of ML for combustion, please refer to these reviews \cite{Zhou201111_ML4combus, Ihme202204_combusML, Ihme202408_AI-catalyst-combus}. Data-driven ML has greatly advanced the data-to-model paradigm, enabling the exploitation of big data in combustion and the resolution of many longstanding challenges in the field.

\begin{figure}[ht]
    \centering
    \includegraphics[width=0.6\textwidth]{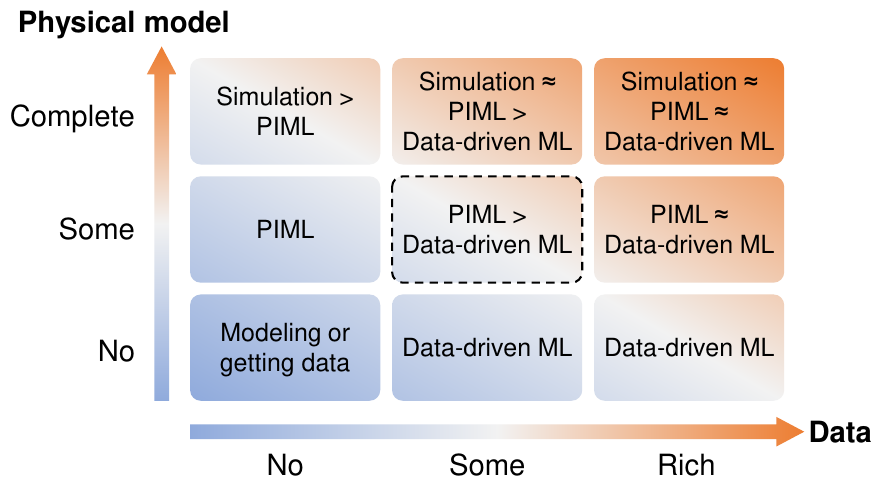}
    \caption{Scenarios under different conditions of physical models and data. PIML is naturally suitable for scenarios where both some physical models and some data are available, which is the most common scenario in practice (indicated by the dashed part). The "$\approx$" sign means that which of the two is superior depends on the specific situation. Here only correct physical models and high-fidelity data are considered.}
    \label{fig: physicalmodel-data}
\end{figure}

Although data-driven ML has many successful applications in combustion, it has inherent limitations, including high data dependency, lack of physical consistency, and limited interpretability--and therefore reduced reliability. Consequently, purely data-driven ML is difficult to apply in scenarios with scarce or low-quality data, making it unsuitable for establishing a unified framework connecting physics, models, and data. To address this issue, the paradigm of physics-informed machine learning (PIML, /\textprimstress p\i m\textschwa l/ or /\textprimstress pa\i m\textschwa l/) has been proposed \cite{Karniadakis202105_PIML}, in which ML is integrated with various forms of physical priors, such as symmetries, conservation laws, and governing equations. Physics can be incorporated through data, models, and optimization, which are the three components of ML. PIML has become an important branch of SciML because it is naturally suited to scenarios where both some physical models and some data are available, which is the most common scenario in practice \cite{Karniadakis202105_PIML}, as illustrated in Figure \ref{fig: physicalmodel-data}. A representative work of PIML is the physics-informed NNs (PINNs) \cite{Raissi201811_PINN}, which employ neural networks as models and integrate differential equations (DEs) as physical constraints within the loss function. Before its former publication \cite{Raissi201811_PINN}, PINNs were first published as two preprints, solving the forward problems \cite{Raissi201711_PINN1} and inverse problems \cite{Raissi201711_PINN2} of DEs, respectively. Here, forward and inverse problems differ slightly from the earlier definitions in Figure \ref{fig: physics-model-data}: forward problems refer to well-posed DEs without observational data, whereas inverse problems refer to ill-posed DEs with possible observational data. PINNs map the space-time coordinates to solutions and use automatic differentiation (AD) \cite{Baydin201804_ADinML} to compute the derivatives in DEs. In fact, the use of NNs to solve DEs dates back to the 1990s \cite{Lee199011_neural-DE, Dissanayake199403_NN-PDE, Lagaris199809_ANN-O/PDE}, but their resurgence was enabled by modern DL, which provides efficient AD techniques. Compared to traditional numerical methods such as the finite difference method (FDM), finite volume method (FVM), and finite element method (FEM), PINNs are mesh-free and differentiable, exhibit no discretization errors, and more naturally suited to leveraging data for solving inverse problems, including field reconstruction and parameter inference. Therefore, PINNs serve as an important complement to traditional numerical methods for solving DEs. In short, PINN is an excellent example of integrating physics into ML, and PIML, as its broader framework, offers a unified approach for incorporating physics, building models, and utilizing data.

Due to the prevalence of ordinary DEs (ODEs), partial DEs (PDEs), and other physical laws across diverse phenomena in many disciplines, PINNs and PIML have garnered significant attention and have developed rapidly in recent years. Combustion, as a field inherently rich in physical laws, has experienced an exponential increase in PIML applications over the same period. While numerous reviews have addressed PIML in general and its applications across various domains (see Section \ref{sec: review-scope}), a comprehensive survey focusing specifically on combustion is still lacking. This paper seeks to fill that gap by providing an extensive review of PIML for combustion--introducing fundamental principles, summarizing major works, highlighting key advancements, proposing a taxonomy of applications and methods, and offering practical suggestions and guidelines for researchers and engineers. In reviewing the existing literature, a central focus of this paper is to present how combustion laws and machine learning can be integrated, either through soft or hard constraints, via loss functions or representation models, and within coordinate-to-variable or field-to-field paradigms. More importantly, as illustrated in Figure \ref{fig: physics-model-data}(b), this paper aims to provide evidence that PIML can act as a unified bridge connecting physics, model, and data in combustion, thereby inspiring greater engagement from both combustion and AI research communities.

\begin{figure}[ht]
    \centering
    \includegraphics[width=1.0\textwidth]{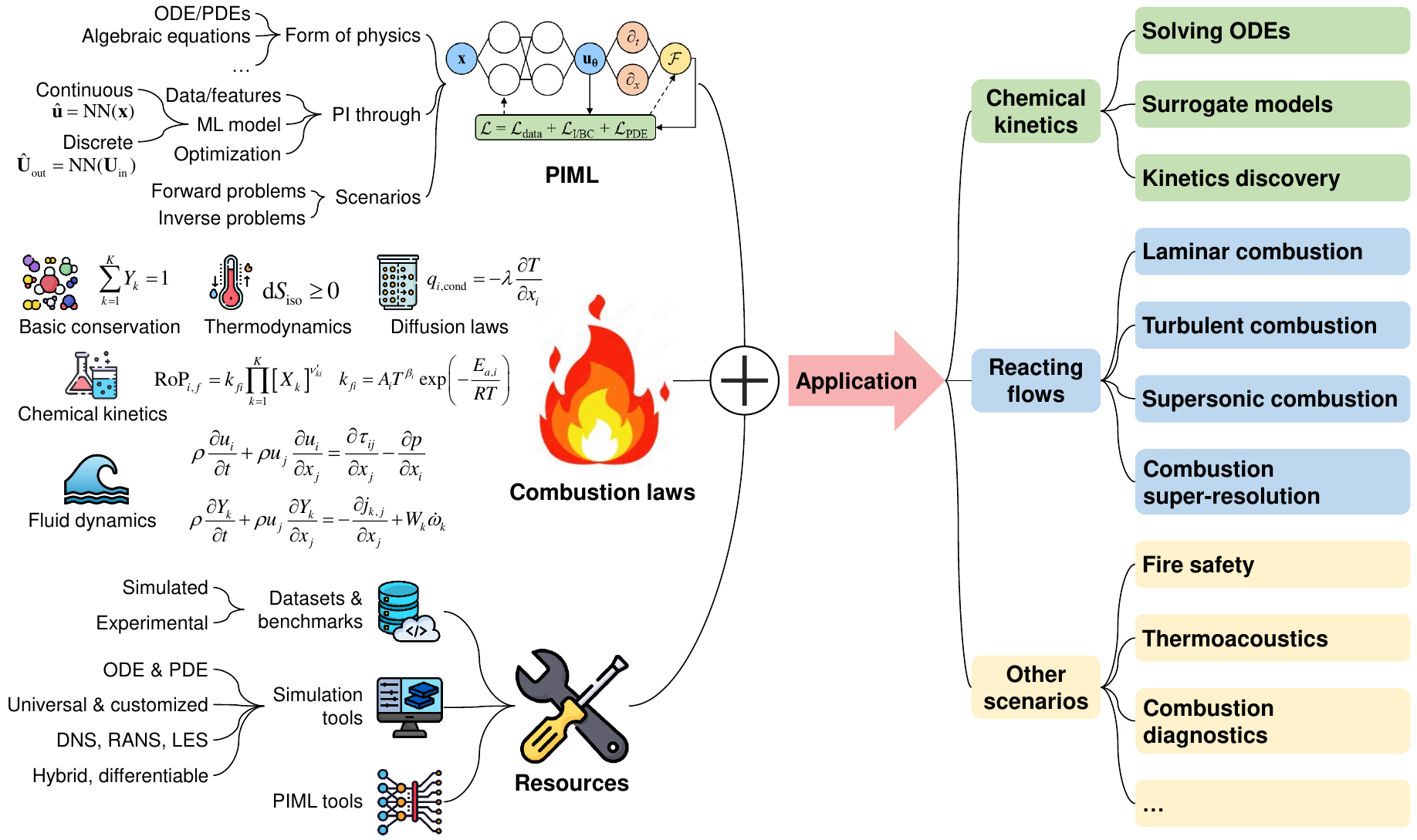}
    \caption{Overview of PIML for combustion. Section \ref{sec: PIML}, Section \ref{sec: combus_laws}, and Section \ref{sec: resource} introduce PIML, combustion laws, and resources, respectively. Sections \ref{sec: PIML_0Dcombus} to \ref{sec: PIML_other-combus} introduce the PIML for combustion chemical kinetics, combustion reacting flows, and other combustion problems, respectively. (Some icons are from Flaticon)}
    \label{fig: overviewPIML4Combus}
\end{figure}

The structure of this paper is shown in Figure \ref{fig: overviewPIML4Combus}. Section \ref{sec: PIML} outlines the fundamental principles, existing reviews, key developments, and related applications of PIML, and also defines the scope and taxonomy adopted in this paper. Section \ref{sec: combus_laws} presents the fundamental laws of combustion commonly used in the literature. Sections \ref{sec: PIML_0Dcombus} to \ref{sec: PIML_other-combus} provide detailed discussions on PIML applications for combustion chemical kinetics, combustion reacting flows, and other combustion-related problems, respectively. Section \ref{sec: resource} compiles publicly available resources, including datasets and benchmarks, simulation tools, and PIML tools. Section \ref{sec: discuss-outlook} offers discussions and future outlooks, while Section \ref{sec: conclusion} concludes the paper.

\section{Physics-informed machine learning}
\label{sec: PIML}

This section will give a detailed introduction to PIML. We first present the fundamentals of PIML, and then provide a review of reviews on PIML, from which the scope of this paper can be drawn. The developments and related applications of PIML are then presented.

\subsection{Fundamentals}
\label{sec: PIML_fundamental}

\begin{figure}[ht]
	\centering
	\includegraphics[width=0.7\textwidth]{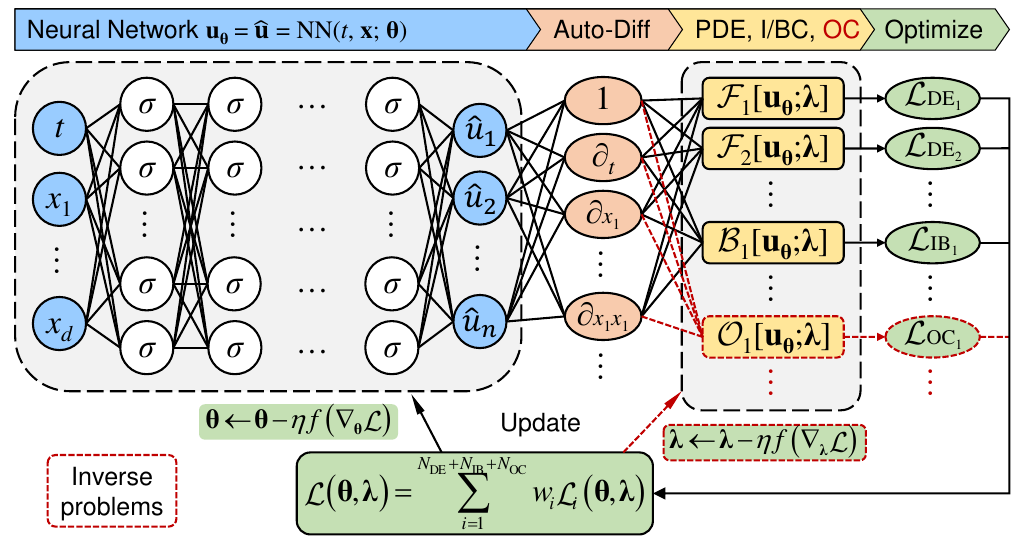}
	\caption{Schematic of PINNs for solving forward and inverse problems of PDEs. The red dashed parts only exist in inverse problems. The blue, green, yellow, and orange parts represent the representation model, optimization process, PDE system, and derivative computation, respectively.}
	\label{fig: PINN}
\end{figure}

This subsection will present the fundamentals of PINNs for solving ODE/PDEs, which is the most basic part of PIML. Figure \ref{fig: PINN} gives a schematic of PINNs for solving forward and inverse problems of PDEs. As mentioned in Section \ref{sec: intro}, forward and inverse problems refer to well-posed PDE problems without observation data and ill-posed PDE problems with possible observation data, respectively. The ill-posed problems include incomplete equations, unknown coefficients, and unknown initial and boundary conditions (IC/BCs). A general form of a PDE problem is given by:
\begin{align}
	\mathcal{F}_i [\mathbf{u}(\mathbf{z}); \bm{\uplambda}] &= 0, \ \  \mathbf{z} \in \Omega, \ \  i=1,...,N_{\mathrm{DE}}
	\label{eq: DE}
	\\
	\mathcal{B}_i [\mathbf{u}(\mathbf{z}); \bm{\uplambda}] &= 0, \ \  \mathbf{z} \in \partial \Omega, \ \  i=1,...,N_{\mathrm{IB}}
	\label{eq: IB}
\end{align}
where $\mathbf{u}$ is the solution, $\mathbf{z} = (t, \mathbf{x}) = (t, x_1,..., x_d)$ is the space-time coordinate vector defined on domain $\Omega$, $\mathcal{F}_i[\cdot]$ is the operator for the $i$-th PDE, and $\mathcal{B}_i[\cdot] $ is the operator for the $i$-th IC/BC. For time-independent PDEs, $\mathbf{z} = \mathbf{x}$. For ODEs, $\mathbf{z} \in \mathbb{R}$, i.e., $\mathbf{z} = t$ or $x$. The operators are parameterized by $\bm{\uplambda}$, such as equation coefficients and boundary values. Note that the initial state is also considered as a part of $\partial \Omega$. For inverse problems, there may be observation conditions (OC):
\begin{align}
	\mathcal{O}_i [\mathbf{u}(\mathbf{z}); \bm{\uplambda}] &= 0, \ \  \mathbf{z} \in \Omega, \ \  i=1,...,N_{\mathrm{OC}}
	\label{eq: OC}
\end{align}
In PINNs, an NN is used to approximate the solution, denoted by
\begin{align}
	\mathbf{u}_{\bm{\uptheta}} (\mathbf{z}) = \widehat{\mathbf{u}} (\mathbf{z}) = \mathrm{NN}(\mathbf{z}; \bm{\uptheta})
	\label{eq: u_theta}
\end{align}
where $\bm{\uptheta}$ is the parameters of the NN. Using AD, the derivatives of $\mathbf{u}_{\bm{\uptheta}}$ to $\mathbf{z}$ can be computed conveniently and then substituted into the operators to get corresponding residuals. The mean squares of the residuals form the loss terms:
\begin{align}
	\mathcal{L}_{\mathrm{DE},i}(\bm{\uptheta}, \bm{\uplambda}) &= \frac{1}{|\mathcal{S}_{\mathrm{DE},i}|} \sum_{\mathbf{z} \in \mathcal{S}_{\mathrm{DE},i}} \left| \mathcal{F}[\mathbf{u}_{\bm{\uptheta}} (\mathbf{z}); \bm{\uplambda}] \right|^2, \ \   \mathcal{S}_{\mathrm{DE},i} \subset \Omega
	\label{eq: loss_DE}
	\\
	\mathcal{L}_{\mathrm{IB},i}(\bm{\uptheta}, \bm{\uplambda}) &= \frac{1}{|\mathcal{S}_{\mathrm{IB},i}|} \sum_{\mathbf{z} \in \mathcal{S}_{\mathrm{IB},i}} \left| \mathcal{B}[\mathbf{u}_{\bm{\uptheta}} (\mathbf{z}); \bm{\uplambda}] \right|^2, \ \   \mathcal{S}_{\mathrm{IB},i} \subset \partial \Omega
	\label{eq: loss_IB}
	\\
	\mathcal{L}_{\mathrm{OC},i}(\bm{\uptheta}, \bm{\uplambda}) &= \frac{1}{|\mathcal{S}_{\mathrm{OC},i}|} \sum_{\mathbf{z} \in \mathcal{S}_{\mathrm{OC},i}} \left| \mathcal{O}[\mathbf{u}_{\bm{\uptheta}} (\mathbf{z}); \bm{\uplambda}] \right|^2, \ \   \mathcal{S}_{\mathrm{OC},i} \subset \Omega
	\label{eq: loss_OC}
\end{align}
where $\mathcal{S}_{\mathrm{DE}}$ and $\mathcal{S}_{\mathrm{IB}}$ are the coordinate point sets sampled on the domain and its boundaries, respectively, and $\mathcal{S}_{\mathrm{OC}}$ is the coordinate point set for observed data. The total loss function is the weighted sum of all loss terms:
\begin{align}
	\mathcal{L}(\bm{\uptheta}, \bm{\uplambda})
	= \sum_{i=1}^{N_{\mathrm{DE}} + N_{\mathrm{IB}} + N_{\mathrm{OC}}} w_i \mathcal{L}_i(\bm{\uptheta}, \bm{\uplambda})
	\label{eq: loss_total}
\end{align}
where $w_i$ is the weight for the $i$-th loss term. The NN parameters are updated by gradient-based optimization algorithms:
\begin{align}
	\bm{\uptheta}_{k+1} = \bm{\uptheta}_k - \eta_k f(\nabla_{\bm{\uptheta}} \mathcal{L}_k)
	\label{eq: update_theta}
\end{align}
where $\eta_k$ is the learning rate at the $k$-th iteration and $f$ is a function related to the optimizer. For inverse problems where $\bm{\uplambda}$ is unknown, it can also be learned by:
\begin{align}
	\bm{\uplambda}_{k+1} = \bm{\uplambda}_k - \eta_k f(\nabla_{\bm{\uplambda}} \mathcal{L}_k)
	\label{eq: update_lambda}
\end{align}
The training will stop after predefined iterations/epochs or when the convergence condition is met.

\subsection{Reviews and scoping}
\label{sec: review-scope}

This subsection will provide a brief review of reviews on PIML and PIML for some related fields. We point out two motivations for this subsection. First, these reviews can provide some important facts and perspectives that support some of the views, scoping, and taxonomy of this paper. Second, the reviews listed in this subsection provide a recommended list for readers interested in PIML itself. Readers who wish to learn more about the progress of PIML can refer to these reviews. We can save space on reviewing PIML itself and focus on PIML for combustion.

We first introduce the reviews on PIML in a chronological sequence. Karniadakis \textit{et al.} \cite{Karniadakis202105_PIML} first provided a comprehensive review of PIML. They pointed out three ways to embed physics into machine learning: observational biases, inductive biases, and learning biases, which correspond to preprocessing the data, designing the model architecture, and modifying the training algorithm, respectively. The classical PINNs belong to the third category, which impose the physical constraints in the loss function, which is a soft manner. Inductive biases is another popular way, with CNNs being the most celebrated example, which respect the invariance in images. Other examples include encoding symmetries, PDEs \cite{Rao202307_PeRCNN, Liu202401_PPNN}, and chemical laws \cite{Ji202101_CRNN} in model architectures. Shukla \textit{et al.} \cite{Shukla202206_scalable-PINN-GNN} reviewed not only the classical PINNs but also how physics-informed learning can be accomplished with graph NNs (GNNs). Cuomo \textit{et al.} \cite{Cuomo202207_SciML-PINN} provided a comprehensive review of PINNs and their variants, covering the NN architectures, learning approaches, theories, applications, and software. Hao \textit{et al.} \cite{Hao202211_PIML-survey} presented a survey on PIML from the perspective of machine learning researchers. The physics priors were categorized into three types: differential equations, symmetry constraints, and intuitive physics. Various approaches to incorporating physical prior into ML were identified: data, model, objective, optimizer, and inference. They categorized the PIML tasks into two classes: using PIML to solve scientific problems and to solve ML problems. For solving scientific problems, the neural simulation and inverse problems were introduced, where the neural simulation were classified into the neural solvers (i.e., PINNs) and neural operators (NOs). For solving ML problems, the PIML for computer vision and reinforcement learning was summarized. Wang \textit{et al.} \cite{Wang202308_PINNGuide} provided an expert's guide to training PINNs, presenting a series of best practices to improve the efficiency and accuracy of PINNs, such as non-dimensionalization, NN architectures, respecting temporal causality, loss balancing, and hard constraints. Faroughi \textit{et al.} \cite{Faroughi202401_PgPiPeNN} presented a review of four NN frameworks: physics-guided NNs, physics-informed NNs, physics-encoded NNs \cite{Rao202307_PeRCNN}, and NOs. The former three frameworks correspond to the observational, learning, and inductive biases in Reference \cite{Karniadakis202105_PIML}, respectively. Kim \textit{et al.} \cite{Kim202402_PINN-Mscale-inverse} reviewed the PINNs on both forward and inverse problems and gave special attention to the multiscale analysis using PINNs. Farea \textit{et al.}  \cite{Farea202408_understandPINN} reviewed the techniques, applications, trends, and challenges of PINNs. The three ways of integrating prior knowledge into PINNs were pointed out: feature engineering, model construction, and regularization, which is consistent with the taxonomy in References \cite{Karniadakis202105_PIML, Faroughi202401_PgPiPeNN}. Zhang \textit{et al.} \cite{Zhang202411_PINN-intelligent} summarized the limitations of PINNs and identified three root causes for their failure: poor multiscale approximation ability and ill-conditioning, weak mathematical rigor, and inadequate integration of physical information. Corresponding future directions and prospects were also provided. More recently, Toscano \textit{et al.} \cite{Toscano202503_PINN2PIKAN} reviewed the recent advances in PIML. They identified three key components of the PIML framework: representation model, governing equation, and optimization process, which correspond to the blue, yellow, and green part of Figure \ref{fig: PINN}, respectively, and is consistent with the taxonomy in Reference \cite{Wu202401_VLT-PINN}. They also reviewed the applications, uncertainty quantification (UQ), theoretical advances, and computational frameworks of PIML, and provided a chronological overview of key advancements in PIML. Meng \textit{et al.} \cite{Meng202505_phy-meet-ML} surveyed a wide variety of recent works in PIML. They presented four categories of physics knowledge in PIML: (1) classical mechanics and energy conservation laws, (2) symmetry, invariant and equivariant functions, (3) numerical methods for PDEs, and (4) Koopman theory. Consistent with References \cite{Karniadakis202105_PIML, Faroughi202401_PgPiPeNN, Farea202408_understandPINN}, they also identified three ways to integrate physics knowledge into ML: data enhancement, NN architecture design, and optimization.

Some reviews focus on the PIML for fluid mechanics \cite{Cai202201_PINN4fluid, Sharma202302_PIML-fluid, Zhao202410_PINN-complex-fluid}, which can provide some insights for PIML for combustion flows. Note that the classification of PIML in Reference \cite{Sharma202302_PIML-fluid} is again consistent with the References \cite{Karniadakis202105_PIML, Faroughi202401_PgPiPeNN, Farea202408_understandPINN, Meng202505_phy-meet-ML}, including the physics-informed features and labels, architectures, and loss functions. However, these reviews rarely include combustion or reacting flows. Recently, Sophiya \textit{et al.} \cite{Sophiya202506_PINN4IGT} reviewed the PINNs in industrial gas turbines (IGTs) research, highlighting their diverse applications in aerodynamics and aeromechanics. However, there are few combustion applications introduced. A more combustion-related review is Reference \cite{Wu202301_PIML-chem-eng}, which focuses on the PIML in multiphysics modeling in chemical engineering. The applications of PIML in momentum transfer/fluid flow, heat transfer, mass transfer, and chemical reaction were first reviewed separately, then the PIML for multiphysics coupling, surrogate model construction, and design optimization were presented. PIML for chemical engineering has also been briefly reviewed in Reference \cite{Toscano202503_PINN2PIKAN}. However, it should be noted that combustion and chemical engineering do not contain each other, but intersect with each other. They both involve chemical reactions but the range of combustion is narrower: a high-temperature exothermic redox chemical reaction by definition. On the other hand, combustion is not limited to chemical engineering, but also involves power generation, transportation, fire safety, and others. Deng \textit{et al.} \cite{Deng202508_SciML-combus} reviewed the SciML in combustion, presenting how SciML can assist: (1) modeling, such as in parameter estimation and reaction mechanism generation, (2) simulation, such as via DL and knowledge-guided ML, and (3) reconstruction, such as for sparse reconstruction and mutual inference. Compared to this review, our review focuses more on PIML, elaborating on how physics and ML can be integrated in diverse ways.

In a broader context, the classical PIML belongs to the category of AI for PDEs (AI4PDEs), which also includes the operator learning. Some researchers have reviewed the AI4PDEs, including DL for PINN and related parameter identification problems \cite{Tanyu202308_DL4PDE-para}, promising directions of AI4PDEs \cite{Brunton202406_dir-ML4PDE}, and AI4PDEs in computational mechanics \cite{Wang202411_AIPDE-compu-mech}. Function learning approximates the mapping between finite-dimensional tensor spaces, while operator learning approximates the mapping between infinite-dimensional function spaces. A representative operator is the solution operator of PDEs, which maps the IC/BCs or parameters (e.g., operating, geometrical, physical) to the solution functions. Function learning can only learn a single PDE solution per training, so the model needs to be retrained once conditions or parameters change. As a progress, operator learning can learn parametric PDE solutions at once, providing a method for rapid PDE-based predictions in practice. When the model of operator learning is an NN, it is referred to as neural operator (NO). Two common NOs are the deep operator network (DeepONet) \cite{Lu202103_DeepONet} and the Fourier neural operator (FNO) \cite{Li202105_FNO}. However, they are both purely data-driven, so their physics-informed versions have also been proposed, i.e., the PI-DeepONet \cite{Wang202109_PI-DeepONet} and PI-FNO \cite{Li202405_PI-FNO}.

Based on these reviews, some conclusions can be drawn. (1) The scope of PIML has been expanded from the classical PINNs to a broader sense. Table \ref{table: narrow-broad} compares the PIML in a narrow and in a broad sense from four aspects: model architecture, ways to inform physics, form of physics, and differentiation. In this paper, both types of PIML will be introduced. Unless otherwise specified, PIML in this paper refers to PIML in a broad sense, while PINNs refer to PIML using NNs as the representation model. Besides, some related terms, such as "physics-guided", "physics-embedded", "physics-encoded", "physics-enforced", "physics-constrained", and "physics-preserved", will all be viewed as "physics-informed" in a broad sense in this paper. (2) Many reviews have consensus on three ways to inform physics \cite{Karniadakis202105_PIML, Faroughi202401_PgPiPeNN, Farea202408_understandPINN, Meng202505_phy-meet-ML, Sharma202302_PIML-fluid}: through data, (ML) model, and optimization, as summarized in Table \ref{table: PIways}. Although the data is obtained by scientific means, ML based solely on physics-informed data is still purely data-driven, so it is a weak mechanism for informing physics, as stated in Reference \cite{Karniadakis202105_PIML}. Therefore, this paper mainly introduces the latter two methods, and the first way will be introduced only if it has a strong embedding. (3) There is a lack of systematic reviews of PIML for combustion, which is the aim of this paper. (4) Many reviews introduce PIML and NOs together. However, most NOs are purely data-driven. This paper will only introduce PIML and physics-informed NOs (PINOs) in detail. By our definition, the latter can also be regarded as a kind of PIML.

\begin{table}[ht]
	\small
	\centering
	\caption{Comparisons between PIML in a narrow and broad sense.}
	\label{table: narrow-broad}
	\begin{tabular}{>{\raggedright\arraybackslash}m{3cm}>{\raggedright\arraybackslash}m{2.5cm}>{\raggedright\arraybackslash}m{9cm}}
		\toprule[1.5pt]
		Aspect & In a narrow sense & In a broad sense \\
		\midrule
		Model architecture & MLP & All models: CNN \cite{Gao202012_PhyGeoNet, Gao202107_PI-CNN}, Transformer \cite{Zhao202405_PINNsFormer}, DeepONet \cite{Wang202109_PI-DeepONet, Kumar202312_PI-DeepONet-reduce, Kumar202507_PI-DeepONet-reduce2, Zanardi202112_PI-DeepONet-noneq, Zanardi202309_adap-PI-DeepONet, Taassob202406_PI-DeepONet-TC-closure, Liu202406_recons-soot, Lu202406_PI-DeepONet-radia}, KAN \cite{Shukla202408_KANvsMLP, Wang202411_KINN}... \\
		How to inform physics & Loss function & All ways: Model architecture \cite{Rao202307_PeRCNN, Liu202401_PPNN, Ji202101_CRNN}, input transformation, output correction \cite{Chen202108_TgHCP, Wu202501_FlamePINN-1D, Wang202503_ANN-hard, Wan202007_CNN-reduce, Wan202007_NN-reduce}, numerical solver \cite{Zhang202208_Mscale-sample, Mao202307_DeepFlame, Wang202503_ANN-hard, Sharma202505_PiMAPNet}... \\
		Form of physics & PDEs & All forms: algebraic equation, inequality, known coefficients \cite{Sorourifar202309_PE-NODE, Fedorov202310_KC-NODE, Kircher202402_GRNN, Kircher202406_GRNN2}... \\
		Differentiation & AD & All methods, such as ND: convolution kernels for space \cite{Gao202012_PhyGeoNet, Gao202107_PI-CNN}, classical schemes for time \cite{Ren202112_PhyCRNet, Rao202307_PeRCNN}... \\
		\bottomrule[1.5pt]
	\end{tabular}
\end{table}

\begin{table}[ht]
    \small
    \centering
    \caption{Three ways to inform physics into ML.}
    \label{table: PIways}
    \begin{tabular}{>{\raggedright\arraybackslash}m{2cm}>{\raggedright\arraybackslash}m{7cm}>{\raggedright\arraybackslash}m{4cm}}
        \toprule[1.5pt]
        Aspect & Sub-aspect & Physics \\
        \midrule
        Data & Generation (simulation or experiment), feature selection, cleansing... & Usually weak \\
        Model & Architecture, input/output transformation, embedded numerical solver... & Usually hard constrained \\
        Optimization & Loss addition, loss weights, multi-stage design... & Usually soft constrained \\
        \bottomrule[1.5pt]
    \end{tabular}
\end{table}

\subsection{Developments}
\label{sec: PIML_develop}

This subsection will briefly introduce the developments of PIML, focusing mainly on PINNs that use ODE/PDEs as physical constraints, as this is the main and classical part of PIML. Most introductions are brief; for detailed introduction, please refer to the reviews in Section \ref{sec: review-scope}. Consistent with the taxonomy of References \cite{Toscano202503_PINN2PIKAN, Wu202401_VLT-PINN}, this paper divides PIML into three components: representation model, optimization process, and PDE system, which will be introduced separately in this subsection. Then some miscellaneous important topics will be introduced.

\subsubsection{Representation model}
\label{sec: PINN_representation}

Representation model is used to represent the solution/field, and corresponds to Equation \eqref{eq: u_theta} and the blue part of Figure \ref{fig: PINN}. Modifying the representation model can be achieved by designing input/output transformations, activation functions, and the whole architecture.

\textbf{Input transformation}. Input transformation, also known as feature embedding, can be denoted by $\mathbf{u}_{\bm{\uptheta}} (\mathbf{z}) = \mathrm{NN}(f_{\mathrm{input}}(\mathbf{z}); \bm{\uptheta})$, where $f_{\mathrm{input}}$ is an input function. It is proven that NNs are more difficult to learn high-frequency features than low-frequency, known as spectral bias or frequency principle \cite{Rahaman201905_spec-bias, Xu202011_F-principle, Xu202409_overview-Fprin-spec-bias}, so multiscale embedding has been proposed to mitigate this issue. In this case, $f_{\mathrm{input}}$ can be linear functions \cite{Liu202011_MscaleDNN} or Fourier functions \cite{Tancik202012_Fourier-let, Wang202106_eigen-bias-Fourier, Wang202405_multi-mag-loss}. Fourier embeddings can also be used to impose hard constraints on periodic BCs \cite{Lu202111_hard-constraint, Dong202106_exact-periBC}.

\textbf{Output transformation}. Output transformation can be denoted by $\mathbf{u}_{\bm{\uptheta}} (\mathbf{z}) = f_{\mathrm{output}}( \mathrm{NN}(\mathbf{z}; \bm{\uptheta}) )$, where $f_{\mathrm{output}}$ is an output function and can also be viewed as the activation function of the output layer. $f_{\mathrm{output}}$ can be used to constrain the range of $\mathbf{u}_{\bm{\uptheta}}$. For example, $f_{\mathrm{output}}$ can be a positive function for temperature and the Sigmoid function for species mass fractions, which should be in the range of $[0,1]$. Another important use of output transformation is to impose hard constraints on BCs \cite{Lu202111_hard-constraint, Sukumar202202_exactBC}, which was adopted as early as the 1990s \cite{Lagaris199809_ANN-O/PDE}. As a simple example, to ensure $u_{\bm{\uptheta}} (x_0) = u_0$, one can let $u_{\bm{\uptheta}} = (x - x_0) \mathrm{NN}(x; \bm{\uptheta}) + u_0$. The theory of functional connections (TFC) can systematically achieve the hard constraints of BCs through output transformation. The model is referred to as Deep-TFC \cite{Leake202103_Deep-TFC} and X-TFC \cite{Schiassi202106_X-TFC} when the free function is a deep NN and an extreme learning machine (ELM), respectively.

\textbf{Activation function}. Activation functions ($\sigma$ in Figure \ref{fig: PINN}) introduce nonlinearity to an NN. In order to implement AD to solve PDEs, the activation functions of PINNs need to have good differentiability. The $\tanh()$ and $\sin()$ are two common choices. Some adaptive activation functions have been proposed to accelerate the PINNs \cite{Jagtap201911_adap-acti, Jagtap202007_adap-acti-recov}.

\textbf{Architecture}. Based on the representation way of the NN, PINNs can be divided into two categories: continuous and discrete PINNs, which are compared in Table \ref{table: continuous-discrete}. Continuous PINNs follow the coordinate-to-variable (query-to-value) mapping way, i.e., $\widehat{\mathbf{u}} = \mathrm{NN} (\mathbf{z})$, while discrete PINNs follow the field-to-field mapping way, i.e., $\widehat{\mathbf{U}}_{\mathrm{out}} = \mathrm{NN} (\mathbf{U}_{\mathrm{in}})$, where $\mathbf{U}_{\mathrm{in}}$ and $\widehat{\mathbf{U}}_{\mathrm{out}}$ are the input and output fields of the NN, respectively. Continuous PINNs provide continuous representations because they can directly predict the field values at any given query coordinates without interpolation, while discrete PINNs provide discrete representations because $\mathbf{U}_{\mathrm{in}}$ and $\widehat{\mathbf{U}}_{\mathrm{out}}$ are always discrete fields on meshes. When $\mathbf{U}_{\mathrm{in}}$ and $\widehat{\mathbf{U}}_{\mathrm{out}}$ are the fields defined on coarse and fine grids, respectively, the task is called super-resolution (SR) and has been applied to combustion (see Section \ref{sec: combus_SR}). When $\mathbf{U}_{\mathrm{in}}$ and $\widehat{\mathbf{U}}_{\mathrm{out}}$ are fields at time $t$ and $t + \Delta t$, respectively, the task is usually referred to as autoregressive (AR) learning \cite{Luo202309_CFDBench, Koehler202410_APEBench}. In this case, the NN can be called a time stepper, next-step predictor, forward propagator \cite{Takamoto202210_PDEBench}, or emulator \cite{Takamoto202210_PDEBench, Koehler202410_APEBench}. On the contrary, the classical continuous PINNs belong to the paradigm of non-autoregressive (NAR) learning \cite{Luo202309_CFDBench}. The AR paradigm has been widely used in surrogate modeling of chemical kinetics. See Section \ref{sec: surr} for details. Note that the AR and NAR learning can also be combined to form a hybrid paradigm \cite{Luo202309_CFDBench, Goswami202312_DeepONet-stiff, Kumar202402_DeepONet-acce}, with DeepONet being a typical method. Instead of using AD in continuous PINNs, discrete PINNs employ numerical differentiation (ND) to compute derivatives. In other words, the PDE residual is $\mathcal{F}[\widehat{\mathbf{U}}_{\mathrm{out}} (\mathbf{U}_{\mathrm{in}})]$ computed on $\mathbf{U}$ using ND, instead of $\mathcal{F}[\widehat{\mathbf{u}} (\mathbf{z})]$ in Equation \eqref{eq: loss_DE} computed on $\widehat{\mathbf{u}}$ and $\mathbf{z}$ using AD. For spatial derivatives, ND can be achieved using differential convolution kernels (similar to the gradient operators in computer vision \cite{Coleman201010_grad-op}), so convolutional NNs (CNNs) are usually adopted. For temporal derivatives, ND can be achieved via classical numerical schemes, such as the Euler scheme. In this case, recurrent NNs (RNN) can be used for modeling the temporal evolution. The choice between continuous and discrete PINNs will be discussed in Section \ref{sec: choices}.

\begin{itemize}
	\item \textbf{\textit{Continuous PINNs}}. The architectures of continuous PINNs mainly include the MLP, ELM \cite{Huang200605_ELM}, radial basis network (RBN) \cite{Ghosh200100_overviewRBFN}, PointNet \cite{Qi201707_PointNet}, DeepONet \cite{Lu202103_DeepONet}, and Kolmogorov-Arnold network (KAN) \cite{Liu202404_KAN, Liu202408_KAN2.0}. Their physics-informed versions have been proposed: PI-ELM \cite{Dwivedi201912_PI-ELM}, PI-RBN \cite{Bai202308_PI-RBN}, PI-PointNet \cite{Kashefi202207_PI-PointNet}, PI-DeepONet \cite{Wang202109_PI-DeepONet}, and PI-KAN \cite{Shukla202408_KANvsMLP, Wang202411_KINN}. Moreover, the basic MLP has been modified in PINNs, including the multiple MLPs \cite{Haghighat202103_inv-surr-solid}, modified MLP \cite{Wang202308_PINNGuide}, and the physics-informed residual adaptive networks (PirateNets) \cite{Wang202412_PirateNet}, which tackles the issue that PINNs have difficulty using very deep NNs. Some generative models, such as the generative adversarial network (GAN) \cite{Goodfellow201406_GAN} and variational autoencoder (VAE) \cite{Kingma201312_VAE}, can also use MLPs as the building blocks and be extended to the PI-GAN \cite{Yang202002_PI-GAN} and PI-VAE \cite{Zhong202210_PI-VAE}.

	\item \textbf{\textit{Discrete PINNs}}. Basic architectures of discrete PINNs include the RNN (represented by the long short-term memory (LSTM) network \cite{Hochreiter199711_LSTM}), CNN \cite{LeCun199811_LeNet, He201606_ResNet}, recurrent CNN (RCNN), GNN \cite{Zhou202104_GNNreview}, and Transformer \cite{Vaswani201706_Transformer}. Some generative models usually use them as the backbone, such as the VAE, GAN, and diffusion model (DM) \cite{Sohl201503_DPM, Ho202006_DDPM, Yang202311_reviewDM}. The basic architectures can also serve as the building blocks of NOs, such as the Fourier neural operator (FNO) \cite{Li202105_FNO}. Most of these NNs have their physics-informed extensions, including the PI-LSTM \cite{Zhang202006_PhyLSTM}, PI-CNN \cite{Gao202012_PhyGeoNet, Gao202107_PI-CNN}, PI-RCNN (or PhyCRNet) \cite{Ren202112_PhyCRNet}, PI-GNN \cite{Gao202201_PI-GNN}, PI-Transformer \cite{Lorsung202402_PI-TT, Zhao202405_PINNsFormer}, PI-DM \cite{Shu202302_PI-DM}, and PI-FNO \cite{Li202405_PI-FNO}. Note that the architectures mentioned in continuous PINNs can also be used in discrete PINNs as long as the mapping is a field-to-field way. The same goes for the generative models, which are architecture-independent. See examples in Section \ref{sec: surr} and Section \ref{sec: combus_SR}. In fact, the discrete time models proposed in the original paper of PINN can be viewed as a discrete PINN on the temporal dimension \cite{Raissi201811_PINN}, which adopted the MLP and Runge-Kutta scheme. However, the architectures just mentioned (such as the CNNs) are rarely applied to continuous PINNs because they typically do not take coordinates as network inputs.
\end{itemize}

\begin{table}[ht]
    \small
    \centering
    \caption{Comparisons between continuous and discrete PINNs. HR and LR represent high-resolution and low-resolution, respectively.}
    \label{table: continuous-discrete}
    \begin{tabular}{>{\raggedright\arraybackslash}m{4cm}>{\raggedright\arraybackslash}m{5.5cm}>{\raggedright\arraybackslash}m{5.5cm}}
        \toprule[1.5pt]
         & Continuous PINNs & Discrete PINNs \\
        \midrule
        Definition & $\widehat{\mathbf{u}} = \mathrm{NN} (\mathbf{z})$ (coordinate-to-variable) & $\widehat{\mathbf{U}}_{\mathrm{out}} = \mathrm{NN} (\mathbf{U}_{\mathrm{in}})$ (field-to-field) \\
        Representation & Continuous (direct query at any points) & Discrete (interpolation at non-grid points) \\
        Differentiation & AD (zero differentiation error, but slower) & ND (faster, but has error) \\
        Learning difficulty & More difficult & Easier \\
        Impose hard constraints & More difficult, especially for PDEs & Easier \\
        Handle large/infinite gradients & Less suitable & More suitable \\
        Handle non-grid data & More suitable & Less suitable \\
        Common architectures & MLP, DeepONet... & CNN, RNN, MLP, DeepONet... \\
        Application & NAR learning, such as solving PDE problems (forward and inverse) & (1) AR learning: $\widehat{\mathbf{U}}_{t + \Delta t} = \mathrm{NN} (\mathbf{U}_{t})$, such as solving PDE problems and surrogate modeling of chemical kinetics; (2) SR: $\widehat{\mathbf{U}}_{\mathrm{HR}} = \mathrm{NN} (\mathbf{U}_{\mathrm{LR}})$ \\
        \bottomrule[1.5pt]
    \end{tabular}
\end{table}

\subsubsection{Optimization process}
\label{sec: PINN_optimization}

Optimization process corresponds to Equations \eqref{eq: loss_DE} to \eqref{eq: update_lambda} and the green part of Figure \ref{fig: PINN}, during which the loss function will be evaluated and minimized. Therefore, ways to improve the optimization process mainly include modifying the loss definition, loss evaluation, and loss minimization.

\textbf{Sampling}. Sampling corresponds to the modification of $\mathcal{S}$ in Equations \eqref{eq: loss_DE} and \eqref{eq: loss_IB}. Basic sampling methods include the uniform, random, and Latin hypercube sampling. To account for the real-time information on the distribution of residuals, thus enhancing the convergence, some adaptive sampling algorithms have been proposed \cite{Lu202102_DeepXDE, Nabian202104_importance-sample, Wu202210_comprehensive-sample}.

\textbf{Loss weights}. Since PINNs belong to the multi-task learning, the imbalance/conflict among different loss terms is a common failure mode in PINNs \cite{Wang202109_grad-flow}. Therefore, the modification of the loss weights, i.e., $w$ in Equation \eqref{eq: loss_total}, is important.
Some examples include designing learning rate annealing \cite{Wang202109_grad-flow}, utilizing the neural tangent kernel (NTK) theory \cite{Wang202110_NTK-PINN}, the self-adaptive loss balanced PINNs (lbPINNs) \cite{Xiang202205_lbPINN}, and the self-adaptive PINNs (SA-PINNs) \cite{McClenny202211_SA-PINN}. For time-dependent problems, it is important to respect the temporal causality, which can be achieved by assigning different weights to the residual points at different time \cite{Wang202402_causalPINN, Penwarden202311_dtPINN}.

\textbf{Loss definition}. Some algorithms add loss terms to Equation \eqref{eq: loss_total} to enhance regularization. Examples are the gradient-enhanced PINNs (gPINNs) \cite{Yu202203_gPINN} and region optimized PINNs (RoPINNs) \cite{Wu202405_RoPINN}. Some algorithms replace the $L^2$ loss in Equations \eqref{eq: loss_DE} to \eqref{eq: loss_OC} with $L^{\infty}$ loss \cite{Wang202206_Linf-PINN} and $L^p$ loss \cite{Wang202405_multi-mag-loss}.

\textbf{Derivative calculation}. Although AD has no differentiation errors, it is time-consuming, especially for high-order derivatives. In contrast, ND is more efficient. ND can not only be used in discrete PINNs (see Section \ref{sec: PINN_representation}), but also in continuous PINNs \cite{Lim202210_PINN-FDM}. Furthermore, AD and ND can be used together to form a coupled-automatic-numerical differentiation (can-PINNs) \cite{Chiu202204_can-PINN}. Although the modifications of derivative calculation directly correspond to the orange part of Figure \ref{fig: PINN}, it will influence the green part (optimization process) through the changes of $\mathcal{F}(\mathbf{u}_{\bm{\uptheta}})$ and $\mathcal{B}(\mathbf{u}_{\bm{\uptheta}})$ in Equations \eqref{eq: loss_DE} and \eqref{eq: loss_IB}.

\textbf{Domain decomposition}. Domain decomposition is important when the solution exhibits distinct behaviours in different domains. For time-dependent problems, the time-marching strategy that decomposes the temporal domain can improve the long-time prediction accuracy. In most cases, separate NNs are assigned to different subdomains and the interface conditions between adjacent subdomains need to be considered. Examples of PINNs using domain decomposition include distributed PINNs (DPINNs) \cite{Dwivedi201907_DPINN}, conservative PINNs (cPINNs) \cite{Jagtap202004_cPINN}, extended PINNs (XPINNs) \cite{Jagtap202011_XPINN}, parareal PINNs (PPINNs) \cite{Meng202007_PPINN}, parallel PINNs \cite{Shukla202109_parallelPINN}, spatiotemporal parallel PINNs (STPINNs) \cite{Xu202306_ST-PINN}, augmented PINNs (APINNs) \cite{Hu202311_APINN}, finite basis PINNs (FBPINNs) \cite{Moseley202307_FBPINN, Dolean202406_multilevel-FBPINN}, and dtPINNs \cite{Penwarden202311_dtPINN}.

\textbf{Optimizer}. The most used optimizers in Equations \eqref{eq: update_theta} and \eqref{eq: update_lambda} are Adam \cite{Kingma201505_Adam} and L-BFGS \cite{Liu198908_L-BFGS}. Some improved optimizers for PINNs include the MultiAdam \cite{Yao202306_MultiAdam} and the NNCG \cite{Rathore202402_loss-landscape}.

\subsubsection{PDE system}
\label{sec: PINN_PDE}

An ODE/PDE system consists of ODE/PDEs and IC/BC/OCs, which corresponds Equations \eqref{eq: DE} to \eqref{eq: OC} and the yellow part of Figure \ref{fig: PINN}. The development of this part includes the formal transformations and scope extensions of the PDE system.

\textbf{Normalization}. Normalization is essential for both numerical simulation and ML, which can mitigate the magnitude difference between different variables, thus stabilizing the convergence. The normalization of PINNs include the normalization of data, PDEs, IC/BC/OCs, and network inputs and outputs, so it contains the topic of input/output transformation in Section \ref{sec: PINN_representation}. Normalization is usually a linear transformation, including the non-dimensionalization \cite{Wang202308_PINNGuide}, standardization \cite{Xu202502_preprocessPINN}, thin-layer normalization \cite{Wu202401_VLT-PINN}, variable scaling \cite{Ko202502_VS-PINN}, and general linear transformation \cite{Wu202401_VLT-PINN}. The unknown physical parameters (i.e., $\bm{\uplambda}$) should also be normalized in inverse problems \cite{Wu202401_VLT-PINN}.

\textbf{Variational form}. The variational form of PDEs is essential for traditional FEM solvers and has been applied to PINNs \cite{Kharazmi201911_VPINN, Kharazmi202012_hp-VPINN}. Similar to the variational form, the energy form of PDEs is important in solid mechanics \cite{Wang202411_AIPDE-compu-mech} and has also been adopted in PINNs \cite{Samaniego202001_DEM, Sheng202012_PFNN, Wang202208_CENN, Wang202409_DCEM}. An advantage of variational form is the reduction of derivative order compared to the original PDEs, thus accelerating the computation.

\textbf{Other formal transformations}. For example, the stream function can be used in 2D incompressible flow problems \cite{Rao202005_PINN-incomp-lami, Wang202109_grad-flow}. In this way, the continuity/incompressibility equation can be exactly satisfied, but higher-order derivatives are introduced. Besides, the common velocity-pressure form of incompressible NS equations can be transformed to the vorticity-velocity form \cite{Jin202011_NSFnet}. For inverse problems with sparse and noisy data, the filtered PDEs can serve as an effective substitute for the original PDEs \cite{Zhang202411_FPDE}.

\textbf{Other DE types}. PINNs have also been applied to non-traditional types of DEs, such as fractional DEs \cite{Pang201908_fPINN} and stochastic DEs \cite{Yang202002_PI-GAN, Zhong202210_PI-VAE}.

\subsubsection{Other important topics}
\label{sec: PINN_other-topic}

There are some other important topics that either do not fall under any of the above topics or may involve more than one of them.

\textbf{Hard constraint}. The classical PINNs impose the physical constraints in a soft manner, so there is still a possibility that physical constraints will be violated if the corresponding residuals do not converge. In contrast, hard constraints can ensure the physical laws are exactly satisfied, thus reducing the number of loss terms in most cases. Hard constraints are also referred to as physics "-embedded", "-encoded", "-enforced", "-preserved" in many papers. For IC/BCs, they can be enforced by input and output transformations as mentioned in Section \ref{sec: PINN_representation}. BCs can also be strictly encoded in discrete PINNs via padding operations \cite{Gao202012_PhyGeoNet, Ren202112_PhyCRNet}. For PDEs, one example is to enforce the continuity equation using the stream function as mentioned in Section \ref{sec: PINN_PDE}. PDEs can also be constrained by linear projection \cite{Chen202108_TgHCP}. For discrete PINNs, PDEs can be encoded using convolutional kernels, such as the physics-encoded RCNN (PeRCNN) \cite{Rao202307_PeRCNN} and the PDE-preserved NN (PPNN) \cite{Liu202401_PPNN}, and can be constrained using traditional numerical methods/solvers \cite{Zhang202208_Mscale-sample, Mao202307_DeepFlame, Wang202503_ANN-hard, Sharma202505_PiMAPNet}. In a broad sense, any explicitly written physical laws that are always satisfied can be regarded as hard constraints. For combustion, the species and element conservations can also be enforced. See Section \ref{sec: surr} and Equation \eqref{eq: hard_species} for details. The laws of chemical kinetics can also be embedded, which will be introduced in Section \ref{sec: NODE}. The choice between soft and hard constraints will be discussed in Section \ref{sec: choices}.

\textbf{Parametric/surrogate modeling}. Parametric/surrogate modeling aims to overcome the limitation of PINNs that can only solve a single PDE at a time. The simplest and most common way is to directly add the parameters to the network inputs, i.e., Equation \eqref{eq: u_theta} becomes $\mathbf{u}_{\bm{\uptheta}} (\mathbf{z}, \bm{\uplambda}) = \mathrm{NN}(\mathbf{z}, \bm{\uplambda}; \bm{\uptheta})$. See some examples in Section \ref{sec: PIML_application}. Other algorithms include utilizing the metalearning \cite{Penwarden202301_meta-PINN-para}, generative pre-trained PINNs (GPT-PINNs) \cite{Chen202310_GPT-PINN}, and the P$^2$INN \cite{Cho202407_P^2INN}. Furthermore, the PINOs mentioned in Section \ref{sec: review-scope} are specifically designed for parametric/surrogate modeling, so they can be regarded as a type of parametric PINNs as well.

\textbf{Ill-conditioning and preconditioning}. Similar to traditional numerical methods, ill-conditioning is also present in PINNs and may also be referred to as pathology \cite{Wang202109_grad-flow} or failure mode \cite{Krishnapriyan202111_failure-mode} in some cases. In a narrow sense, ill-conditioning in PINNs has been attributed to the PDE system/problem or differential operator (i.e., $\mathcal{F}$) itself. Examples in combustion include the stiffness in ODEs of chemical kinetics (see Section \ref{sec: solveODE}), the chaoticity of turbulence, and any other intrinsic complexities. Ill-conditioning in PINNs has been analyzed from different perspectives: a derived operator \cite{DeRyck202310_operator-precond}, loss landscape \cite{Rathore202402_loss-landscape}, error-residual ratio \cite{Liu202402_precondPINN}, and control system \cite{Cao202410_illPINN}, in which different definitions of the condition number have also been proposed to evaluate the ill-conditioning. Corresponding preconditioning strategies include scaling the model parameters (i.e., $\uptheta$) \cite{DeRyck202310_operator-precond}, using second-order optimizers \cite{Rathore202402_loss-landscape}, discretizing the PDE system \cite{Liu202402_precondPINN}, and leveraging the pseudo time-stepping method \cite{Cao202410_illPINN}. The normalization strategies mentioned in Section \ref{sec: PINN_PDE} can also be viewed as a kind of preconditioning for the PDE system. In a broad sense, ill-conditioning in PINNs can also originate from the representation model and optimization process, such as the aforementioned spectral bias of NNs (Section \ref{sec: PINN_representation}) and the issue of loss imbalance/conflict (Section \ref{sec: PINN_optimization}).

\textbf{Uncertainty quantification (UQ)}. UQ is important for reliable modeling of practical systems due to the intrinsic incompleteness of data and model as well as the discrepancies between them \cite{Psaro202301_UQinSciML}. UQ has been applied in PINNs for quantifying both data uncertainty \cite{Yang201905_advUQ, Zhang201907_totalUQ, Yang202010_B-PINN} and model uncertainty \cite{Zou202403_model-misspe}.

\textbf{Theory}. The total error of PINNs can be decomposed into three parts \cite{Shin202010_convergencePINN}: approximation error, estimation error, and optimization error. Many studies have estimated the errors of PINNs, such as for forward problems \cite{Mishra202201_gene-error}, inverse problems \cite{Mishra202106_gene-error-inv}, Kolmogorov PDEs \cite{DeRyck202211_error-Kolmo}, and NS equations \cite{DeRyck202301_error-NS}. The aforementioned NTK theory for PINNs \cite{Wang202110_NTK-PINN} and ill-conditioning in PINNs are also theoretical advances in PIML.

\subsection{Related applications}
\label{sec: PIML_application}

PIML has been widely applied to fluid mechanics, ranging from incompressible flows \cite{Rao202005_PINN-incomp-lami, Jin202011_NSFnet} to supersonic flows \cite{Mao201912_PINN-highspeed, Jagtap202206_PINN-inv-supersonic, Wassing202312_para-Euler}, from laminar flows \cite{Rao202005_PINN-incomp-lami} to turbulent flows \cite{vonSaldern202211_mean-assi, Eivazi202207_PINN-RANS, Pioch202301_turb-model-PINN, Luo202006_para-iden-RANS, Hou202211_kepsNet} and transitional flows \cite{Wu202503_KH-PINN}, from single-phase flows to two-phase flows \cite{Buhendwa202103_infer2phase, Jalili202312_heat-2phase, Qiu202409_paraPINN-2phase}, from internal flows to external flows such as boundary layer flows \cite{Bararnia202202_PINN-BLthermal, Arzani202211_BL-PINN, Wu202401_VLT-PINN, Huang202401_highReBL}, and from analytical and computational fluid mechanics \cite{Jin202011_NSFnet} to experimental fluid mechanics \cite{Cai202103_espresso, Wang202201_dense-vel-recons, Hasanuzzaman202301_enhancePIV, Molnar202301_PI-BOS, Zhu202402_insight-stratified, Cai202405_PINN-PTV, Zhang202411_FPDE}. Compared to forward problems, solving inverse problems using PINNs serves as a new paradigm in fluid mechanics which compensates for the shortcomings of traditional algorithms. One representative method is the hidden fluid mechanics (HFM) \cite{Raissi202002_HFM}, which can reconstruct the velocity and pressure fields based on the observations of concentration. Flow field reconstruction is quite important for experimental fluid mechanics because the data are always sparse, noisy, and partial. For that, PINNs have been widely applied to enhance the experimental data obtained from particle image velocimetry (PIV) \cite{ Wang202201_dense-vel-recons, Hasanuzzaman202301_enhancePIV}, particle tracking velocimetry (PTV) \cite{Wang202201_dense-vel-recons, Cai202405_PINN-PTV}, and other measurements \cite{Cai202103_espresso, Molnar202301_PI-BOS, Zhu202402_insight-stratified, Zhang202411_FPDE}. Inverse problems are also important for turbulence modeling based on Reynolds-averaged Navier-Stokes (RANS) equations and large eddy simulations (LES), where PINNs have been applied to both the task of field reconstruction \cite{vonSaldern202211_mean-assi, Eivazi202207_PINN-RANS, Pioch202301_turb-model-PINN} and parameter inference \cite{Luo202006_para-iden-RANS, Hou202211_kepsNet}. The parametric/surrogate modeling mentioned in Section \ref{sec: PINN_other-topic} has also been applied to fluid mechanics, including the cardiovascular flows \cite{Sun201911_surr-fluid}, Euler equations \cite{Wassing202312_para-Euler}, and flows around airfoils \cite{Cao202407_para-airfoil-invisicd, Cao202501_para-airfoil-RANS, Cao202501_para-airfoil-lami}. For more details of PINNs for fluid mechanics, readers can refer to these reviews \cite{Cai202201_PINN4fluid, Sharma202302_PIML-fluid, Zhao202410_PINN-complex-fluid}. In addition to pure fluid flows, PINNs have also been applied to heat transfer \cite{Cai202104_PINN4heat, Mishra202104_PINN-radi-trans, Zhang202304_paraPINN-thermal}, mass transfer, and multi-species flows \cite{Laubscher202108_PINN-multi-spe}. These are all subprocesses of combustion reacting flows. For a detailed introduction, please refer to this review \cite{Wu202301_PIML-chem-eng}.

\section{Physical laws in combustion}
\label{sec: combus_laws}

Combustion is a typical convection-diffusion-reaction (CDR) process, where reaction is the main feature that distinguishes it from other types of flow. Therefore, many physical laws exist in combustion and can be integrated into PIML, either in a soft or hard manner. This section will present the basic physical laws in combustion, which are summarized from References \cite{Kee2005_react-flow, Poinsot2012_theo-nume-comb, Law2010_comb-phy, web_Chemkin-theory, web_Cantera-science} and are categorized into the laws of multicomponent system, thermodynamics, diffusion "kinetics", chemical kinetics, and fluid dynamics. Although this section is common knowledge for combustion-side researchers, we believe it can serve as an important reference for AI-side researchers.

\subsection{Multicomponent system}
\label{sec: multicomponent}

Consider a gaseous multicomponent system consisting of $K$ species. The molecular weight, mass fraction, mole fraction, and molar concentration of species $k$ are denoted as $W_k$, $Y_k$, $X_k$, and $[X_k]$, respectively. Species conservation is expressed by:
\begin{align}
	\sum_{k=1}^{K} Y_k = 1, \quad \quad \sum_{k=1}^{K} X_k = 1
	\label{eq: species_conservation}
\end{align}
Element conservation demands the total mass or amount of each element $i$ keep constant in a closed system:
\begin{align}
	m_{e,i} &= m \sum_{k=1}^{K} \frac{\alpha_{ik} W_{e,i}}{W_k} Y_k = \mathrm{const}, \quad i = 1,...,N_e
	\label{eq: element_conservation}
	\\
	n_{e,i} &= n \sum_{k=1}^{K} \alpha_{ik} X_k = \mathrm{const}, \quad i = 1,...,N_e
\end{align}
where $m$ denotes the mass, $n$ denotes the amount of substance, the subscript $e$ represents element, and $\alpha_{ik}$ is the number of element $i$ in the chemical formula of species $k$. The mean molecular weight of the mixture is calculated by:
\begin{align}
	W = \sum_{k=1}^{K} X_k W_k = (\sum_{k=1}^{K} Y_k / W_k)^{-1}
\end{align}
and the mean density is given by:
\begin{align}
	\rho = \sum_{k=1}^{K} [X_k] W_k = W \sum_{k=1}^{K} [X_k]
\end{align}
Note that $Y_k$, $X_k$, and $[X_k]$ are dependent:
\begin{align}
	X_k &= \frac{W}{W_k} Y_k
	\\
	[X_k] &= \frac{\rho}{W_k} Y_k
\end{align}

\subsection{Thermodynamics}

The basic thermodynamic properties include the pressure $p$, temperature $T$, volume $V$, entropy $S$, and four energy functions: internal energy $U$, enthalpy $H=U+pV$, Helmholtz free energy $A=U-TS$, and Gibbs free energy $G=H-TS$. The species and their mixture must satisfy a state equation. Take the simplest ideal gas equation for example:
\begin{align}
	pV = nRT \quad \mathrm{or} \quad pW = \rho R T
\end{align}
where $n$ is the amount of substance and $R = 8.314 \mathrm{J}/(\mathrm{mol} \cdot \mathrm{K}) $ is the universal gas constant.

The first law of thermodynamics for a closed system, which is equivalent to the energy conservation, is given by:
\begin{align}
	\mathrm{d}U = \delta Q - \delta W
\end{align}
where $\delta Q $ is the heat supplied to the system from its surroundings and $\delta W$ is the work done by the system. The second law of thermodynamics states that the entropy of an isolated system will not decrease:
\begin{align}
	\mathrm{d}S_{\mathrm{iso}} \geqslant 0
\end{align}
where "$=$" holds only for reversible processes and equilibrium states. For closed systems, many inequalities can be derived: $(\mathrm{d}S)_{U, V} \geqslant 0$, $(\mathrm{d}S)_{H, p} \geqslant 0$, $(\mathrm{d}U)_{S, V} \geqslant 0$, $(\mathrm{d}H)_{S, p} \geqslant 0$, $(\mathrm{d}A)_{T, V} \geqslant 0$, $(\mathrm{d}G)_{T, p} \geqslant 0$, where the last one is always used for chemical equilibrium. Based on the first and second laws, the basic differential relations of thermodynamics can be derived: $\mathrm{d}U = T\mathrm{d}S - p\mathrm{d}V$, $\mathrm{d}H = T\mathrm{d}S + V\mathrm{d}p$, $\mathrm{d}A = -S\mathrm{d}T - p\mathrm{d}V$, and $\mathrm{d}G = -S\mathrm{d}T + V\mathrm{d}p$,
which hold for systems with constant compositions.

In combustion simulations, the temperature-dependent relations of thermodynamic properties are needed, which are often provided for the heat capacity $C_p$, $H$, and $S$ at standard pressure $p^{\circ}$: $C_{p,k}^{\circ}(T)$, $H_k^{\circ}(T)$, and $S_k^{\circ}(T)$, where the superscript $\circ$ denotes standard pressure. The relations must satisfy the definition of $C_p$ and the aforementioned differential relation of $H$: 
\begin{align}
	C_{p,k}^{\circ} &= \frac{\mathrm{d}H_k^{\circ}}{\mathrm{d}T} = T\frac{\mathrm{d}S_k^{\circ}}{\mathrm{d}T}
\end{align}
The relations are often in the form of polynomials of $T$, such as the NASA 7-coefficient polynomials. At pressure $p$, the corresponding properties are: $C_{p,k} = C_{p,k}^{\circ}$, $H_k = H_k^{\circ}$, and $S_k = S_k^{\circ} - R\ln(p/p^{\circ})$. Compared to mole-based properties, mass-based properties are more commonly used: $c_{p,k} = C_{p,k}/W_k$, $h_k = H_k/W_k$, $s_k = S_k/W_k$. The mean thermodynamic properties can be calculated by \cite{web_Chemkin-theory}:
\begin{align}
	&C_p = \sum_{k=1}^K C_{p,k} X_k, \quad c_p = \frac{C_p}{W} = \sum_{k=1}^K c_{p,k} Y_k
	\\
	&H = \sum_{k=1}^K H_k X_k, \quad h = \frac{H}{W} = \sum_{k=1}^K h_k Y_k
	\\
	&S = \sum_{k=1}^K (S_k - R\ln X_k) X_k, \quad s = \frac{S}{W}
\end{align}
Other properties such as $U$, $A$, and $G$ can be obtained by their definitions.

\subsection{Diffusion "kinetics"}

Combustion involves the basic transport phenomena, including the diffusion of momentum, heat, and mass. The corresponding transport properties are the viscosity $\mu_k$, thermal conductivity $\lambda_k$, and binary diffusion coefficients $D_{kj}$ between species $k$ and $j$. These properties can be calculated either by theories such as the molecular theory or by temperature-dependent relations \cite{Kee2005_react-flow, web_Chemkin-theory, web_Cantera-science}. Diffusion laws describe the relationship between momentum flux (i.e., viscous tensor) $\bm{\uptau}$, energy flux $\mathbf{q}$, and species mass flux $\mathbf{j}_k$ with respect to their driving forces. The momentum flux is given by the definition of Newtonian fluid:
\begin{align}
	\tau_{ij} = \mu \left(\frac{\partial u_i}{\partial x_j} + \frac{\partial u_j}{\partial x_i} - \frac{2}{3} \frac{\partial u_k}{\partial x_k} \delta_{ij} \right)
\end{align}
where $u$ is the velocity and the subscripts $i$, $j$ denote the spatial dimensions. The energy flux is given by:
\begin{align}
	q_i = -\lambda \frac{\partial T}{\partial x_i} + \sum_{k=1}^K j_{k,i} h_k - \sum_{k=1}^K \frac{RT}{W_k X_k} D_k^T d_{k,i}
\end{align}
where the three terms represent the contributions of temperature gradient (Fourier's law), species diffusion, and Dufour effect, respectively. $d_{k,i}$ is the diffusion driving force, and $D_k^T$ is the thermal diffusion coefficient. $d_{k,i}$ and $j_{k,i}$ are given by:
\begin{align}
	d_{k,i} &= \frac{\partial X_k}{\partial x_i} + (X_k - Y_k) \frac{1}{p} \frac{\partial p}{\partial x_i}
	\\
	j_{k,i} &= \rho Y_k V_{k,i}
	\\
	V_{k,i} &= \frac{1}{X_k W} \sum_{j=1}^K W_j D_{kj} d_{j,i} - \frac{D_k^T}{\rho Y_k} \frac{1}{T} \frac{\partial T}{\partial x_i}
\end{align}
where $V_{k,i}$ is the diffusion velocity of species $k$ on dimension $i$ and its second term represents the Soret effect.

The mean properties can be calculated either by the multicomponent method or by mixture-averaged approximations. In the latter case, for example, the mixture-averaged properties are calculated by:
\begin{align}
	\mu &= \sum_{k=1}^K \frac{\mu_k X_k}{\sum_{j=1}^K \Phi_{kj} X_j}
	\\
	\Phi_{kj} &= \frac{1}{\sqrt{8}} \left(1 + \frac{W_k}{W_j} \right)^{-\frac{1}{2}} \left[1 + \left(\frac{\mu_k}{\mu_j} \right)^{\frac{1}{2}} \left(\frac{W_j}{W_k} \right)^{\frac{1}{4}} \right]^2
	\\
	\lambda &= \frac{1}{2} \left( \sum_{k=1}^K X_k \lambda_k + \frac{1}{\sum_{k=1}^K X_k / \lambda_k} \right)
	\\
	D_{km} &= \frac{1 - Y_k}{\sum_{j \neq k}^K X_j / D_{jk}}
\end{align}
where $D_{km}$ is the mean diffusion coefficient of species $k$ to other species. In an isobaric process, the diffusion velocity can be simplified to:
\begin{align}
	V_{k,i} = -\frac{1}{X_k} D_{km} \frac{\partial X_k}{\partial x_i} - \frac{D_k^T}{\rho Y_k} \frac{1}{T} \frac{\partial T}{\partial x_i}
\end{align}
where the first term represents the Fick's law. Note that the mixture-averaged approximations violate the mass flux conservation:
\begin{align}
	\sum_{k=1}^{K} \mathbf{j}_k = \rho \sum_{k=1}^{K} Y_k \mathbf{V}_k = 0
\end{align}
A common trick to address this is to further correct $\mathbf{j}_k$ by $\mathbf{j}_k \leftarrow \mathbf{j}_k - Y_k \sum_{k=1}^{K} \mathbf{j}_k$.

\subsection{Chemical kinetics}
\label{sec: chemical}

Here only gas-phase chemical reactions are considered. Practical reactions consist of multiple elementary reactions. A reaction system involving $K$ species and $I$ reactions can be represented as:
\begin{align}
	\sum_{k=1}^{K} \nu'_{ki} \chi_k \Leftrightarrow \sum_{k=1}^{K} \nu''_{ki} \chi_k, \quad i = 1,...,I
\end{align}
where $\chi_k$ is the symbol of species $k$, $\nu'$ and $\nu''$ indicate forward and reverse stoichiometric coefficients, respectively. The net stoichiometric coefficient is defined as:
\begin{align}
	\nu_{ki} = \nu''_{ki} - \nu'_{ki}
\end{align}
The rate-of-progress variable of reaction $i$ (unit: mol/(m$^3\cdot$s)) can be computed by the law of mass action:
\begin{align}
	\mathrm{RoP}_i = k_{fi} \prod_{k=1}^{K} [X_k]^{\nu'_{ki}} - k_{ri} \prod_{k=1}^{K} [X_k]^{\nu''_{ki}}
	\label{eq: RoP}
\end{align}
where $k_{fi}$ and $k_{ri}$ are forward and reverse rate constants of reaction $i$. Then the production rate of species $k$ is:
\begin{align}
	\dot{\omega}_k = \sum_{i=1}^{I} \nu_{ki} \mathrm{RoP}_i
	\label{eq: omega_dot}
\end{align}
The mass conservation of chemical reactions is written as:
\begin{align}
	\sum_{k=1}^{K} \nu_{ki} W_k = 0, \quad i = 1,...,I
\end{align}
so that
\begin{align}
	\sum_{k=1}^{K} W_k \dot{\omega}_k = \sum_{i=1}^{I} \left( \mathrm{RoP}_i \sum_{k=1}^{K} \nu_{ki} W_k \right) = 0
\end{align}

The forward rate constants are usually computed by the Arrhenius law:
\begin{align}
	k_{fi} = A_i T^{\beta_i} \exp \left( -\frac{E_{a,i}}{RT} \right)
	\label{eq: k_fi}
\end{align}
where $A$, $\beta$, and $E_a$ are the pre-exponential factor, temperature exponent, and activation energy, respectively. The reverse rate constants are computed by chemical equilibrium:
\begin{align}
	k_{ri} &= k_{fi} / K_{c,i}
	\label{eq: k_ri}
	\\
	K_{c,i} &= K_{p,i} \left( \frac{p^{\circ}}{RT} \right) ^ {\sum_{k=1}^{K} \nu_{ki}}
	\\
	K_{p,i} &= \exp \left( -\frac{\Delta G_i^{\circ}}{RT} \right) = \exp \left( \frac{\Delta S_i^{\circ}}{T} - \frac{\Delta H_i^{\circ}}{RT} \right)
	\\
	\Delta S_i^{\circ} &= \sum_{k=1}^{K} \nu_{ki} S_k^{\circ}
	\\
	\Delta H_i^{\circ} &= \sum_{k=1}^{K} \nu_{ki} H_k^{\circ}
\end{align}
where $K_c$ and $K_p$ are concentration-based and pressure-based equilibrium constants, respectively.

The above discussions are the most basic models. In practical reaction mechanisms, some exceptions exist, such as specified reaction order, three-body reactions, pressure-dependent reactions, and irreversible reactions \cite{Kee2005_react-flow, web_Chemkin-theory, web_Cantera-science}. In these cases, Equations \eqref{eq: RoP} and \eqref{eq: k_fi} may be modified. For irreversible reactions (i.e., "$\Leftrightarrow$" become "$\Rightarrow$"), $k_{ri} = 0$ instead of Equation \eqref{eq: k_ri}.

\subsection{Fluid dynamics}
\label{sec: fluid-dynamics}

The governing equations for reacting flows include the conservations of mass, momentum, energy, and species \cite{Poinsot2012_theo-nume-comb, Law2010_comb-phy}:
\begin{align}
	\frac{\partial \rho}{\partial t} + \frac{\partial \rho u_i}{\partial x_i} &= 0
	\label{eq: mass}
	\\
	\rho \frac{\partial u_i}{\partial t} + \rho u_j \frac{\partial u_i}{\partial x_j} &= \frac{\partial \tau_{ij}}{\partial x_j} - \frac{\partial p}{\partial x_i}
	\\
	\rho \frac{\partial e}{\partial t} + \rho u_j \frac{\partial e}{\partial x_j} &= -\frac{\partial q_j}{\partial x_j} - (\tau_{ij} - p \delta_{ij}) \frac{\partial u_i}{\partial x_j}
	\\
	\rho \frac{\partial Y_k}{\partial t} + \rho u_j \frac{\partial Y_k}{\partial x_j} &= - \frac{\partial j_{k,j}}{\partial x_j} + W_k \dot{\omega}_k, \quad k = 1,...,K
	\label{eq: Y}
\end{align}
where the Einstein summation convention is adopted and $e$ is the energy, containing both sensible and chemical energy. Note that the external forces and heat sources are not considered. The energy equation is usually rewritten as the equation of $T$ \cite{Poinsot2012_theo-nume-comb}:
\begin{equation}
\begin{aligned}
	\rho c_p \frac{\partial T}{\partial t} + \rho c_p u_j \frac{\partial T}{\partial x_j}
	=& \frac{\partial}{\partial x_j} \left( \lambda \frac{\partial T}{\partial x_j} \right)
	- \left( \sum_{k=1}^{K} j_{k,j} c_{p,k} \right) \frac{\partial T}{\partial x_j}
	\\
	&- \sum_{k=1}^{K} h_k W_k \dot{\omega}_k
	+ \left( \frac{\partial p}{\partial t} + u_j \frac{\partial p}{\partial x_j} \right)
	+ \tau_{ij}\frac{\partial u_i}{\partial x_j}
	\label{eq: T}
\end{aligned}
\end{equation}
where the five terms of the RHS represent the contributions of thermal conduction, species diffusion, reaction heat release, pressure change, and viscous heating, respectively.

Equations \eqref{eq: mass} to \eqref{eq: T} are the unsimplified governing equations for reacting flows and are used for direct numerical simulation (DNS). To save computational resources, simplified models are used in many cases. For example, the concepts of reaction progress variable ($c$) and mixture fraction ($Z$) can be derived in the contexts of premixed and non-premixed combustion, respectively \cite{Poinsot2012_theo-nume-comb}. Under certain assumptions, their governing equations can also be derived, which are fewer than the original equations. For turbulent combustion, the governing equations for Reynolds-averaged Navier-Stokes simulation (RANS) and large eddy simulation (LES) are also different from Equations \eqref{eq: mass} to \eqref{eq: T}. For two types of turbulent combustion models, namely the transport probability density function (TPDF) method \cite{Pope198500_TPDF-reactive} and the flamelet-like method \cite{Peters198400_flamelet, Zhang202304_SOTA-flamelet}, including the flamelet generated manifolds (FGM) \cite{vanOijen200006_FGM, vanOijen201610_SOTA-FGM} and flamelet/progress variable (FPV) method \cite{Pierce2001_FPV, Pierce200404_FPV}, the corresponding governing equations are also different from Equations \eqref{eq: mass} to \eqref{eq: T}. For the sake of brevity, these simplified methods are not described in detail here. For details, readers can refer to the corresponding references.

\section{PIML for combustion chemical kinetics (0D combustion)}
\label{sec: PIML_0Dcombus}

Combustion chemical kinetics can be regarded as a kind of 0D combustion, because it mainly involves combustion state variables and their time derivatives and does not involve spatial processes such as flow and diffusion. Chemical reactions are probably the most central feature of combustion, so there is a lot of research on PIML for the chemical kinetics of combustion, which will be presented in this section. Based on the research status, the introduction in this section is not limited to PIML in the narrow sense and chemical kinetics in combustion. On the one hand, these studies not only inform physics through loss functions, but also in other ways, such as network structures. As mentioned in Section \ref{sec: PIML_develop}, these physics-guided, physics-embedded, and physics-constrained ML can be regarded as PIML in a broad sense, so the related work will also be introduced in this section. On the other hand, chemical kinetics is not only present in combustion but also in other chemical processes. The studies of PIML for general chemical kinetics is instructive and can even be directly applied to combustion chemical kinetics, so some of the related work is also introduced in this section. Some representative studies on PIML for combustion chemical kinetics are listed in Table \ref{table: chemical-kinetics}, which can be divided into three categories: solving ODEs of chemical kinetics, surrogate modeling of chemical kinetics, and applications of neural ODEs to surrogate modeling and kinetics discovery, which are detailed in the next three subsections, respectively.

\begin{table}
	\small
	\centering
	\caption{Recent studies on PIML for combustion and combustion-like chemical kinetics. $\mathcal{C}_{\mathrm{ele}}$, $\mathcal{C}_{\mathrm{spe}}$, and $\mathcal{C}_{\mathrm{ene}}$ represent the conservations of elements (atoms), species (total mass), and energy, respectively. PVG and MA mean progress variable generation and manifold approximation, respectively.}
	\label{table: chemical-kinetics}
	\begin{tabular}{>{\raggedright\arraybackslash}m{3cm}>{\raggedright\arraybackslash}m{2cm}>{\raggedright\arraybackslash}m{4cm}>{\raggedright\arraybackslash}m{4.5cm}>{\raggedright\arraybackslash}m{0.8cm}}
		\toprule[1.5pt]
		Study & Method abbreviation & Aim & How to inform/embed physics & Codes \\
		\midrule
		Ji \cite{Ji202108_stiff-PINN}, 2021.8 & stiff-PINN & Solve stiff ODEs & Loss of ODEs & \href{https://github.com/DENG-MIT/Stiff-PINN}{\underline{URL}} \\
		De Florio \cite{DeFlorio202206_stiff-X-TFC}, 2022.6 & -- & Solve stiff ODEs & Loss of ODEs & -- \\
		Wang \cite{Wang202207_ROBER_Z-eq}, 2022.7 & -- & Solve stiff ODEs & Loss of ODEs & -- \\
		Galaris \cite{Galaris202108_PI-RPNN, Fabiani202304_parsi-PI-RPNN}, 2021.8 & PI-RPNN & Solve stiff ODEs & Loss of ODEs & \href{https://github.com/GianlucaFabiani/RPNN_for_Stiff_ODEs}{\underline{URL}} \\
		Li \cite{Li202110_PINN-coal}, 2021.10 & -- & Solve ODEs & Loss of ODEs & -- \\
		Weng \cite{Weng202211_MscalePINN-stiff}, 2022.11 & M-PINN & Solve stiff ODEs (inverse) & Loss of ODEs & -- \\
		Gusmao \cite{Gusmao202204_kINN}, 2022.4 & KINN & Solve ODEs (forward and inverse) & Loss of ODEs, $\mathcal{C}_{\mathrm{spe}}$ by trigonometric transformation & \href{https://github.com/gusmaogabriels/kinn/tree/paper_reg}{\underline{URL}} \\
		Bibeau \cite{Bibeau202312_PINN-biodiesel}, 2023.12 & -- & Solve inverse ODEs (parameter inference) & Loss of ODEs & \href{https://github.com/chaos-polymtl/bio-pinn}{\underline{URL}} \\
		\midrule
		Almeldein \cite{Almeldein202307_acce-chem}, 2023.7 & -- & Surrogate model & Loss of $\mathcal{C}_{\mathrm{ele}}$ & -- \\
		Zhang \cite{Zhang202408_CRK-PINN}, 2024.8 & CRK-PINN & Surrogate model & Losses of ODEs, $\mathcal{C}_{\mathrm{ele}}$, $\mathcal{C}_{\mathrm{spe}}$, and $\mathcal{C}_{\mathrm{ene}}$ & -- \\
		Wan \cite{Wan202007_CNN-reduce}, 2020.7 & -- & Surrogate model of reduced chemistry & $\mathcal{C}_{\mathrm{ele}}$ by correction & -- \\
		Wang \cite{Wang202503_ANN-hard}, 2025.3 & ANN-hard & Surrogate model & $\mathcal{C}_{\mathrm{ele}}$ by correction, $\mathcal{C}_{\mathrm{ene}}$ by using equations & \href{https://github.com/tianhanz/DNN-Models-for-Chemical-Kinetics}{\underline{Dataset}} \\
		Kumar \cite{Kumar202312_PI-DeepONet-reduce, Kumar202507_PI-DeepONet-reduce2}, 2023.12 & -- & Surrogate model & Losses of $\mathcal{C}_{\mathrm{ele}}$ and $\mathcal{C}_{\mathrm{spe}}$ & -- \\
		Zanardi \cite{Zanardi202112_PI-DeepONet-noneq}, 2021.12 & -- & Surrogate model & Loss of ODEs & -- \\
		Zanardi \cite{Zanardi202309_adap-PI-DeepONet}, 2023.9 & CG-DeepONet & Surrogate model & Loss of ODEs, Boltzmann layer, hierarchical architecture & -- \\
		Weng \cite{Weng202411_EFNO-stiff}, 2024.11 & EFNO & Surrogate model & Losses of $\mathcal{C}_{\mathrm{ele}}$ and $\mathcal{C}_{\mathrm{spe}}$ & -- \\
		Readshaw or Ding \cite{Readshaw202103_LES-PDF, Ding202105_MLtab-hybrid, Ding202210_MLtab-blend}, 2021.3 & HFRD-MMP & Surrogate model & $\mathcal{C}_{\mathrm{spe}}$ by normalization & -- \\
		Readshaw \cite{Readshaw202304_MLtab-conserve}, 2023.4 & -- & Surrogate model & $\mathcal{C}_{\mathrm{ele}}$ by recalculating residual species & -- \\
		Salunkhe \cite{Salunkhe202206_ChemTab, Salunkhe202303_ChemTab-extend}, 2022.6 & ChemTab & Surrogate model (PVG and MA) & Constraints on the projection matrix and progress variables & -- \\
		\midrule
		Vijayarangan \cite{Vijayarangan202311_dyn-informed-NODE}, 2023.11 & -- & Surrogate model by NODE & Dynamics-informed discovery of latent space & -- \\
		Kumar \cite{Kumar202311_PC-NODE, Kumar202503_PC-NODE2}, 2023.11 & PC-NODE & Surrogate model by NODE & Loss of $\mathcal{C}_{\mathrm{ele}}$ & -- \\
		Kumar \cite{Kumar202412_Phy-ChemNODE}, 2024.12 & Phy-ChemNODE & Surrogate model by NODE & Dynamics-informed discovery of latent space, loss of $\mathcal{C}_{\mathrm{ele}}$ & -- \\
        Koenig \cite{Koenig202507_ChemKAN}, 2025.7 & ChemKAN & Surrogate model by NODE & Architecture that mimics the chemical kinetics, loss of $\mathcal{C}_{\mathrm{ele}}$ & -- \\
		Sorourifar \cite{Sorourifar202309_PE-NODE}, 2023.9 & r-NODE, k-NODE & Surrogate model by NODE & Embedded stoichiometric and rate information & -- \\
		Fedorov \cite{Fedorov202310_KC-NODE}, 2023.10 & KC-NODE & Surrogate model by NODE & Embedded stoichiometry and rate expression & \href{https://github.com/LIKAT-Rostock/kcnode-paper}{\underline{URL}} \\
		Kircher \cite{Kircher202402_GRNN, Kircher202406_GRNN2}, 2024.2 & GRNN & Surrogate model by NODE & Embedded stoichiometry and thermodynamics & -- \\
		\midrule
		Ji \cite{Ji202101_CRNN}, 2021.1 & CRNN & Discover kinetics & Embedded rate expression & \href{https://github.com/DENG-MIT/CRNN}{\underline{URL}} \\
		Su \cite{Su202303_NODE-para-opt}, 2023.3 & -- & Learn kinetics parameters & Embedded rate expression & -- \\
		Doeppel \cite{Doeppel202407_AC-CRNN}, 2024.7 & AC-CRNN & Discover kinetics & Embedded rate expression, $\mathcal{C}_{\mathrm{ele}}$ by a NN layer & -- \\
		\bottomrule[1.5pt]
	\end{tabular}
\end{table}

\subsection{Solving ODEs of chemical kinetics}
\label{sec: solveODE}

As shown in Equations \eqref{eq: mass} to \eqref{eq: T}, the governing PDEs become ODEs of chemical kinetics when the spatial gradients are all zero, and the reaction source terms become the only driving force for the combustion system. The ODEs are given by:
\begin{align}
	\frac{\mathrm{d} T}{\mathrm{d} t} &= -\frac{1}{\rho c_p} \sum_{k=1}^{K} h_k W_k \dot{\omega}_k
	\\
	\frac{\mathrm{d} Y_k}{\mathrm{d} t} &= \frac{1}{\rho} W_k \dot{\omega}_k, \quad k = 1,...,K
\end{align}
An general ODE system can be written as:
\begin{align}
	&\frac{\mathrm{d} \mathbf{y}}{\mathrm{d} t} = \mathbf{f}(\mathbf{y}, t), \quad t \in [0, T]
	\\
	&\mathbf{y}(t=0) = \mathbf{y}_0
\end{align}
which is an initial value problem (IVP) and $\mathbf{y} = [y_1,...,y_N]^T$ is the vector of dependent variables. One of the most challenging aspects of solving ODEs is the stiffness, where an ODE system is characterized by dynamics with widely separated time scales, which is also common in combustion chemical kinetic problems. A commonly used indicator to quantify the stiffness is the stiffness ratio. Consider a linearized ODE system: $\mathrm{d} \mathbf{y}/\mathrm{d} t = \mathbf{J} \mathbf{y}$, where $\mathbf{J}$ is the Jacobian matrix, then the stiffness ratio is defined as:
\begin{equation}
	S = \frac{\max_j \{|\Re(\lambda_j)|\}}{\min_j \{|\Re(\lambda_j)|\}} = \frac{\tau_{\mathrm{slowest}}}{\tau_{\mathrm{fastest}}}
\end{equation}
where $\lambda_j \in \mathbb{C}$ are the eigenvalues of $\mathbf{J}$. The reciprocals of the eigenvalues can represent the time scales, $\tau$, of the system \cite{Vijayarangan202311_dyn-informed-NODE}. To validate the algorithms (traditional and PINNs) for solving stiff ODEs, some classical problems are widely tested, including the ROBER and POLLU problems, which consist of 3 species and 3 reactions, and 20 species and 25 reactions, respectively. As an example, the ROBER problem describes the following reaction:
\begin{align}
	A &\xrightarrow{k_1}  B
	\\
	B + B &\xrightarrow{k_2}  C + B
	\\
	B + C &\xrightarrow{k_3}  A + C
\end{align}
So the evolution of the species concentrations are described by the following ODEs:
\begin{align}
	&\frac{\mathrm{d} y_1}{\mathrm{d} t} = -k_1 y_1 + k_3 y_2 y_3
	\\
	&\frac{\mathrm{d} y_2}{\mathrm{d} t} = k_1 y_1 - k_2 y_2^2 - k_3 y_2 y_3
	\\
	&\frac{\mathrm{d} y_3}{\mathrm{d} t} = k_2 y_2^2
\end{align}
where $y_1$, $y_2$, and $y_3$ are the concentrations of $A$, $B$, and $C$, respectively. The default IC is $\mathbf{y}(0) = [1, 0, 0]^T$. The reaction rate constants are $k_1 = 0.04$, $k_1 = 3 \times 10^7$, and $k_3 = 10^4$, which vary in a range of nine orders of magnitude, thus resulting in the strong stiffness of the system.

\begin{figure}[ht]
    \centering
    \includegraphics[width=1\textwidth]{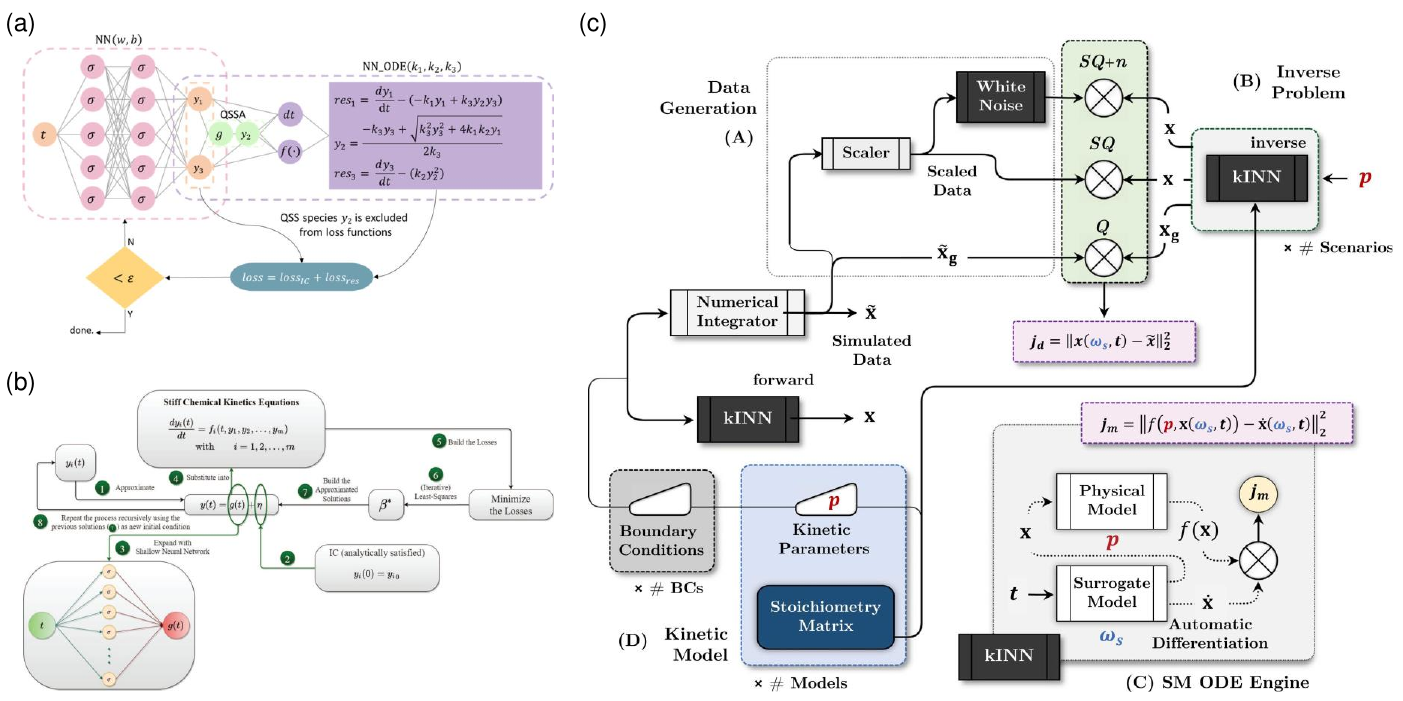}
    \caption{PIML for solving ODEs of chemical kinetics. (a) Stiff-PINN for the ROBER problem \cite{Ji202108_stiff-PINN}. (b) Using the X-TFC algorithm to solve stiff chemical kinetics equations \cite{DeFlorio202206_stiff-X-TFC}. (c) Kinetics-informed neural networks \cite{Gusmao202204_kINN}. Figures are reprinted with permission.}
    \label{fig: chemical_ODE}
\end{figure}

Figure \ref{fig: chemical_ODE} shows some studies on PIML for solving ODEs of chemical kinetics. Ji \textit{et al.} \cite{Ji202108_stiff-PINN} first proposed the stiff-PINNs, using PINNs to solve stiff chemical kinetic ODEs, where the quasi-steady-state assumption (QSSA) was employed to reduce the stiffness of the system. In QSSA, QSS species are identified by assuming that their net production rates are negligible compared to their consumption and production rates, then the ODEs of QSS species become differential-algebraic equations (DAEs). For example, $B$ can be chosen as the QSS species in the ROBER problem, so its algebraic equation can be obtained by assuming $\mathrm{d} y_2 / \mathrm{d} t = 0$. Because QSS species are usually radicals and unstable intermediates with short time scales, the elimination of their ODEs can greatly reduce the stiffness. The stiff-PINNs were used to solve the ROBER and POLLU problems with one and ten QSS species, respectively, and yielded satisfactory results. De Florio \textit{et al.} \cite{DeFlorio202206_stiff-X-TFC} used X-TFC \cite{Schiassi202106_X-TFC} (see Section \ref{sec: PINN_optimization}) to solve stiff chemical kinetic ODEs, where the TFC can guarantee the ICs are exactly satisfied, as illustrated in Figure \ref{fig: chemical_ODE}(b). The temporal domain was decomposed into several intervals to deal with the long-time issue. The X-TFC showed great performance for 4 tested problems, and even outperformed the traditional numerical methods in some cases. Wang \textit{et al.} \cite{Wang202207_ROBER_Z-eq} used PINNs to solve the ROBER problem. Hard constraints on ICs via logarithmic functions were employed, yielding accurate results. Galaris \textit{et al.} \cite{Galaris202108_PI-RPNN} proposed the physics-informed random projection NNs (PI-RPNNs) to solve stiff ODEs, where the single-layer NN had fixed weights between the input and hidden layer so only the parameters from the hidden to the output layer needed to be optimized. The loss functions were minimized by the Gauss-Newton method and the hard constraints on ICs were designed. Four stiff ODE problems were tested and PI-RPNN yielded good accuracy and can outperform the traditional solvers in the cases with steep gradients. Later, Fabiani \textit{et al.} \cite{Fabiani202304_parsi-PI-RPNN} extended the PI-RPNNs to other stiff ODEs and three index-1 DAEs. For more practical applications, Li \textit{et al.} \cite{Li202110_PINN-coal} applied PINNs to solve the coal gasification chemical kinetic problems. The unreacted-core shrinking model was adopted for the gas-solid reactions, the temperature was simplified to be constant, and the ICs were hardcoded. Results showed that PINNs could accurately predict the production rates of the gaseous species.

Although the above studied focus on solving the chemical kinetic ODEs in a forward way without data supervision, inverse problems can also be solved when the available data are utilized. Weng \textit{et al.} \cite{Weng202211_MscalePINN-stiff} developed the multiscale PINNs for stiff chemical kinetic problems, where the species with different time scales were classified and trained by multiple NNs. By introducing ground truth data points, the proposed method showed superior ability on solving three stiff chemical kinetic problems. Gusmao \textit{et al.} \cite{Gusmao202204_kINN} proposed the kinetics-informed NNs (KINNs) to solve forward and inverse problems of chemical kinetics, as shown in Figure \ref{fig: chemical_ODE}(c). Practical species and reaction types were considered, including three types of species (gas, absorbed molecules, and surface radicals) and five types of reactions (denoted by \textit{g}, \textit{a}, \textit{d}, \textit{c}, and \textit{s}). ICs were enforced by an IC operator coupled with a shallow NN and species conservation was enforced by trigonometric transformations. In forward problems, no data were provided, while in inverse problems, synthetic data were available and three scenarios were designed based on the information about chemical intermediates and adsorbed molecules, and the presence of noise. The unknown kinetic parameters were optimized together with the NN parameters in inverse problems. The results showed that KINNs can accurately solve the ODEs in forward problems and estimate the kinetic parameters in inverse problems. Later, KINNs were applied to model the data from temporal analysis of products \cite{Nai202410_kINN-TAP} and extended to a robust version by incorporating the maximum likelihood approach \cite{Gusmao202312_kINN-MLE}. Bibeau \textit{et al.} \cite{Bibeau202312_PINN-biodiesel} used PINNs to predict the kinetics of biodiesel production in microwave reactors. The system contained five species and three reactions. The Arrhenius equation was linearized and the temperature evolution was simplified to a first-order ODE, resulting in 15 unknown parameters related to the reaction kinetics and energy balance, which were estimated by PINNs from experimental data. Results showed that the reaction kinetics of the biodiesel process could be identified successfully by PINNs with satisfactory extrapolation ability.


\subsection{Surrogate models of chemical kinetics}
\label{sec: surr}

In combustion simulations, the direct integration of the chemical reaction source terms is computationally expensive and is the main bottleneck for fast simulations of industrial combustion. Therefore, many methods have been proposed to accelerate the computation \cite{Zhou201111_ML4combus, Ihme202204_combusML, Wang202503_ANN-hard, Weng202411_EFNO-stiff}, such as tabulation methods, reduced chemistry, and manifold methods. ML methods have also been adopted for computation acceleration, usually by constructing NN surrogate models of the chemical kinetics \cite{Zhou201111_ML4combus}. The NNs are usually first trained using kinetic data, and then used to replace the integration of reaction source terms, so the overall computation efficiency can be improved. Based on the type of NN outputs, the NN surrogate models can be classified into two paradigms: discrete and continuous models. Specifically, the discrete models learn the mapping between the state variables at the current time and next time (i.e., following the AR paradigm), or the corresponding changes:
\begin{align}
	\mathbf{y}(t + \Delta t) &= \mathrm{NN} (\mathbf{y}(t))
	\label{eq: surr_next}
	\\
	\mathrm{or} \quad \Delta \mathbf{y}(t) =  \mathbf{y}(t + \Delta t) - \mathbf{y}(t) &= \mathrm{NN} (\mathbf{y}(t))
	\label{eq: surr_change}
\end{align}
For continuous models, the continuous dynamics can be learned by predicting the derivatives or source terms of the state variables:
\begin{align}
	\frac{\mathrm{d} \mathbf{y} (t)}{\mathrm{d} t} &= \mathrm{NN} (\mathbf{y}(t))
	\label{eq: surr_grad}
	\\
	\mathrm{or} \quad \dot{\bm{\upomega}}_{\mathbf{y}} (t) &= \mathrm{NN} (\mathbf{y}(t))
	\label{eq: surr_source}
\end{align}
where $\dot{\bm{\upomega}}_{\mathbf{y}}$ is the source terms of $\mathbf{y}$. If only the data of $\mathbf{y}$ are available, then Equation \eqref{eq: surr_grad} needs to be integrated to approximate $\mathbf{y}$. This is the paradigm of neural ODEs, which will be discussed in Section \ref{sec: NODE_surr}.

To enhance the robustness and physical consistency of the surrogate models, many studies consider the constraints of physical conservations to train the models in a physics-informed way. As an example, the total mass, mass of each element, mass of non-reacting species, and energy at time $t + \Delta t$ should be equal to the corresponding values at time $t$, so corresponding loss terms can be easily added when using Equation \eqref{eq: surr_next} or \eqref{eq: surr_change} as surrogate models. Theoretically, the total mass conservation will be automatically satisfied if the conservation of each element is satisfied. The species conservation is given by Equation \eqref{eq: species_conservation}, which is a more strict constraint on total mass conservation than the aforementioned one ($\sum Y (t + \Delta t) = \sum Y (t)$). Note that this kind of physics-informed surrogate NNs are different from the PINNs in the narrow sense in that they take $\mathbf{y}(t)$ as inputs instead of $t$.

\begin{figure}[ht]
    \centering
    \includegraphics[width=1\textwidth]{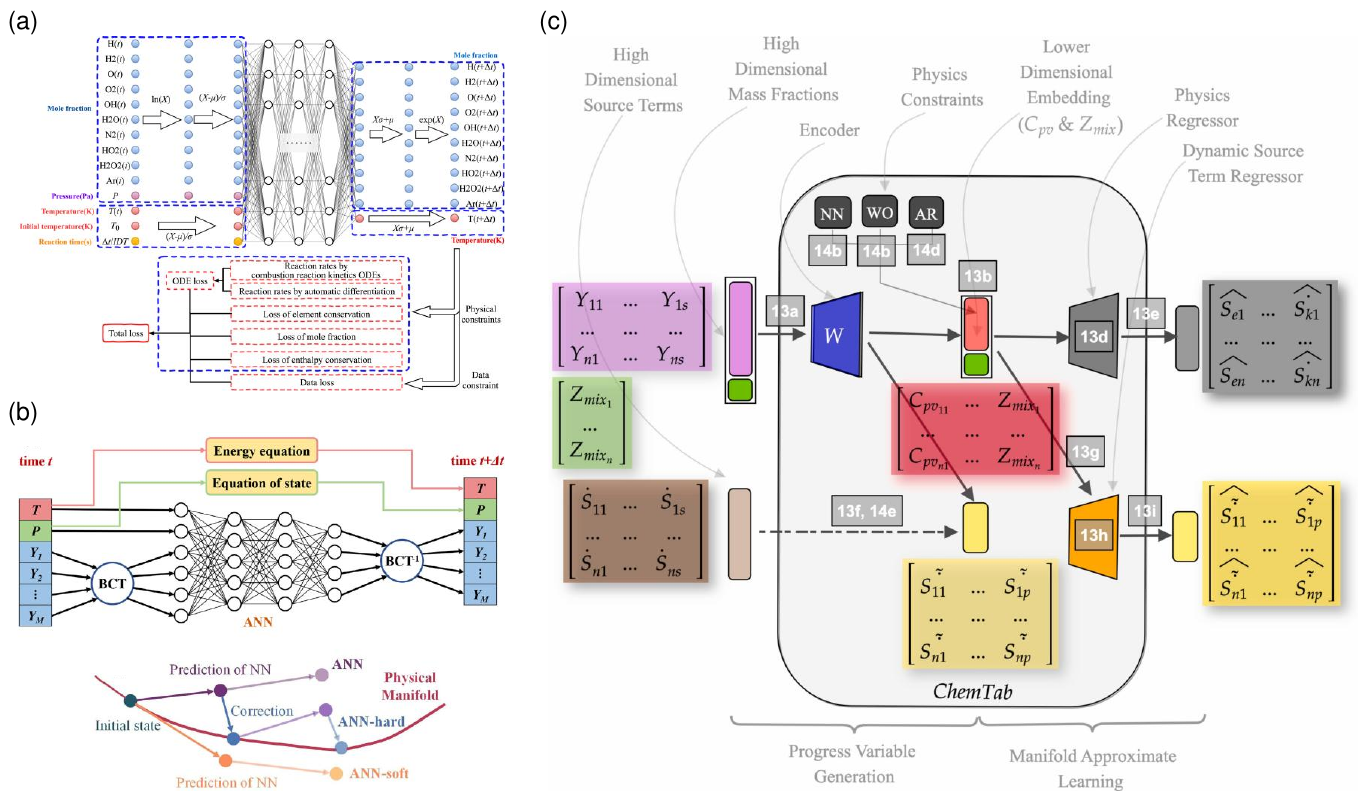}
    \caption{PIML for surrogate modeling of chemical kinetics. (a) The architecture of CRK-PINN \cite{Zhang202408_CRK-PINN}. (b) ANN-hard: schematic and comparisons with other approaches \cite{Wang202503_ANN-hard}. (c) ChemTab extended architecture and training procedure \cite{Salunkhe202303_ChemTab-extend}. Figures are reprinted with permission.}
    \label{fig: chemical_surrogate}
\end{figure}

Some studies on PIML for surrogate modeling of chemical kinetics are illustrated in Figure \ref{fig: chemical_surrogate}. Almeldein \textit{et al.} \cite{Almeldein202307_acce-chem} developed the physics-informed surrogate NNs for accelerating chemical kinetics calculations following Equation \eqref{eq: surr_next}, where the conservation of four elements was added in the loss function. The mixture-of-experts architecture was adopted to allow multiple subnetworks to specialize in different regimes, such as pre-ignition low-temperature chemistry, high-temperature reactions, and post-combustion equilibriums. The 0D CH$_4$-air combustion was tested, demonstrating the effectiveness of element conservation and the acceleration ability of the surrogate models on both CPUs and GPUs. Zhang \textit{et al.} \cite{Zhang202408_CRK-PINN, Zhang202406_CRK-PINN2} proposed the parametric physics-informed surrogate NNs for combustion reaction kinetics (CRK-PINNs) based on the paradigm of Equation \eqref{eq: surr_next}. As illustrated in Figure \ref{fig: chemical_surrogate}(a), the kinetic ODEs and the conservations of elements, species, and energy were considered in the loss function to improve the physical completeness and low data dependence. A logarithmic normalization strategy was adopted to mitigate the multiscale of species and stiffness of the system. Based on the H$_2$-air mechanism with 10 species and 21 reactions, the CRK-PINN was validated in three scenarios: solving forward and inverse problems, surrogate chemical kinetics, and accelerating simulations of combustion reacting flows. In the first scenario, CRK-PINN can solve the ignition process without data, reconstruct the process with sparse data, and infer intermediate species and temperature with the data of major species. In the second scenario, CRK-PINN can surrogate chemical kinetics in wide operating conditions and exhibited higher accuracy and lower data dependence than data-driven NNs for both time-continuous and time-advanced predictions. In the last scenario, CRK-PINN can replace the direct integration solver to predict the reaction source terms in the DNS of a 2D laminar Bunsen flame and a 3D turbulent jet flame, showing superior accuracy than data-driven NNs and great acceleration compared to the direct integration method.

In addition to the soft constraints in the form of loss functions, there are also studies that impose hard constraints on physical conservations. Zhang \textit{et al.} \cite{Zhang202208_Mscale-sample} proposed a multiscale sampling method to enhance the accuracy and robustness of NNs to surrogate chemical kinetics, showing superior ability than the manifold, Monte Carlo, and GAN sampling. The Box-Cox transformation (BCT) was adopted to preprocess the data of species mass fractions:
\begin{equation}
	f(x) =
	\left\{ \begin{aligned}
		&\frac{x^{\lambda} - 1}{\lambda} \quad &\lambda \neq 0
		\\
		&\ln x \quad &\lambda = 0
	\end{aligned} \right.
	\label{eq: BCT}
\end{equation}
which is a good alternative to the logarithmic normalization \cite{Zhang202408_CRK-PINN}. Although the paradigm of Equation \eqref{eq: surr_change} was adopted, only the species were directly predicted by the NNs, while the temperature and pressure were updated by transport fluxes, thus ensuring the mass and energy conservation. The algorithm has been validated on various CFD cases of combustion and been applied to DeepFlame \cite{Mao202307_DeepFlame}, a deep learning empowered open-source platform for reacting flow simulations. Wang \textit{et al.} \cite{Wang202503_ANN-hard} extended the method by imposing constraints on element conservations in both soft and hard ways. As illustrated in Figure \ref{fig: chemical_surrogate}(b), the ANN-soft penalized the element conservations by adding a loss term, while ANN-hard enforced the element conservations by adding a correction term, $\Delta \mathbf{Y}_c$, to the NN-predicted mass fractions. The correction term can be easily derived by the element conservations and the changes of species mass fractions, and was first computed in the Box-Cox transformed space to avoid the multiscale issue. One advantage of ANN-hard over ANN-soft is that there is no cumbersome adjustment of the loss weights. In terms of prediction accuracy and error of element conservations, ANN-hard exhibited superior performance than ANN-soft and they all outperformed the regular ANN on various flame cases. The computational overhead caused by the correction term was negligible, demonstrating the efficiency of ANN-hard. In fact, the correction method to ensure element conservations has been adopted in earlier studies \cite{Wan202007_CNN-reduce, Wan202007_NN-reduce}, where the NNs were trained to surrogate the kinetics of reduced chemistry, following the paradigm of Equation \eqref{eq: surr_source}.

In addition to the paradigm of function learning, some studies follow the paradigm of operator learning to learn combustion chemical kinetics, using neural operators such as the DeepONet \cite{Lu202103_DeepONet} and FNO \cite{Li202105_FNO}. Some data-driven DeepONets have been used to surrogate chemical kinetics \cite{Venturi202211_flexDeepONet, Goswami202312_DeepONet-stiff, Kumar202402_DeepONet-acce}, where the input of branch nets and the final output are similar to Equation \eqref{eq: surr_next}, while the input of truck nets is related to the temporal coordinates or increments, enabling the models to learn kinetics with variable $\Delta t$, which is often fixed in classical surrogate models. To some extent, they can be viewed as a hybrid paradigm of AR and NAR learning. One insight that these works can provide is that learning kinetics in the latent space with reduced dimension can improve computational efficiency, which can be achieved by integrating a self-supervised autoencoder (AE) \cite{Goswami202312_DeepONet-stiff, Kumar202402_DeepONet-acce}. Kumar \textit{et al.} \cite{Kumar202402_DeepONet-acce} also accelerated the computation by selecting representative species for kinetics learning. The other scalars can be reconstructed by a pre-trained correlation NN. This strategy was also adopted in Reference \cite{Kumar202312_PI-DeepONet-reduce, Kumar202507_PI-DeepONet-reduce2}, where the species and element conservations were integrated into the loss functions of both correlation NN and DeepONet, thus constituting a physics-informed learning. The framework was tested for CH$_4$ oxidation, demonstrating that the physical constraints can reduce the conservation errors without loss of prediction accuracy. Zanardi \textit{et al.} \cite{Zanardi202112_PI-DeepONet-noneq} used the PI-DeepONets to learn the solution operator of non-equilibrium chemistry in hypersonic flows, where the coarse-grained (CG) kinetic ODEs were considered. The input of the branch and truck nets were the ICs and temporal coordinates, respectively. The framework was tested on the O$_2$--O mixture, exhibiting two orders of magnitude in speedup compared to numerical integrators. Later, Zanardi \textit{et al.} \cite{Zanardi202309_adap-PI-DeepONet} extended the framework to be more physics-informed by many strategies, including a Boltzmann transformation layer to enforce local equilibrium distributions between states in the same group, the hierarchical architecture and transfer learning between different temporal scales, and the adaptive online pruning technique. Weng \textit{et al.} \cite{Weng202411_EFNO-stiff} applied FNO to learn stiff chemical kinetics, following the paradigm of Equation \eqref{eq: surr_next}. The loss of element and species conservations were considered and the BCT (Equation \eqref{eq: BCT}) was also adopted. Based on the focal loss \cite{Lin201710_focal-loss}, a balanced loss function was proposed to address the unbalanced sampling issue in complex reacting flow problems. The results on four test cases (ROBER, POLLU, H$_2$ autoignition, and H$_2$--NH$_3$ turbulent jet flame) demonstrated the effectiveness of physical constraints, BCT, and the balanced loss function.

Some studies used NNs to learn chemical tabulations, which can be viewed as surrogate modeling of chemical kinetics as well. In References \cite{Readshaw202103_LES-PDF, Ding202105_MLtab-hybrid, Ding202210_MLtab-blend}, NNs were used to learn the tabulation of thermochemistry following the paradigm of Equation \eqref{eq: surr_change}. The hybrid flamelet/random data were generated for training and multiple NNs were used to predict each species with different magnitudes of concentration changes. The species conservation (Equation \eqref{eq: species_conservation}) was enforced by normalizing the predicted species concentrations. Readshaw \textit{et al.} \cite{Readshaw202304_MLtab-conserve} extended this framework by imposing constraints on element conservation, which was achieved by recalculating the mass fractions of selected residual species after the initial prediction. The results demonstrated that this conserved NN exhibited higher accuracy and physical consistency than the non-conserved NNs. Salunkhe \textit{et al.} \cite{Salunkhe202206_ChemTab} developed the framework of ChemTab to jointly learn the progress variable generation (PVG) and manifold approximation (MA). For the PVG, the high dimensional species mass fractions are linearly embedded to the progress variables: $\mathbf{C_{pv}} = \mathbf{YW}$, while MA means the mapping between $\mathbf{C_{pv}} \oplus Z$ and the source terms is learned by a NN, where $Z$ is the mixture fraction. Therefore, the overall framework is similar to Equation \eqref{eq: surr_source}. Inspired by the principal component analysis (PCA), some constraints were added in the loss function, including the orthogonality of $\mathbf{W}$ and $\mathbf{C_{pv}} \oplus Z$, and the unit-norm constraint on $\mathbf{W}$. It was demonstrated that removing any of these constraints will make the results worse. Salunkhe \textit{et al.} \cite{Salunkhe202303_ChemTab-extend} extended the ChemTab by adding another NN to learn the relationship between $\mathbf{C_{pv}} \oplus Z$ and lower dimensional source terms, which was computed using the same projection method and the same $\mathbf{W}$, as illustrated in Figure \ref{fig: chemical_surrogate}(c). The unit-norm constraint on $\mathbf{W}$ was replaced with the non-negativity constraint, which was achieved in a hard manner by truncating the negative values.

\subsection{Neural ODEs: surrogate models and kinetics discovery}
\label{sec: NODE}

\begin{figure}[ht]
    \centering
    \includegraphics[width=1\textwidth]{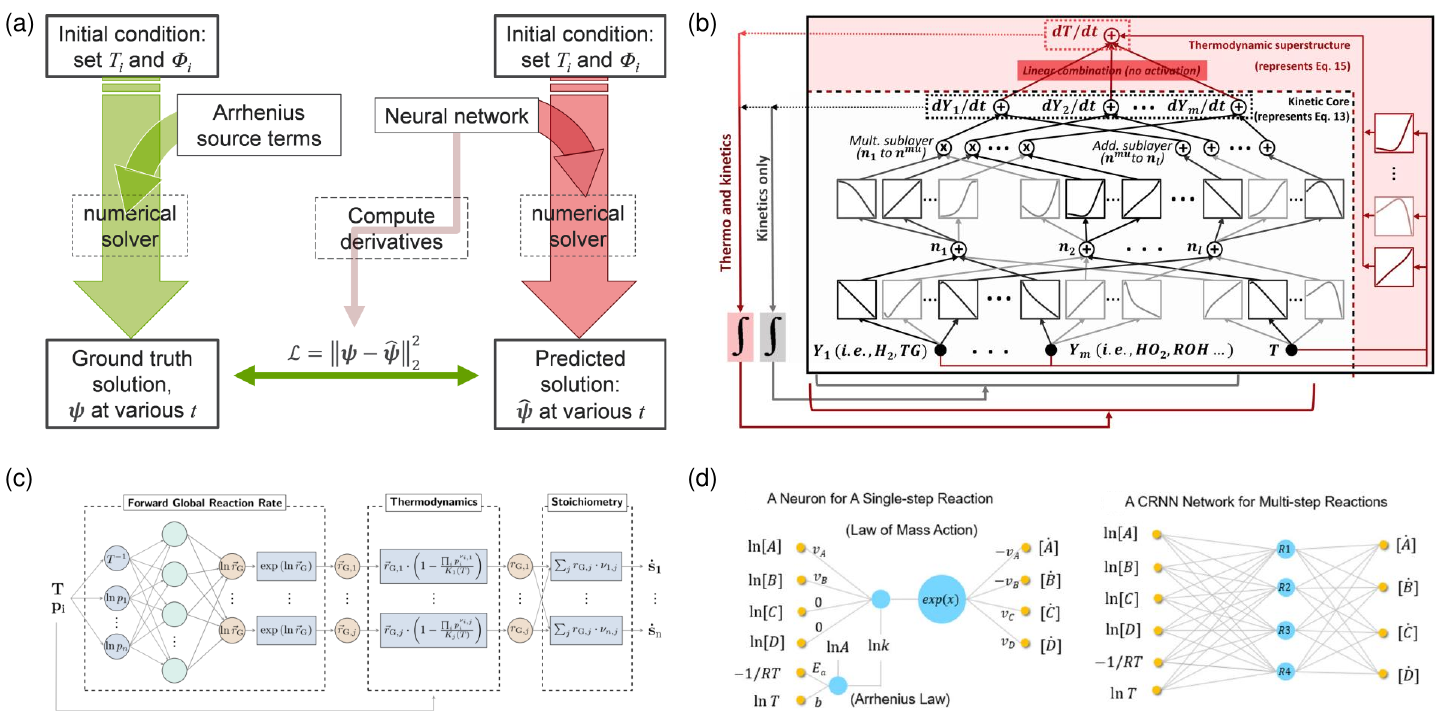}
    \caption{Neural ODEs for continuous surrogate modeling and discovering kinetics. (a) The ChemNODE approach \cite{Owoyele202109_ChemNODE}. $\bm{\psi}$ indicates the thermochemical scalars. (b) The structure of ChemKAN \cite{Koenig202507_ChemKAN}. (c) GRNN architecture with embedded stoichiometry and thermodynamics \cite{Kircher202402_GRNN}. (d) CRNN illustrated for a reaction system with four species and four reaction steps \cite{Ji202101_CRNN}. Figures are reprinted with permission.}
    \label{fig: chemical_NODE}
\end{figure}

Neural ODE (NODE) was first proposed by Chen \textit{et al.} \cite{Chen201812_NODE} to develop a new family of NNs that parameterize the derivative of the hidden state using NNs, which is different from the traditional approach that defines discrete sequence of hidden layers, such as the ResNet and RNNs. Equation \eqref{eq: surr_grad} is the basic expression of NODEs, while state variables are computed by integrating the Equation \eqref{eq: surr_grad}:
\begin{align}
	\mathbf{y} (t_1) = \mathbf{y} (t_0) + \int_{t_0}^{t_1} \mathrm{NN} (\mathbf{y}(t)) \mathrm{d}t = \mathrm{ODESolve} (\mathbf{y}(t_0), t_0, t_1, \mathrm{NN})
\end{align}
Compared to discrete approaches, NODEs can better learn the continuous dynamics of a system, thus showing better reconstruction and extrapolation performance \cite{Chen201812_NODE}. Since NODE is proposed for learning the first-order dynamics of sequential systems, it is natural to apply it to chemical kinetic systems. Extensive studies have applied NODEs to the modeling of chemical kinetics for both building surrogate models and discovering kinetics (some are shown in Figure \ref{fig: chemical_NODE}), where many of them have informed/embedded physics to the NODEs.

\subsubsection{Continuous surrogate models}
\label{sec: NODE_surr}

As mentioned in Section \ref{sec: surr}, NODEs can be viewed as continuous surrogate models of chemical kinetics following the paradigm of Equation \eqref{eq: surr_grad}. Owoyele \textit{et al.} \cite{Owoyele202109_ChemNODE} first applied NODEs to model chemical kinetics, termed as ChemNODE and shown in Figure \ref{fig: chemical_NODE}(a). To improve the training stability, a progressive training strategy was employed, where the species were trained sequentially. For a homogeneous hydrogen-air reactor, ChemNODE showed accurate results across various operating conditions and yielded a speedup of 2.3x over the full chemical mechanism method. Kim \textit{et al.} \cite{Kim202109_stiffNODE} used NODEs to learn stiff ODEs. The tests on the ROBER and POLLU problems demonstrated that the scaling of equations and loss functions was essential for NODEs to learn stiff ODEs.

Although the name contains the "ODE", the vanilla NODE is indeed a data-driven method because no information of the ODEs is utilized. Recall that PINNs predict $\mathbf{y}$ by $\mathbf{y} = \mathrm{NN} (t)$ and differentiate it to get $\dot{\mathbf{y}}$, while NODEs, in contrast, predict $\dot{\mathbf{y}}$ by $\dot{\mathbf{y}} = \mathrm{NN} (\mathbf{y}(t))$ and integrate it to get $\mathbf{y}$. Therefore, if the ODEs can be explicitly added to the loss functions of NODEs, they can be viewed as PINNs in a broad sense, or an integral version of PINNs. Some studies have made the NODEs physics-informed in different ways. Sholokhov \textit{et al.} \cite{Sholokhov202306_PI-NODE} developed the physics-informed NODE (PI-NODE) to build reduced-order models (ROMs) for dynamical systems, where an AE was employed to perform model reduction and the NODE was used to learn the nonlinear dynamics in the latent space. Based on the ODEs in the physical space, the latent-space dynamics was derived and constitute a physics-informed loss, which was evaluated on a set of collocation points. The addition of the physics-informed loss improved the long-term performance, the robustness to data scarcity and noise, and the compressibility of the ROMs. Vijayarangan \textit{et al.} \cite{Vijayarangan202311_dyn-informed-NODE} developed an AE-based ROM for stiff chemical kinetics using dynamics-informed training, where the NODE was used to learn the latent-space dynamics. The "dynamics-informed" means that the discovery of the latent space is informed by the latent dynamical evolution rule during the training procedure, which is different from the definition of "physics-informed" in a narrow sense. Two additional losses were added to ensure the mapping of the encoder and decoder was bijective. Results demonstrated that the proposed ROM could serve as an effective stiffness-removal framework. Kumar \textit{et al.} \cite{Kumar202311_PC-NODE, Kumar202503_PC-NODE2} developed the physics-constrained NODE (PC-NODE) to learn stiff chemical kinetics, where the conservations of each element were constrained by including corresponding loss terms. Here the element conservation was constrained between the predicted and true values, instead of the values at time $t$ and $t + \Delta t$ in the paradigm of discrete surrogate models. Compared to the purely data-driven NODE, results demonstrated that the PC-NODE framework improved not only the physical consistency but also the training efficiency. Later, Kumar \textit{et al.} \cite{Kumar202412_Phy-ChemNODE} combined the NODE with AE to develop the Phy-ChemNODE. The four loss terms can be considered as a combination of the bijective constraints of the dynamics-informed NODE \cite{Vijayarangan202311_dyn-informed-NODE} and the element conservation constraints of the PC-NODE \cite{Kumar202311_PC-NODE, Kumar202503_PC-NODE2}. Phy-ChemNODE was tested on a CH$_4$-O$_2$ combustion kinetics, where the 33 physical variables were reduced to 4 latent variables. The element conservation constraints were shown to help the model to obey the mass conservation better.

Although the vanilla NODEs are purely data-driven, their interpretability can be enhanced by using the KANs as the network architecture. Koenig \textit{et al.} \cite{Koenig202409_KAN-ODEs} proposed the KAN-ODEs, combining the advantage of NODEs that do not require prior knowledge and the advantages of KANs that have increased interpretability and faster neural scaling laws compared to MLPs. The KAN-ODE was validated on learning  Lotka-Volterra ODEs, modeling source terms of the Fisher-KPP PDE, and learning 1D PDEs. The scaling performance was shown superior to MLPs and the model performance could be further enhanced by sparsification and using symbolic regression. Subsequently, based on the KAN-ODEs and LeanKAN \cite{Koenig202507_LeanKAN}, Koenig \textit{et al.} \cite{Koenig202507_ChemKAN} proposed the ChemKAN for modeling and acceleration of combustion chemical kinetics. The physics was informed in two ways. First, the architecture strongly mimicked the chemical kinetics, as illustrated in Figure \ref{fig: chemical_NODE}(b). A KAN was first used to represent the species production rates (core structure), then they were linearly combined to represent the temperature rate (thermodynamic superstructure) with another KAN to reflect the thermophysical parameter variation. This serial architecture offers the flexibility that one can only learn the species when temperature is not a variable. Second, an optional loss function for element conservation was introduced, which is similar to the PC-NODE and Phy-ChemNODE. The first test case was a biodiesel production modeling, where the thermodynamic superstructure and physical loss were "turned off". Compared with DeepONet, ChemKAN exhibited stronger scaling behaviour on test sets, less tendency toward overfitting, and stronger robustness to noise. In the second case of H$_2$-air combustion, the ability of ChemKAN for computation acceleration was validated. Compared to MLP-based ChemNODE, ChemKAN showed similar speed-up ratios, but solved more species.

In addition to inform physics in the form of loss functions, some studies embed the information of stoichiometry and/or rate expressions into the NODEs to enhance the physical consistency. According to Equation \eqref{eq: omega_dot}, the rates of species and reactions are linearly related by the stoichiometric matrix. Thus, if the stoichiometric matrix is known, one can use NODEs to directly predict $\mathbf{RoP}$ and then transform it to $\dot{\bm{\upomega}}$, or vice versa. Based on the known information, Sorourifar \textit{et al.} \cite{Sorourifar202309_PE-NODE} developed two versions of physics-enhanced NODEs for chemical reaction systems: r-NODE and k-NODE. The r-NODE approximates $\mathbf{RoP}$ when only the stoichiometry is known and embedded, while the k-NODE approximates the kinetic constants when the mass-action parts of $\mathbf{RoP}$ are further known and embedded. Results suggest that a ground truth model that accurately reflects the data can be learned only when the embedded information accurately reflects the underlying mechanics. Fedorov \textit{et al.} \cite{Fedorov202310_KC-NODE} developed the kinetics-constrained NODEs (KC-NODEs) for kinetic modeling, where the kinetic and thermodynamic knowledge were embedded into the NN architecture, including the stoichiometric matrix, the law of mass action, and the Arrhenius dependency of the rates on temperature. The KC-NODE was successfully applied to the kinetic modeling of the hydrogenation of CO$_2$ to methane using numerical data and to higher hydrocarbons using real experimental data. Kircher \textit{et al.} \cite{Kircher202402_GRNN, Kircher202406_GRNN2} proposed the global reaction neural network (GRNN) to learn kinetics from reactor data, where the stoichiometry and thermodynamics were explicitly embedded as the network layers, as shown in Figure \ref{fig: chemical_NODE}(c). The GRNN can be trained directly when source data are provided and trained in a NODE way when only reactor data are available, which is a 1D plug flow reactor model in the study. It was demonstrated that the embedded knowledge enhanced the model robustness to significant noise in the data.

\subsubsection{Kinetics discovery}

In terms of embedding the knowledge of chemical kinetics into NNs, the chemical reaction neural network (CRNN), proposed by Ji \textit{et al.} \cite{Ji202101_CRNN} for kinetics discovery, is an earlier work than References \cite{Sorourifar202309_PE-NODE, Fedorov202310_KC-NODE, Kircher202402_GRNN} and is also trained in a NODE-way. As shown in Figure \ref{fig: chemical_NODE}(d), a neuron of CRNN represents a single-step reaction, where the law of mass action (Equation \eqref{eq: RoP}) and Arrhenius law (Equation \eqref{eq: k_fi}) are embedded as the nodes and weights. For the input layer, the weights represent the reaction orders and the bias represents the rate constant in the logarithmic scale. The weights of the output layer correspond to the stoichiometric coefficients. A multi-step reaction system can be represented by stacking the neurons to formulate a CRNN with one hidden layer, where the number of hidden neurons is equal to the number of reactions. Due to the physical interpretability, once the CRNN has been trained, the kinetic parameters can be easily recovered by checking the weights and biases, then the reaction pathways can be fully revealed. The number of hidden nodes/reactions, which is a hyperparameter, can be determined by grid searching and a threshold pruning strategy is suggested to clip the NN weights below a certain threshold \cite{Ji202101_CRNN}. The CRNN was first tested on three cases with increasing complexity: an elementary reaction network without temperature dependence, biodiesel production with temperature dependence, and an enzyme reaction network in which some species are present in both reactants and products. Results demonstrated that CRNN could accurately learn the reaction pathways and kinetic parameters from noisy data and even from incomplete data with missing species.

Later, CRNN has been applied to many reaction systems, such as the biomass pyrolysis \cite{Ji202201_CRNN-biomass} and the decomposition of energetic materials \cite{Tang202208_CRNN-nanoAl, Wang202301_CRNN-CL-20, Sun202404_CRNN_HMX, Chen202501_HyChem}. Li \textit{et al.} \cite{Li202301_Bayesian-CRNN} developed the Bayesian CRNN to simultaneously infer the reaction pathways and perform uncertainty quantification on the identified reaction pathways, kinetic parameters, and concentration predictions with unseen initial conditions. Su \textit{et al.} \cite{Su202303_NODE-para-opt} used NODEs for kinetic parameter optimization of hydrocarbon fuels, where the framework can be viewed as CRNN with only kinetic parameters unknown, all pathways fixed and all other unnecessary connections trimmed. Considering that CRNN does not ensure mass and element/atom conservations, Doeppel \textit{et al.} \cite{Doeppel202407_AC-CRNN} developed the atom conserving CRNN (AC-CRNN), where the atom conservation was enforced by adding a NN layer that multiplicated the key species coefficients with a conservation matrix. The standard cases in Reference \cite{Ji202101_CRNN} was tested. Compared to CRNN, the AC-CRNN exhibited improved stability and robustness against noisy and missing data, and was more efficient due to the reduced number of trainable parameters.

\section{PIML for combustion reacting flows}
\label{sec: PIML_reacting-flow}

The previous section focuses on reaction problems, but most of the practical combustion is a convection-diffusion-reaction (CDR) process, where the laws in Section \ref{sec: fluid-dynamics} need to be considered to describe the spatial processes such as flow and diffusion. This section will present the studies on PINNs for combustion reacting flows, with representative studies listed in Table \ref{table: reacting-flow}, which provides the dimension, problem type, adopted physical losses, treatment of reaction, and other information. Based on the combustion problem type, the studies in Table \ref{table: reacting-flow} are classified into four parts: laminar combustion, turbulent combustion, supersonic combustion, and combustion SR via discrete PINNs, which will be introduced in four subsections respectively. Before that, some studies on PINNs for diffusion-reaction (DR) and advection-diffusion-reaction (ADR) problems are first introduced.

\begin{table}
	\small
	\centering
	\caption{Recent studies on PINNs for combustion reacting flows. The four parts in this table represent the laminar combustion, turbulent combustion, supersonic combustion, and combustion SR via discrete PINNs, respectively. $\mathbf{U}$ is the velocity vector. FRC means finite-rate-chemistry. "Sim." and "exp." represent "simulation" and "experiment", respectively. The tilde ("$\sim$") indicates (Favre) averaged values in turbulent conditions.}
	\label{table: reacting-flow}
	\begin{tabular}{>{\raggedright\arraybackslash}m{2cm}>{\raggedright\arraybackslash}m{1cm}>{\raggedright\arraybackslash}m{1.5cm}>{\raggedright\arraybackslash}m{2cm}>{\raggedright\arraybackslash}m{1cm}>{\raggedright\arraybackslash}m{2cm}>{\raggedright\arraybackslash}m{1cm}>{\raggedright\arraybackslash}m{1.5cm}>{\raggedright\arraybackslash}m{0.8cm}}
		\toprule[1.5pt]
		Study & \makecell{Dimen-\\sion} & Problem type & Physical loss (equation of) & \makecell{Turbu-\\lence} & Reaction in physical loss & Ref. data & NN architecture & Codes \\
		\midrule
		Choi \cite{Choi202201_CSTR}, 2022.1 & 3D+t & Inverse & Mass, $\mathbf{U}$, $T$, $Y_{1-3}$ & No & Van de Vusse & Sim. & MLP & -- \\
		Wang \cite{Wang202207_ROBER_Z-eq}, 2022.7 & 2D & Forward & $Z$ & No & Absent (fast chemistry) & Sim. & MLP & -- \\
		Sitte \cite{Sitte202208_poolfire}, 2022.8 & 2D+t & Inverse & Mass, $\mathbf{U}$ & No & Absent & Sim. & MLP & -- \\
		Liu \cite{Liu202309_surr-flame}, 2023.9 & 1D, 2D, 2D+5 & Forward & Mass, $\mathbf{U}$, $T$, $Y_{1-5}$ & No & 1-step CH$_4$ & Sim. & MLP & -- \\
		Song \cite{Song202410_PINN-FPV}, 2024.10 & 2D & Both & Mass, $\mathbf{U}$, $Z$, $c$ & No & Detailed CH$_4$ (FPV table) & Sim. & MLP & -- \\
		Cao \cite{Cao202411_PTS-PINN-flame}, 2024.11 & 2D & Forward & Mass, $\mathbf{U}$, $T$, $Y_{1-5}$ & No & 1-step CH4 & Sim. & MLP & -- \\
		Wu \cite{Wu202501_FlamePINN-1D}, 2025.1 & 1D & (1) Both (2) Forward (3) Inverse & (1) $T$ (2) $T$, $Y_{1-5}$ (3) Mass, $V$, $T$, $Y_{1-5}$ & No & (1) Simplified FRC (2)(3) 1-step CH$_4$ & Sim. & (1) MLP (2)(3) MMLP-RWF & \href{https://github.com/CAME-THU/FlamePINN-1D}{\underline{URL}} \\
		\midrule
        Liu \cite{Liu202312_recons-turb}, 2023.12 & 3D+t & Inverse & Mass, $\mathbf{U}$, $T$, $Y_{1-6}$, $\sum Y$ & Direct & 2-step CH$_4$ & Sim. & MLP & -- \\
        Liu \cite{Liu202509_3Dfrom2D}, 2025.9 & 3D+t & Inverse & $\mathbf{U}$ & Direct & Absent & Sim. & MLP & -- \\
		von Saldern \cite{vonSaldern202211_mean-assi}, 2022.11 & 2D & Inverse & Mass & Filtered & Absent & Sim. & MLP & -- \\
		Peng \cite{Peng202404_recons-flameD}, 2024.4 & 2D+t & Inverse & Mass, $\widetilde{\mathbf{U}}$, $\widetilde{T}$, $k$, $\varepsilon$ & Mean & Absent & Sim. & MLP & -- \\
        Cao \cite{Cao202509_PINN-turb-combus}, 2025.9 & 2D, 2D+1& Both & Mass, $\widetilde{\mathbf{U}}$, $\widetilde{T}$, $\widetilde{Y}_{1-5}$, $k$, $\varepsilon$, $\mu_t$ & Mean & EDM & Sim. & MLP & -- \\
		Taassob \cite{Taassob202310_PINN-TC-closure}, 2023.10 & 2D+1 & Inverse & Mass, $\widetilde{\mathbf{U}}$, $\widetilde{\mathrm{PC}}_{1-2}$, $\widetilde{Z}$, $\mu_t$ & Mean (learned) & Learned & Exp. & MLP & -- \\
		Taassob \cite{Taassob202406_PI-DeepONet-TC-closure}, 2024.6 & 2D+2 & Inverse & Mass, $\widetilde{\mathbf{U}}$, $\widetilde{Y}_{1-6}$, $\widetilde{Z}$, $\mu_t$ & Mean (learned) & Learned & Exp. & DeepONet & -- \\
		Liu \cite{Liu202406_recons-soot}, 2024.6 & (1) 2D+t (2) 2D & Inverse & (1) $\mathbf{U}$, $T$ (2) $Y_{\mathrm{soot}}$ & No & (1) Absent (2) Learned & Exp. & (1) MLP (2) AE-DeepONet & -- \\
		Yadav \cite{Yadav202412_RF-PINN}, 2024.12 & (1) 1D (2)(3) 2D & Inverse & Mass, (1) $c$ (,$\mathbf{U}$) (2) $c$ (3) $\widetilde{c}$ & (1)(2) No (3) Mean (learned) & (1)(2) Pre-learned (3) EBU model & (1)(2) Sim. (3) Exp. & MLP & -- \\
		\midrule
		Wang \cite{Wang202307_PINN-RDC}, 2023.7 & 1D+t & Inverse & Mass, $u$, $E$, $c$ & No & Simplified FRC & Sim. & MLP & -- \\
		Deng \cite{Deng202407_scramjet}, 2024.7 & 2D+t+2 & Inverse & Mass, $\widetilde{\mathbf{U}}$ & Mean & Absent & Sim. & ResUNet & -- \\
		\midrule
        Zhang \cite{Zhang202509_CRF-PINN}, 2025.9 & 2D, 2D+t & Inverse & Mass, $\mathbf{U}$, $T$, $Y_{1-5}$, $\sum Y$ & (1) No (2) Mean (learned) & 1-step CH$_4$ and learned & (1) Sim. (2) Exp. & MRF-CNN & -- \\
		Bode \cite{Bode202101_subfilter}, 2021.1 & 3D & Inverse & Mass & Filtered to full & Absent & Sim. & ESR-GAN & \href{https://git.rwth-aachen.de/Mathis.Bode/PIESRGAN}{\underline{URL}} \\
		Bode \cite{Bode202302_turbulent}, 2023.2 & 3D+t & Inverse & Mass, $\sum Y$ & Filtered to full & Absent & Sim. & ESR-GAN & -- \\
		Wang \cite{Wang202311_PI-GAN-RDC}, 2023.11 & 2D & Inverse & Mass, $\mathbf{U}$ & No & Absent & Sim. & GAN & -- \\
		Wang \cite{Wang202408_PI-RGAN-RDC}, 2024.8 & 2D+t & Inverse & Mass & No & Absent & Sim. & R-GAN & \href{https://github.com/pyrimidine/Physics-Informed-Recurrent-GAN-for-Flow-field-Reconstruction}{\underline{URL}} \\
		\bottomrule[1.5pt]
	\end{tabular}
\end{table}

\subsection{(Advection-)diffusion-reaction problems}

Conceptually, DR and ADR problems are more complex than pure reaction problems due to the increase of problem dimension and existence of diffusion and advection, but are simpler than CDR problems due to the lack of nonlinear convection terms. Nevertheless, they share the feature of spatial processes and reactions in CDR problems, so we first briefly review the PINNs for DR and ADR problems before moving to CDR problems, which may shed light on the studies on PINNs for practical combustion reacting flows.

For DR problems, many DR systems are not encountered in the area of combustion, such as the well-known Allen-Cahn equation:
\begin{align}
	u_t - 0.0001u_{xx} + 5 u (u^2 - 1) = 0
\end{align}
which describes the process of phase separation in multicomponent alloy systems, and has become a benchmark problem for validating PINN algorithms \cite{Raissi201811_PINN, McClenny202211_SA-PINN, Penwarden202311_dtPINN, Wang202402_causalPINN}. It can be seen that the reaction source term is quite different from the combustion ones (Section \ref{sec: chemical}), and only one variable is considered. Other DR systems, such as $\lambda$-$\omega$, FitzHugh-Nagumo, and Gray-Scott equations have been solved by PhyCRNet \cite{Ren202112_PhyCRNet} and PeRCNN \cite{Rao202307_PeRCNN}. Although these problems involve more than one variable, the polynomial reaction kinetics still differ from the chemical kinetics of practical combustion. Nevertheless, these studies have validated the feasibility of PINNs to solve PDEs with both diffusion terms and highly nonlinear reaction source terms. Niaki \textit{et al.} \cite{Niaki202106_PINN-curing} used PINNs to solve the DR PDEs of the thermochemical curing process of composite-tool systems during manufacture, where the cure kinetics contained the Arrhenius law, further validating the applicability of PINNs to solve combustion-like DR problems.

For ADR problems, their difference from CDR problems is that the velocities are constants or functions instead of field variables, so the nonlinearity is mainly caused by reaction instead of convection. Ngo \textit{et al.} \cite{Ngo202110_fixedbed} applied PINNs to the solution and parameter inference of a fixed-bed reactor model for catalytic CO$_2$ methanation. The momentum and energy equations were neglected due to the low pressure drop and isothermal conditions, respectively, so the governing equations were 1D advection-reaction equations of four species, which considered the law of mass action and Arrhenius law. Results showed that the forward PINN could solve the reactor model, whereas the inverse PINN could reveal an unknown effectiveness factor using the pre-trained forward PINN. Hou \textit{et al.} \cite{Hou202207_OG-PINN} applied PINNs to forward and inverse one-dimensional one-component (1D1C) ADR problems, where both location-dependent and linear reaction terms were considered. The adopted first derivative constraint (FDC) method, which is similar to gPINN \cite{Yu202203_gPINN}, has been demonstrated to enhance the model stability and accuracy. To avoid the slow training speed caused by FDC, Sun \textit{et al.} \cite{Sun202307_multicomp} proposed an NN connection structure named deep function family construction, which added residual connections to fully connected NNs. The modified PINN was applied to forward and inverse 1D3C ADR problems, where the reaction source terms were given by the law of mass action for a reaction system of three species and five reactions. Both steady and unsteady problems were solved and the modified PINN outperformed the vanilla PINN, especially for cases with steep fronts. The results of inverse problems indicated the capability of PINNs to learn missing physics (multiple unknown parameters) in ADR systems. This was further demonstrated by Hou \textit{et al.} \cite{Hou202308_PINN-CDR}, where the inverse problem of the 1D3C ADR system was also solved. Laghi \textit{et al.} \cite{Laghi202302_ADR-4PINNs} compared four PINNs on six 1D steady-state ADR problems with analytical solutions. The four PINNs were vanilla PINNs, PI-ELM \cite{Dwivedi201912_PI-ELM}, Deep-TFC \cite{Leake202103_Deep-TFC}, and X-TFC \cite{Schiassi202106_X-TFC}. Comparative analysis demonstrated that X-TFC was the best-performing framework in terms of both accuracy and computational time.

In summary, the above studies on DR and ADR problems have validated the applicability of PINNs to physical processes with diffusion, constant flow, and nonlinear reactions. However, they do not consider not only the velocity variable but also the temperature variable, and thus the reaction rate constants described by the Arrhenius law and the changes in physical properties due to heat release. For practical combustion reacting flows, some representative studies on PINNs for solving them are listed in Table \ref{table: reacting-flow} and will be reviewed in the following subsections.

\subsection{Laminar combustion}

\begin{figure}[ht]
    \centering
    \includegraphics[width=1\textwidth]{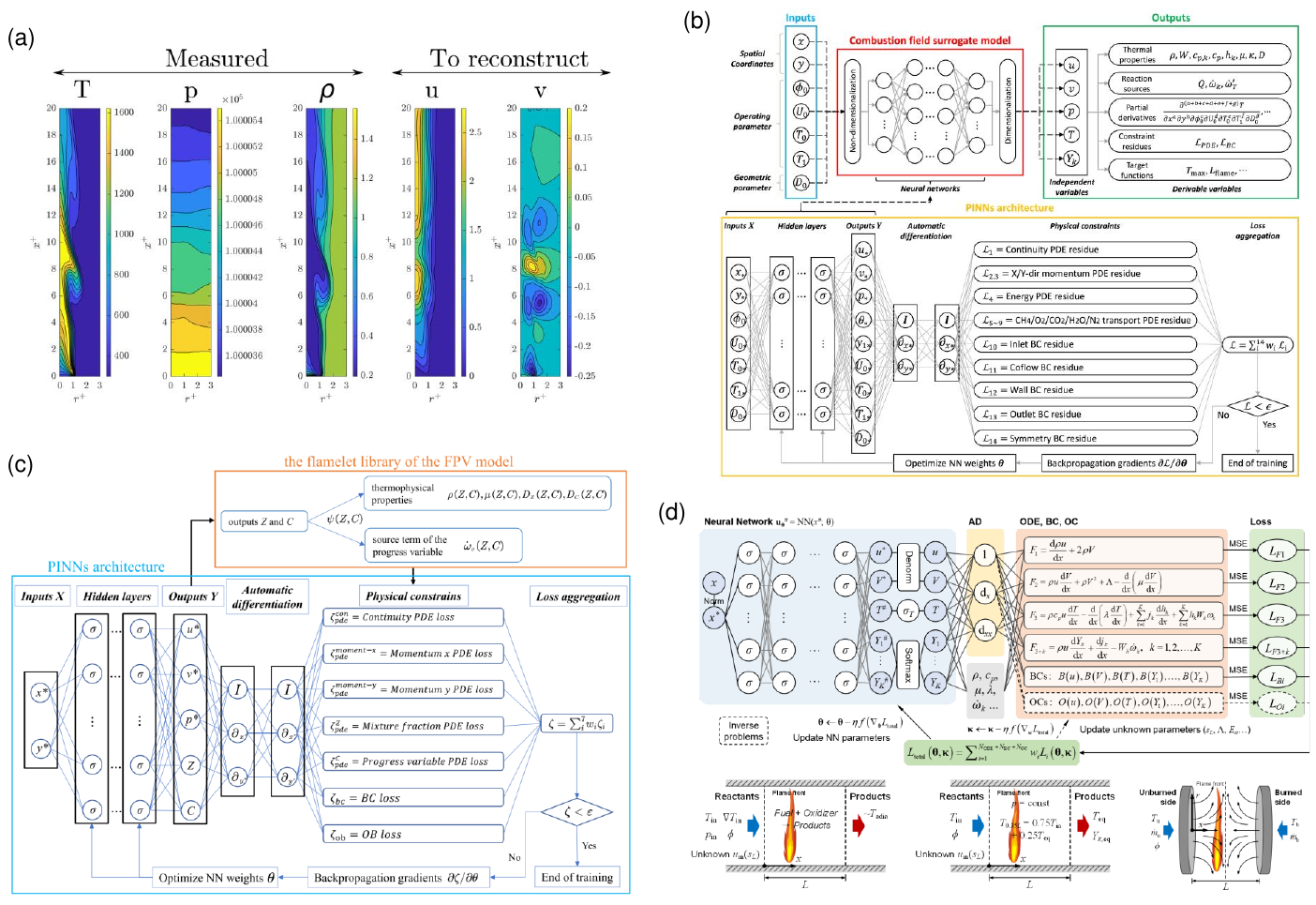}
    \caption{PINNs for laminar combustion. (a) Applying the HFM framework to reconstruct puffing pool fires \cite{Sitte202208_poolfire}. (b) Parameterized combustion field surrogate model based on PINNs \cite{Liu202309_surr-flame}. (c) PINNs coupled with the FPV model \cite{Song202410_PINN-FPV}. (d) FlamePINN-1D for solving forward and inverse problems of 1D laminar flames \cite{Wu202501_FlamePINN-1D}. Figures are reprinted with permission.}
    \label{fig: flow_laminar}
\end{figure}

This subsection will present a chronological overview of studies on PINNs for laminar subsonic combustion, some of which are illustrated in Figure \ref{fig: flow_laminar}. Choi \textit{et al.} \cite{Choi202201_CSTR} used PINNs to reconstruct the 3D transient fields of a continuous stirred tank reactor (CSTR) with van de Vusse reaction. The governing equations of two reference frames were considered to enable the reconstruction of the multi-reference frame system. The fields can be accurately reconstructed from the data of velocity, temperature, and species, and some proposed strategies, such as the similarity-based sampling, have been shown to improve the model performance. Wang \textit{et al.} \cite{Wang202207_ROBER_Z-eq} used PINNs to forward solve the $Z$-equation for a jet diffusion flame and obtained accurate results. Because the assumptions of unity Lewis number and infinite fast chemistry are adopted, the $Z$-equation is free of reaction source terms and the temperature and species can be determined solely by $Z$. Sitte \textit{et al.} \cite{Sitte202208_poolfire} used PINNs to reconstruct the velocity fields in puffing pool fires based on the observations of density, pressure, and temperature, following the HFM framework \cite{Raissi202002_HFM}, as illustrated in Figure \ref{fig: flow_laminar}(a). The continuity and momentum equations are taken as the physical losses and a two-stage training strategy is adopted. The results show that PINNs can provide satisfactory velocity results and can act as a physics-based denoiser to smooth out the noise on the measured data.

Liu \textit{et al.} \cite{Liu202309_surr-flame} applied PINNs to solve forward problems of 1D and 2D premixed combustion with temperature-dependent properties and a one-step global mechanism of CH$_4$. To the best of our knowledge, although some simplifications are adopted for the transport models and the reaction mechanism is simple, this is the first study on PINNs for combustion reacting flows with detailed physical models and chemical kinetics based on the laws of mass action and Arrhenius. More importantly, after validating the model accuracy on 1D and 2D problems, a surrogate model was constructed for the 2D combustion, where five parameters were added to the NN inputs so that the model can learn multi-condition solutions, as shown in Figure \ref{fig: flow_laminar}(b). The parameterized PINN surrogate model was shown to accurately predict the fields, be applicable to sensitive analysis, and be much more efficient for industrial designs than traditional numerical simulations. Song \textit{et al.} \cite{Song202410_PINN-FPV} applied PINNs to 2D laminar premixed flames considering detailed chemistry of CH$_4$, which was accomplished by integrating the flamelet/progress variable (FPV) model. The equations of mass, momentum, $Z$, and $c$ were solved by PINN, during which the thermophysical properties and source term of $c$ were extracted from the pre-tabulated FPV library, as illustrated in Figure \ref{fig: flow_laminar}(c). The model can accurately solve the fields with and without observation data, demonstrating the feasibility of PINNs to integrate detailed reaction mechanisms using tabulation methods. Using pseudo-time stepping PINNs (PTS-PINNs) \cite{Cao202410_illPINN}, Cao \textit{et al.} \cite{Cao202411_PTS-PINN-flame} solved the premixed and non-premixed flames with a one-step mechanism of CH$_4$ and obtained satisfactory results. Vanilla PINNs failed to solve the premixed combustion with large gradients and tended to predict non-reacting solutions, but PTS-PINNs showed great convergence performance not only for the premixed combustion but also for non-premixed combustion where the thin reaction zones introduced a higher level of complexity.

The above studies only involve forward problems or inverse problems of field reconstruction without parameter inference, which is another important aspect of inverse problems. In computational 1D laminar flames, which are important for the study of flame limit, stretch, laminar flame speed ($s_L$), and flamelet tabulation, unknown eigenvalues exist even in forward problems, such as $s_L$ in freely-propagating premixed flames \cite{Kee2005_react-flow, web_Chemkin-theory, web_Cantera-science}. Wu \textit{et al.} \cite{Wu202501_FlamePINN-1D} established the FlamePINN-1D framework to solve both forward and inverse problems of 1D laminar flames in a unified manner, as illustrated in Figure \ref{fig: flow_laminar}(d). Three cases were solved: freely-propagating premixed flames with simplified and detailed physical models, and counterflow premixed flames with detailed physical models. The detailed physical models are the same as those in Section (\ref{sec: multicomponent}) to (\ref{sec: fluid-dynamics}) with a one-step global mechanism of CH$_4$ being considered, except that the governing equations are for 1D reacting stagnation flows. A warmup pretraining strategy and some hard physical constraints were proposed. For example, the predicted species mass fractions are transformed by the Softmax function:
\begin{align}
	Y_k = \mathrm{Softmax}(Y_k^{\#}) = \frac{\exp(Y_k)}{\sum_{k=1}^{K} \exp(Y_k)}
    \label{eq: hard_species}
\end{align}
where $Y_k^{\#}$ is the output of NN. The transformation ensures that the range constraint ($0 \leqslant Y_k \leqslant 1$) and species conservation ($\sum_k Y_k = 1$, Equation \eqref{eq: species_conservation}) can be exactly satisfied, so it can be used for various combustion problems. $Y^{\#}$ can be viewed as the logarithmic normalization of $Y$ \cite{Zhang202408_CRK-PINN}, which can mitigate the multiscale of their magnitudes. The results show that for forward problems, FlamePINN-1D can accurately solve the flame fields and simultaneously infer the $s_L$ of freely-propagating premixed flames. For inverse problems, FlamePINN-1D can accurately reconstruct the continuous fields and infer the transport properties and chemical kinetics parameters from noisy sparse observations, which indicates the possibility of optimizing chemical mechanisms from observations of laboratory 1D flames. In fact, in their inverse problems where the activation energy was unknown, the chemical kinetics part can be considered as a CRNN with other parameters are known. It is proven that the model tends to get unburned solutions without warmup pretraining, the similar situation to Reference \cite{Cao202411_PTS-PINN-flame}. Other adopted strategies, such as hard constraints, thin-layer normalization \cite{Wu202401_VLT-PINN}, and modified MLP with random weight factorization (MMLP-RWF) \cite{Wang202308_PINNGuide}, have also been proven to be essential for the model performance.

In addition to the classical PINNs, the PINNs in a broad sense (Table \ref{table: narrow-broad}) can also be applied to laminar combustion problems. For example, Corrochano \textit{et al.} \cite{Corrochano202506_PA-ROM} developed a physics-aware ROM for an axisymmetric laminar CH$_4$-O$_2$ flame. The simulation data were preprocessed to obtain the proper orthogonal decomposition (POD) modes, then deep learning models were used as a time integrator to advance temporal modes in time, i.e., the models took the AR paradigm instead of the NAR paradigm adopted in classical PINNs (see Section \ref{sec: PINN_representation}). A physics-aware loss function was introduced which constrained the predicted species mass to be equal to the true ones in each time instant. This loss differed from the classical PINN losses in that it was still a data-driven loss without PDE information, so it was called "physics-aware" rather than "physics-informed", but can still be called "physics-informed" in a broad sense in our extended definition. Two models, including the LSTM and 1D CNN, were tested, demonstrating that the physics-aware loss could slightly improve the model performance.

\subsection{Turbulent combustion and experimental data}
\label{sec: turb-exp}

\begin{figure}[ht]
    \centering
    \includegraphics[width=1\textwidth]{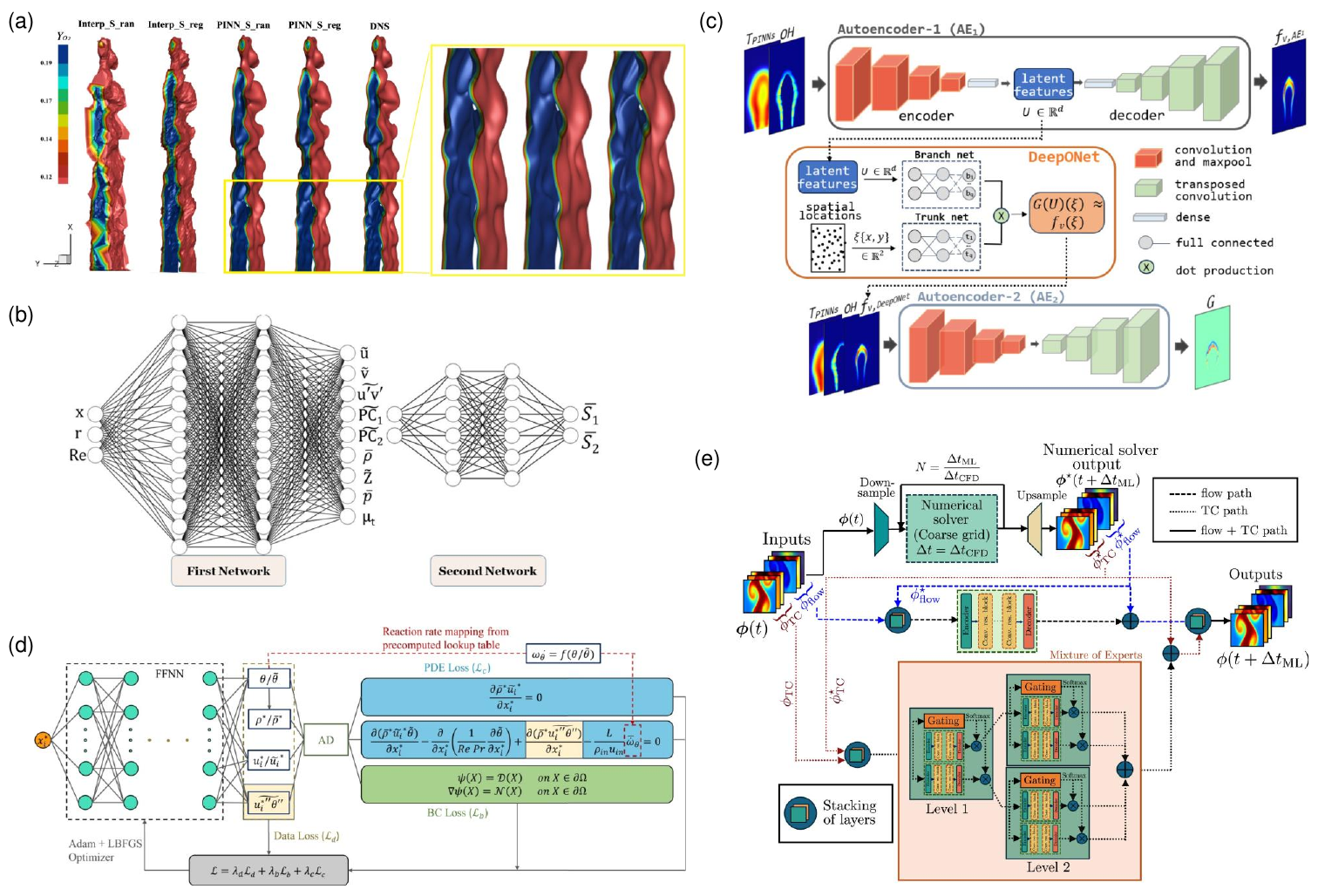}
    \caption{PINNs for turbulent combustion and experimental data. (a) PINNs reconstruct turbulent flames: comparisons of the instantaneous 3D distributions of $Y_{\mathrm{O}_2}$ from linear interpolation, PINNs and DNS data \cite{Liu202312_recons-turb}. (b) PINNs for turbulent combustion closure \cite{Taassob202310_PINN-TC-closure}. (c) The AE-based DeepONet framework to reconstruct the soot fields \cite{Liu202406_recons-soot}. (d) The framework of RF-PINNs \cite{Yadav202412_RF-PINN}. (e) The architecture of PiMAPNet \cite{Sharma202505_PiMAPNet}. Figures are reprinted with permission.}
    \label{fig: flow_turbulent}
\end{figure}

Compared to laminar combustion and using simulation data for learning or validating, the problems are more complex for turbulent combustion and when only practical experimental data are available due to high uncertainty of these situations. Some studies on PINNs for turbulent combustion and experimental data are shown in Figure \ref{fig: flow_turbulent}.

The most straightforward way to apply PINNs to turbulent combustion is directly integrating the original reacting flow equations, i.e., Equations \eqref{eq: mass} to \eqref{eq: T}, without any turbulence models or turbulent combustion models. Liu \textit{et al.} \cite{Liu202312_recons-turb} used PINNs to reconstruct the fields of 3D turbulent flames based on sparse, noisy data sampled from high-fidelity DNS data. Three cases were tested: freely-propagating planar premixed flames with low and high levels of turbulence, and slot-jet flames with a jet Reynolds number of 850. For the physical models, temperature-dependent properties, constant Prandtl number, and unity Lewis number were considered. A two-step mechanism of CH$_4$ was adopted, which consisted of six species and two reactions. Three PINN sub-models were employed to learn momentum, species, and temperature equations, respectively, and the species conservation was also considered in the losses (Equation \eqref{eq: species_conservation}). Self-adaptive linear layers were utilized and a warmup pretraining was conducted to facilitate model convergence. It is shown that the model can accurately reconstruct the fields and get satisfactory results of flow modes, vortical structures, and PDFs of variables and flame curvature, and the results were more accurate than the linear interpolation method, as shown in Figure \ref{fig: flow_turbulent}(a). Subsequently, Liu \textit{et al.} \cite{Liu202509_3Dfrom2D} applied NNs to 3D reconstruction of temperature and velocity from 2D measurements of turbulent flames. First, the 3D temperature field were reconstructed from 2D data using data-driven models: two vector-quantized VAEs first learned the latent spaces of 2D and 3D temperature fields separately, and then a DM parameterized by an unconstrained transformer learned the 2D to 3D translation using the learned vector-quantized space. After that, a wavenumber-based PINN (WN-PINN) was proposed to reconstruct the 3D velocity fields based on the 2D velocity and reconstructed 3D temperature data, where momentum equation was used as the physical constraint. To address the multiscale issue, the velocity field was decomposed into low and high wavenumber components and learned by two separate NNs. Two cases were tested: freely propagating planar premixed flames with three operating conditions, and a swirling premixed flame. Results showed excellent reconstruction for both temperature and velocity. More importantly, the results showed that for unseen cases, both pre-trained models and models trained on these new cases from scratch are not as accurate as the fine-tuned pre-trained models, demonstrating the effectiveness and efficiency of transfer learning.

In addition to direct turbulence, PINNs have also been widely applied to turbulent combustion assuming mean and filtered turbulence. As an early attempt, Nandi \textit{et al.} \cite{Nandi202111_SimNet-react, Nandi202203_Modulus-react} used PINNs to develop digital twins for 2D and 3D industry-scale boiler combustion without observation data (forward problems). The zero-equation model was used for the turbulent Reynolds stresses \cite{Nandi202111_SimNet-react} and various reaction source terms were tested, including the linear form of two species \cite{Nandi202111_SimNet-react} and one-step mechanism of CH$_4$ \cite{Nandi202203_Modulus-react}. Some strategies were proposed to handle the coupling between the source terms of $T$ and $Y$. Von Saldern \textit{et al.} \cite{vonSaldern202211_mean-assi} applied PINNs to the assimilation of turbulent mean flow data of axisymmetric jet flows. In their first example, the mean density fields of a turbulent jet flame were reconstructed based on the data of velocity fields, which were generated by LES. Although only the continuity equation was considered in the physical loss, the velocity and density fields were accurately reconstructed. Peng \textit{et al.} \cite{Peng202404_recons-flameD} used PINNs to reconstruct the heat flow field of Sandia Flame D \cite{Barlow199808_Sandia-flame} based on the data of velocity and temperature, which were obtained by simulation with the partially premixed combustion model. The predicted variables and physical losses did not consider the species and chemical reactions, but considered the turbulence kinetic energy ($k$), its dissipation rate ($\varepsilon$), and their transport equations, which is a common turbulence model. The results showed that PINNs could accurately reconstruct the fields within both small and large time intervals. Compared to data-driven methods, PINNs was shown to yield more accurate results and require less data when accuracy requirements are the same. Cao \textit{et al.} \cite{Cao202509_PINN-turb-combus} applied PINNs for modeling 2D steady turbulent combustion of CH$_4$ and air, where the $k-\varepsilon$ model and eddy dissipation model (EDM) were adopted for modeling turbulence and reaction rates, respectively. For forward problems, PINNs could successfully simulate the non-reacting jet flows but exhibited noticeable discrepancies from reference fields for reacting cases, demonstrating the significant challenges of forwardly solving turbulent combustion problems using PINNs. With a small amount of velocity data, the PINN results achieved a high degree of accuracy. The superiority of PINNs over traditional methods for surrogate modeling was also validated by the results of parametric modeling.

For experimental combustion, there are discrepancies between the data and pre-specified physical models, especially for turbulence and reaction models. As shown in Table \ref{table: reacting-flow}, two strategies are adopted in existing studies to deal with this issue: (1) Reduce or simplify the models; (2) Learn the models, i.e., closure terms. Taassob \textit{et al.} \cite{Taassob202310_PINN-TC-closure} used PINNs to evaluate closure terms for turbulence and chemical source terms in the Sandia turbulent nonpremixed flames \cite{Barlow199808_Sandia-flame}. As illustrated in Figure \ref{fig: flow_turbulent}(b), the NN took coordinates ($x$, $r$) and Reynolds number as input for parametric learning, and output not only the mean flow fields, including $\overline{\rho}$, $\widetilde{\mathbf{U}}$, $\overline{p}$, $\widetilde{Z}$, and two principal components ($\widetilde{\mathrm{PC}}$) of $T$ and $Y$, but also the Reynolds stress ($\widetilde{u'v'}$) and turbulent viscosity ($\mu_t$) to learn the turbulence closure, which was achieved by adding the $\mu_t$ relation to the physical losses:
\begin{align}
	-\overline{\rho} \widetilde{u'v'} = \mu_t \left( \frac{\partial \widetilde{u}}{\partial r} + \frac{\partial \widetilde{v}}{\partial x} \right)
\end{align}
To learn the reaction closure, a second NN was used to learn the mapping from $\widetilde{\mathrm{PC}}$ to their averaged reaction source terms. The PINNs were trained on Sandia flames D and F, and were validated on flames D, F, and E. The results showed that the reconstructed fields matched well with the experimental data and the learned closure terms were physically reasonable. Later, Taassob \textit{et al.} \cite{Taassob202406_PI-DeepONet-TC-closure} extended this method to Sydney flames. The extensions include: (1) Using DeepONet instead of MLP because more parameters are involved than Sandia flames. (2) Using six $Y_k$ instead of $\widetilde{\mathrm{PC}}$ and learn the six reaction source terms, whose estimated observations were available. A two-stage training was conducted to learn the turbulence and reaction closures sequentially. The model was trained on three flames and validated on four flames and could predict satisfactory flow fields and species source terms.

Liu \textit{et al.} \cite{Liu202406_recons-soot} used PINNs to reconstruct the soot fields in acoustically forced laminar sooting flames in two steps. In the first step, since the velocity data measured by particle image velocimetry (PIV) were missing in sooting regions, and the temperature data measured by non-linear excitation regime two-line atomic fluorescence (nTLAF) were missing in low-temperature regions, a PINN was used to reconstruct the velocity and temperature fields. The momentum and temperature equations were used as physical losses where the heat release and radiation were neglected due to low temperatures. Results showed that PINN could learn the available data well and reconstruct the missing fields. Next, a physics-informed AE-based DeepONet (PI-AE-DeepONet) with two AEs was used to reconstruct the soot fields, as shown in Figure \ref{fig: flow_turbulent}(c). The simplified transport equation of soot mass fraction was taken as the physical loss where the unknown source term was learned by the second AE. The first AE was used to learn the latent features of OH images and PINN-reconstructed $T$ images via supervised learning. The DeepONet took the latent features as the branch inputs and output the soot volume fraction. It was shown that PI-AE-DeepONet yielded better predictions than various data-driven models, including U-Net, FNO, the first AE, and AE-DeepONet, and could learn the source terms accurately.

Yadav \textit{et al.} \cite{Yadav202412_RF-PINN} developed the reactive flow PINNs (RF-PINNs) to reconstruct all flow quantities of interest solely based on sparse velocity or temperature data. The framework of RF-PINNs is given in Figure \ref{fig: flow_turbulent}(d). Three cases were tested: 1D freely-propagating flames, 2D laminar slit flames, and 2D experimental turbulent flames. Based on some assumptions, the equations of $T$ and $Y_k$ were reduced to the equation of the progress variable $c$, which was denoted as $\theta$ in the original paper. For laminar cases, the reaction source term of $c$ was pre-learned by a data-driven NN: $\dot{\omega}_c = \mathrm{NN}(c)$, while for turbulent cases, the source term was computed by the eddy break-up (EBU) model:
\begin{align}
	\overline{\dot{\omega}}_c = C \frac{\varepsilon}{k} \overline{\rho} \widetilde{c} (1 - \widetilde{c})
\end{align}
where $C$ was a model constant. In the first case, by comparative analysis, it was shown that using only the mass and $c$ equations was enough to get accurate results, so the momentum equation was not considered in the latter two cases to alleviate the computational burden. In the third case, two scalar turbulent flux terms were added to the NN output so that they can be learned by the NN instead of being explicitly modeled. The results showed that either from sparse velocity or temperature data, the missing fields could be well reconstructed.

Discrete PINNs have also been applied to turbulent combustion, including field reconstruction and AR learning. For field reconstruction, especially for SR reconstruction, details will be introduced in Section \ref{sec: combus_SR}. For AR learning, Sharma \textit{et al.} \cite{Sharma202505_PiMAPNet} applied the PIML approach to predict the dynamic behaviour of hydrogen jet flames in an AR manner (see Section \ref{sec: PINN_representation}). As illustrated in Figure \ref{fig: flow_turbulent}(e), the physics-informed hybrid multiscale and partitioned network (PiMAPNet) was constructed that integrated low-resolution physics-based models (numerical solver) with trainable NNs, so it can be viewed as a hybrid solver. The coarse grid numerical solver served as the physics-informed component of the model, which was similar to the physics-informed way in References \cite{Zhang202208_Mscale-sample, Mao202307_DeepFlame, Wang202503_ANN-hard} that integrated a numerical solver. Three other methods were selected for comparison: a coarse grid numerical solver without ML component, a purely data-driven CNN without physics-informed component, and a PPNN variant. As mentioned in Section \ref{sec: PINN_other-topic}, the original PPNN preserves/embeds the PDEs by discretizing the PDE RHS using convolutions filters instead of a numerical solver. Compared to the PPNN variant, PiMAPNet decomposed the state vector into hydrodynamic/flow variables and thermochemical variables. The flow variables were learned by a single CNN while the thermochemical variables were learned by a mixture-of-experts (MoE) that partitioned the thermochemical state-space so that separate CNNs could be assigned to different regions of the manifold. Results showed that the data-driven CNN yielded unphysical values and the PPNN variant exhibited the hotspot issue. In contrast, PiMAPNet successfully mitigated the issues of bias accumulation and hotspots and showed superior extrapolation performance. It was shown that the MoE effectively partitioned the manifold in a physically meaningful way, distinctly isolating the fuel-rich, fuel-lean, and mixing regions, which was the core reason for its superior performance. Speed analysis also showed that PiMAPNet was faster than a numerical solver with comparable accuracies.

\subsection{Supersonic combustion}

\begin{figure}[ht]
    \centering
    \includegraphics[width=0.9\textwidth]{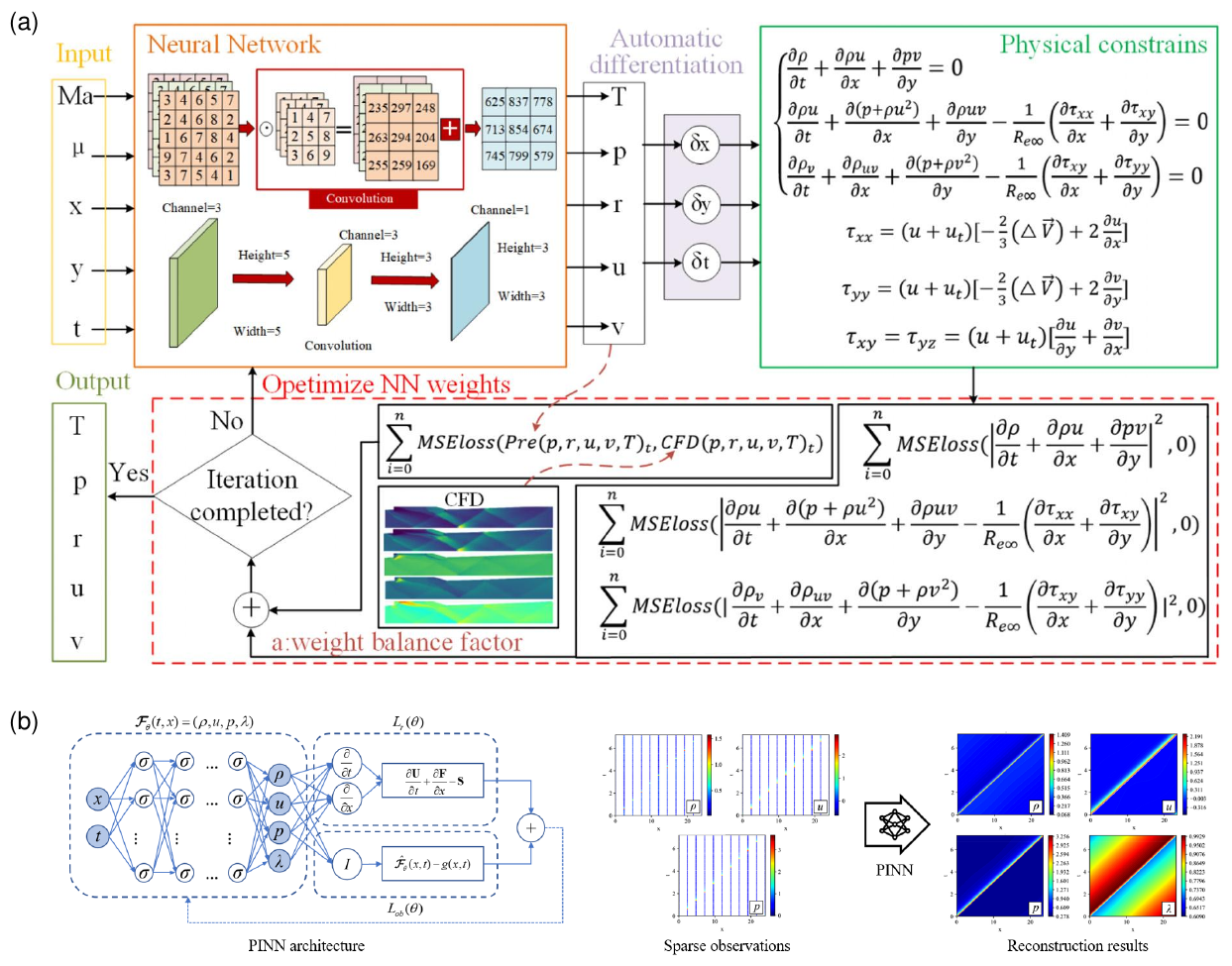}
    \caption{PINNs for supersonic combustion. (a) PI-ResUNet for reconstructing the combustion flow field in a scramjet combustor \cite{Deng202407_scramjet}, where $r$ indicates the density ($\rho$) and $\mu$ includes both the turbulent and laminar viscosity coefficients. (b) Illustration of the flow field reconstruction problem for RDCs \cite{Wang202307_PINN-RDC}. Figures are reprinted with permission.}
    \label{fig: flow_supersonic}
\end{figure}

For supersonic combustion, the core feature is the presence of shock waves and their interaction with combustion, which poses challenges for experimental implementation, numerical solutions, and PINNs. In the early work of Mao \textit{et al.} \cite{Mao202109_DeepMMnet-hypersonic}, they applied the DeepM\&Mnet \cite{Cai202103_DeepMMnet} for hypersonics to predict the coupled flow and finite-rate chemistry behind a normal shock. DeepM\&Mnet was proposed as a data assimilation framework for fast simulating multiphysics and multiscale problems, where the pre-trained DeepONets were used as building blocks to constrain the solutions \cite{Cai202103_DeepMMnet}. In Reference \cite{Mao202109_DeepMMnet-hypersonic}, the global mass conservation was constrained by adding a loss term, thus making the framework physics-informed. Comparative analysis demonstrated the effectiveness of this constraint on accuracy improvement. Guo \textit{et al.} \cite{Guo202310_scramjet-inlet} developed the physics-informed residual CNNs to reconstruct the flow field at the inlet of scramjet with the Mach number $Ma = 10$, where the flow was non-viscous and non-reacting. After this study, Deng \textit{et al.} \cite{Deng202407_scramjet} developed a physics-informed U-shaped residual NN (PI-ResUNet) to reconstruct the unsteady combustion flow field in a scramjet combustion chamber with $Ma = 2$. As shown in Figure \ref{fig: flow_supersonic}(a), the 2D unsteady compressible NS equations without reaction source terms were taken as the physical constraints and the operating parameters (Mach number and effective viscosity coefficient) were taken as the inputs to form a parametric learning. Comparing the PI-ResUNet, ResNet, PI-ResNet, and PI-UNet, it was demonstrated that introducing the physical constrains and residual structures could effectively improve the model robustness. Note that although the hidden layers of PI-ResUNet are convolutional layers, the inputs and outputs are still linear layers. Therefore, it is still a continuous PINN that uses AD for differentiation instead of the convolutional differentiation adopted in discrete PINNs (Section \ref{sec: PINN_representation}). Nasir \textit{et al.} \cite{Nasir202507_PINN-rocket} used PINNs for modeling combustion dynamics in rocket engine chambers, where the 2D governing equations of mass, momentum, energy, and species were considered. With a small amount of data, PINNs were demonstrated to properly model the intricate pressure oscillations and combustion processes.

Wang \textit{et al.} \cite{Wang202307_PINN-RDC} applied PINNs to reconstruct the combustion field in rotating detonation combustors (RDCs) based on partially observed data. As shown in Figure \ref{fig: flow_supersonic}(b), a PINN was built to describe the 1D RDC numerical model and estimate the continuous physical fields, taking the space-time coordinates as inputs and physical fields as outputs. In this study, the observations included only ten sets of points with fixed spatial coordinates, representing measured data with fixed sensors. The studied problem was an inverse problem, where the equations and partial true results were known, but the overall solution of the equations was unknown. Results showed that compared to interpolation-based methods, PINN could better capture different distribution patterns for the flow characteristics before and after the detonation wave, which matched the ground truth satisfactorily. The proposed PINN can serve as a surrogate model of the RDC, which has promising engineering value in the design optimization and real-time monitoring of RDCs. After this study, Wang \textit{et al.} \cite{Wang202311_PI-GAN-RDC, Wang202408_PI-RGAN-RDC} further used the discrete PINNs for SR in RDCs, which will be discussed in Section \ref{sec: combus_SR}.

\subsection{Combustion super-resolution via discrete PINNs}
\label{sec: combus_SR}

\begin{figure}[ht]
    \centering
    \includegraphics[width=1.0\textwidth]{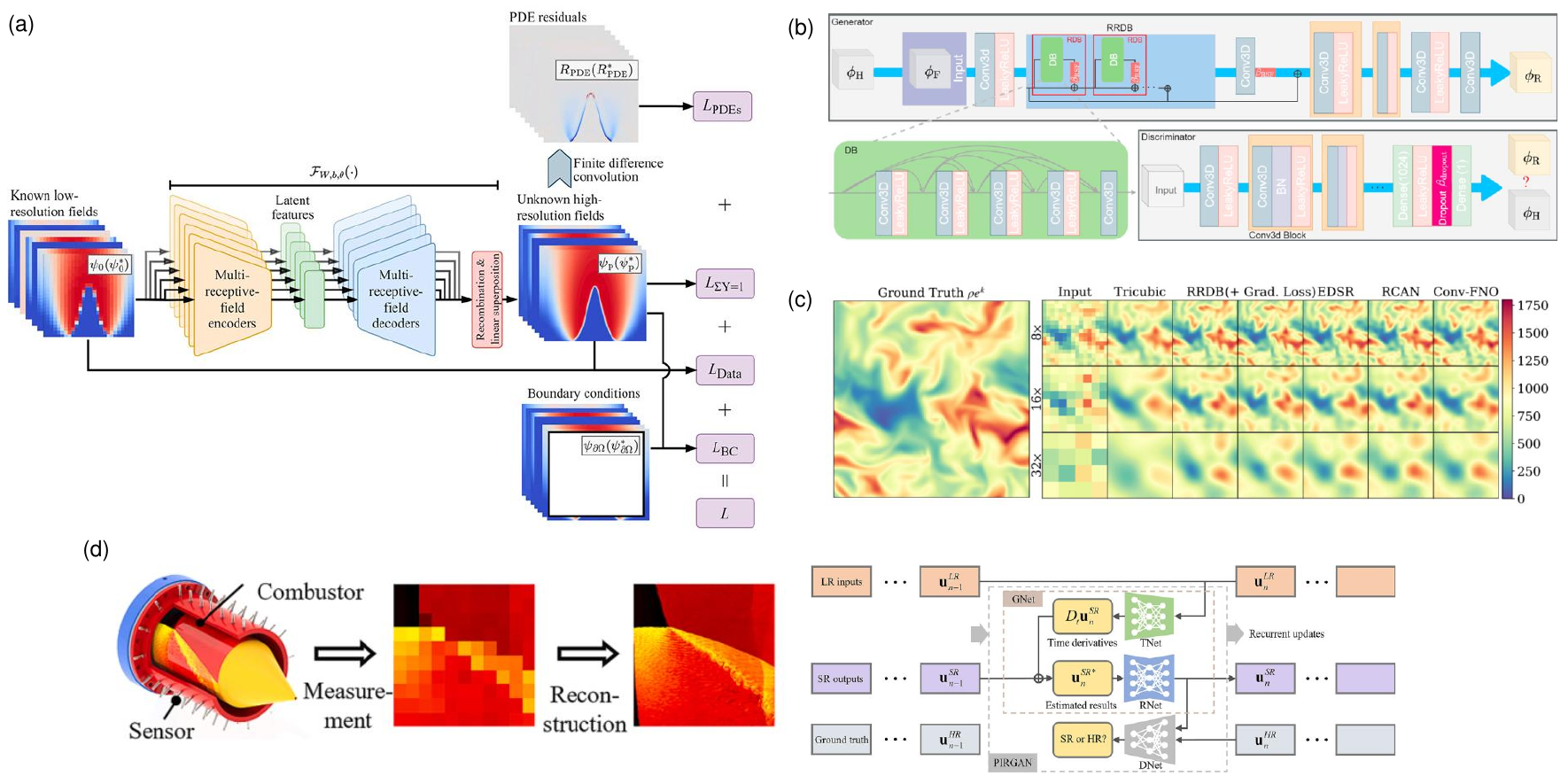}
    \caption{Discrete PINNs for SR reconstruction of laminar, turbulent, and supersonic combustion. (a) CRF-PINN for SR reconstruction \cite{Zhang202509_CRF-PINN}. (b) The architecture of PI-ESR-GAN \cite{Bode202101_subfilter}. (c) 3D volumetric SR with different models \cite{Chung202309_BLASTNet2.0}. (d) PI-RGAN for the SR reconstruction of flow fields in RDCs \cite{Wang202408_PI-RGAN-RDC}. Figures are reprinted with permission.}
    \label{fig: flow_SR}
\end{figure}

As mentioned in Section \ref{sec: PINN_representation}, discrete PINNs can be applied to SR reconstruction tasks. This field-to-field modeling paradigm is particularly suited for SR reconstruction tasks, offering advantages such as enhanced scalability for high-dimensional data, improved generalization to complex working conditions, and efficient handling of structured grid-based simulation results. Some studies on discrete PINNs for SR reconstruction of laminar, turbulent, and supersonic combustion are illustrated in Figure \ref{fig: flow_SR}.

Based on the multi-receptive-field CNN \cite{Zhang202411_MRF-PINN}, Zhang \textit{et al.} \cite{Zhang202509_CRF-PINN} developed the combustion reacting flow PINNs (CRF-PINNs) for field reconstruction of laminar and turbulent combustion. CRF-PINNs can be used in three typical scenarios of field reconstruction: cross-field translation, SR reconstruction (as shown in Figure \ref{fig: flow_SR}(a)), and denoising, where the reacting flow equations in a cylindrical coordinate system and the species conservation serve as the physical constraints, eliminating the need for high-fidelity labeled data during training. For the laminar Bunsen flame, with one of the velocity, temperature, and species fields, CRF-PINNs can accurately reconstruct the other two. The cross-field translation was also demonstrated for unsteady Bunsen flames. For SR reconstruction and denoising, CRF-PINNs showed superior performance over the corresponding data-driven baseline models. For turbulent flames, the turbulent viscosity and the turbulence-chemistry interaction coefficient are learned by the NN along with unknown high-fidelity fields. The Sandia flame F \cite{Barlow199808_Sandia-flame} was tested for sparse reconstruction from partial, experimental measurements. The 13 measurement points along the centerline were used for evaluation, demonstrating satisfactory model accuracy. In summary, this work proves the feasibility of discrete PINNs for various scenarios of field reconstruction, and again showcases the treatment of turbulence and chemistry in PINNs.

Given the inherent complexity of combustion modeling, researchers have further augmented the regular discrete PINNs with generative modeling techniques, such as GANs. Compared to discriminative modeling approaches, generative modeling demonstrates superior capability in capturing small-scale turbulence dynamics \cite{Bode202101_subfilter} and refining predictions of discontinuous features \cite{Wang202311_PI-GAN-RDC}. Bode \textit{et al.} \cite{Bode202101_subfilter} first used the physics-informed enhanced super-resolution generative adversarial network (PI-ESR-GAN) to model turbulent combustion. As shown in Figure \ref{fig: flow_SR}(b), the PI-ESR-GAN framework combines unsupervised deep learning with physics-informed loss functions, employing a two-step training methodology to enhance network generalization, particularly for extrapolation tasks. In decaying turbulence tests involving turbulent mixing, the model demonstrated strong \textit{a priori} and \textit{a posteriori} performance, with the physics-informed continuity loss term proving essential. The PI-ESR-GAN methodology has been extended to diverse combustion configurations, demonstrating remarkable versatility. In spray combustion applications, the model was successfully implemented for different cases, with validation against experimental data revealing superior predictive capability compared to classical LES approaches \cite{Bode202201_spray}. For lean premixed combustion and gas turbine combustors, by developing a modified architecture for finite-rate chemistry flow, the enhanced model demonstrated superior accuracy in both \textit{a priori} and \textit{a posteriori} validations while maintaining computational efficiency \cite{Bode202210_FRC}. For turbulent non-premixed flames, a multi-mesh strategy was developed to optimize prediction accuracy across different grid resolutions, enabling reliable extrapolation to untrained Reynolds numbers \cite{Bode202210_nonuniform}. The approach was subsequently extended to premixed combustion \cite{Bode202302_turbulent}. For that, the physical information in the loss function were adjusted, the training process was smoothed, and especially the effects of density changes were considered. When applied to DNS data of turbulent premixed flame kernels, the enhanced model achieved remarkable accuracy in reproducing fine-scale dynamics on coarse computational meshes. This advancement enables more efficient statistical analysis, while significantly reducing computational costs compared to traditional approaches. Using high-fidelity DNS data, Chung \textit{et al.} \cite{Chung202309_BLASTNet2.0} benchmarked the scaling behaviour of five DL models on the 3D combustion SR task, as shown in Figure \ref{fig: flow_SR}(c). One of the models used the ESR-GAN (referred to as RRDB in the paper) and incorporated a physics-based loss (i.e., physics-informed loss in a broad sense), which penalized the gradients of channel variables in a supervised manner. It was demonstrated that the benefits of this loss can persist with increasing model size.

For supersonic combustion, Wang \textit{et al.} \cite{Wang202311_PI-GAN-RDC} developed the PI-GAN for flow field SR reconstruction on RDCs. The model utilized a GAN architecture enhanced with additional physical loss terms to ensure the outputs adhere to fundamental physical principles. Compared to interpolation and other methods, PI-GAN greatly improved the resolution of the flow field while accurately capturing detailed flow structures. Subsequently, Wang \textit{et al.} \cite{Wang202408_PI-RGAN-RDC} proposed an advanced physics-informed recurrent GAN (PI-RGAN), as shown in Figure \ref{fig: flow_SR}(d). This innovative framework incorporated temporal derivatives as intermediate variables during the data generation process within a recurrent architecture, while simultaneously employing spatial derivatives of the output to establish comprehensive physical constraints. Experimental analysis revealed some features of the proposed model. The PI-RGAN model could well capture the detailed flow field structures and temporal characteristics in RDCs. Moreover, the introduction of a recurrent structure in the PI-RGAN model enhanced the model robustness against noisy inputs.

\section{PIML for other combustion scenarios}
\label{sec: PIML_other-combus}

\begin{figure}[ht]
    \centering
    \includegraphics[width=1\textwidth]{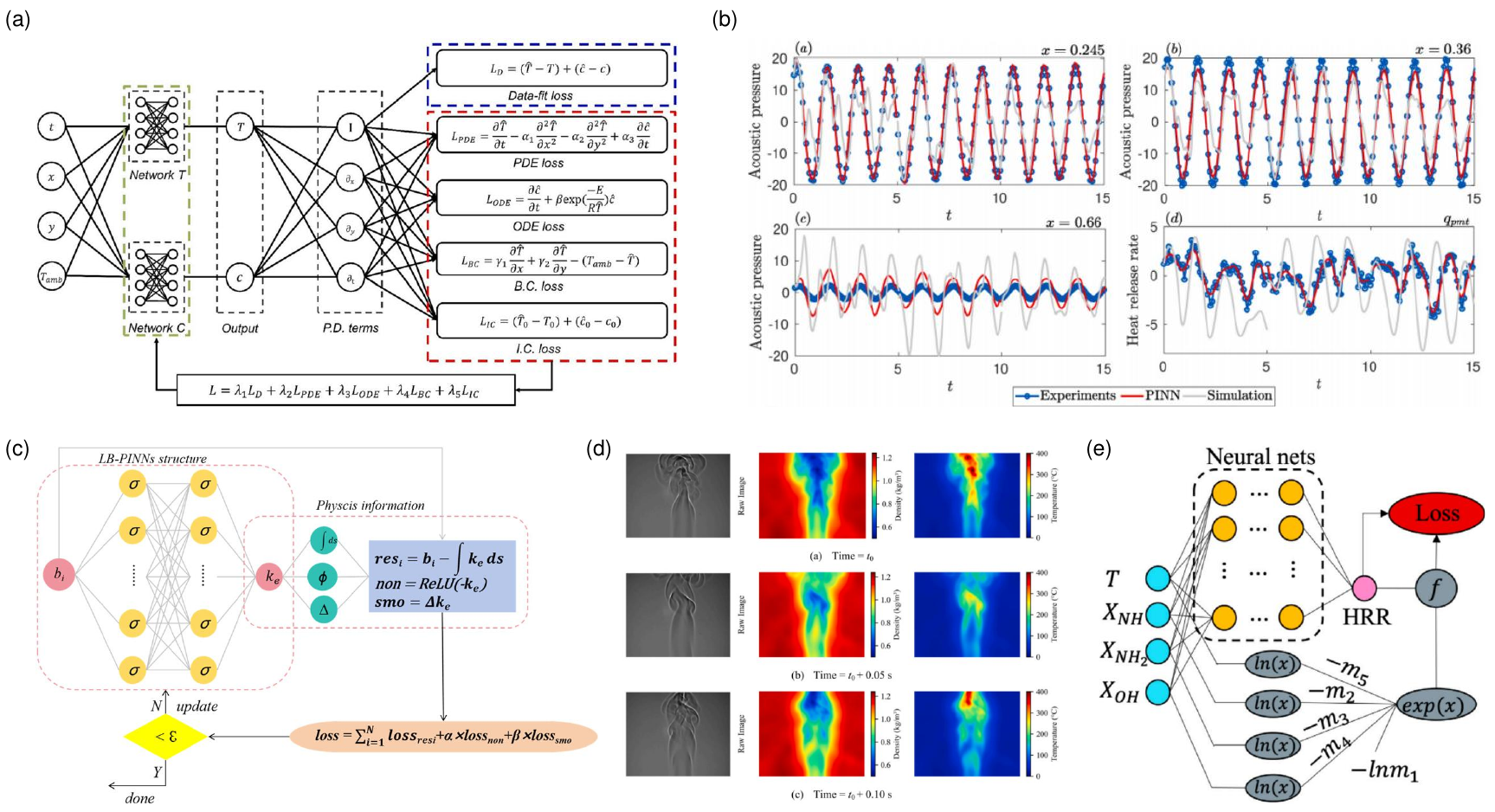}
    \caption{PIML for fire safety, thermoacoustics, combustion diagnostics, and others. (a) PINNs for modeling and prediction of lithium-ion battery thermal runaway \cite{Kim202301_Li-runaway}. (b) PINNs to learn thermoacoustic interactions in combustors \cite{Mariappan202410_PINN-thermoacoustic}. (c) The structure of Lambert-Beer law PINN (LB-PINN) \cite{Wang202410_LB-PINN}. (d) PINNs reconstruct density and temperature fields from shadowgraph images of alcohol burner flames \cite{Wang202507_PI-shadowgraph}. (e) PIML to discover the HRR markers for premixed NH$_3$/H$_2$/air flames \cite{Chi202208_HRR}. Figures are reprinted with permission.}
    \label{fig: other-combus}
\end{figure}

Some studies on PIML for combustion cannot be classified to the previous two sections but are also important and enlightening. Therefore, they will be introduced in this section, which are categorized into four parts: fire safety, thermoacoustics, combustion diagnostics, and others. Some of these studies are illustrated in Figure \ref{fig: other-combus}.

\subsection{Fire safety}

Fire safety constitutes a critical discipline focused on the prevention, detection, and mitigation of uncontrolled fires, aiming to protect life, property, and the environment against fire-related hazards. However, predicting their complex dynamics remains a challenge due to the multiscale and multiphysics processes, including turbulence, heat and mass transfer, soot formation, radiation, and possible explosion. Given the advantages of PIML in integrating data and physics, researchers are exploring its applicability in advancing fire science and enhancing fire safety.

Bottero \textit{et al.} \cite{Bottero202010_PIML-wildfire} replaced the PDE solver in the fire simulator to achieve real-time prediction of wildfire spread, where the level set equation was utilized. Tests on an ideal case and a real wildfire both demonstrated that the PINNs were able to capture the shape of the firelines. Dabrowski \textit{et al.} \cite{Dabrowski202304_BPINN-wildfire} used PINNs to solve the level set equation for modeling the wildfire fire-front. A forecast likelihood was introduced to deal with the issue that PINNs failed to maintain temporal continuity when there are extreme changes in exogenous forcing variables. Moreover, the B-PINN was incorporated to provide UQ in fire-front predictions. The models were successfully applied to modeling two recorded grassland fires. Vogiatzoglou \textit{et al.} \cite{Vogiatzoglou202411_PINN-wildfire} applied PINNs to infer the parameters in the wildfire spread model, consisting of energy and mass fraction equations. The parameters included the dispersion coefficient vector, mean gaseous velocity vector, and the overall heat transfer coefficient, so the number of parameters was three in 1D cases and five in 2D cases. Five cases were tested, including both clean and noisy data, and both synthetic and real-world data, demonstrating the accuracy of parameter inference and wildfire prediction. Some other wildfire-related applications do not focus on the spread prediction. For post-fire applications, Seydi \textit{et al.} \cite{Seydi202408_PIML-VBS-SBS} used PIML to estimate the soil burn severity and reveal the intricate relationships between soil and vegetation burn severities. For firefighting, the AI Firefighter \cite{Emekci202410_AIFirefighter} was proposed to predict the optimal firefighting strategy given an arbitrary terrain and an arbitrary fire distribution, where the NN combined PIML with a decision-making process.

Other applications are not related to wildfire. As illustrated in Figure \ref{fig: other-combus}(a), Kim \textit{et al.} \cite{Kim202301_Li-runaway} used PINNs to model and predict the lithium-ion battery thermal runaway, which is an important topic of fire safety. The PDEs of temperature and lithium concentration served as the governing equations, which was a DR system. The model applicability to semi-supervised learning, unsupervised learning, and surrogate modeling was investigated, showing that PINNs exhibited better performance than purely data-driven NNs. Lu \textit{et al.} \cite{Lu202406_PI-DeepONet-radia} used the PI-DeepONet for surrogate modeling of the radiative heat transfer, where the radiation transfer equation (RTE) and its BCs were considered as the physical losses. The model took the location and propagation direction as trunk net inputs, the parameters (such as the absorption coefficients) as branch net inputs, so the solution operator of the RTE could be approximated. The surrogate model offers significant speedup over traditional methods, thus possessing the potential to replace the traditional RTE solvers in industry-scale fire simulations, where the solution of RTE is often a major performance bottleneck. Parsa \textit{et al.} \cite{Parsa202503_PINN-non-charring} applied PINNs to predict the transient burning rates of non-charring materials, where the energy equations in the temperature rise phase and burning phase were considered in the loss function. Results showed that PINNs could effectively capture the intricate dynamics of transient burning behavior and were more efficient than traditional numerical methods.

\subsection{Thermoacoustics}

Thermoacoustics concerns the two-way coupling between unsteady heat release and acoustic pressure waves, a fundamental phenomenon in combustion engines and combustors. Understanding this interaction is critical for both harnessing its potential in efficient thermoacoustic engines and mitigating the dangerous pressure oscillations that can lead to system failure. The predictive modeling of thermoacoustic instabilities remains challenging due to its nonlinear, multiscale, and multiphysics nature, which are now being addressed by PIML.

Ozan \textit{et al.} \cite{Ozan202208_PA-limit-cycle} used PINNs to learn the nonlinear limit cycles and adjoint limit cycles occurred in thermoacoustic oscillations. The physics was informed in two ways: (1) the governing equations that are derived from mass, momentum and energy conservation were integrated into the loss function, and (2) the periodicity was enforced using periodic activation functions in the NN. It was demonstrated that the PDE losses could promote the model performance in noisy-data and scarce-data situations, and only the NN with periodic activation functions could accurately extrapolate to the validation set. Mariappan \textit{et al.} \cite{Mariappan202410_PINN-thermoacoustic} used PINNs to learn thermoacoustic interactions in a bluff body anchored flame combustor, where three PDEs arising from the acoustic model and the van der Pol (VDP) oscillator model served as the physical losses. Based on the available data of the acoustic pressure fluctuations at three locations and the total flame heat release rate, the model aimed to identify four physical parameters and recover the entire fields of the acoustic pressure, velocity, and total heat release. Five cases were tested in the order of increasing complexity with the final case using the experimental data. As shown in Figure \ref{fig: other-combus}(b), the satisfactory results indicated the potential of PINNs to improve existing or design new thermoacoustically stable and structurally efficient combustors. Xie \textit{et al.} \cite{Xie202410_PINN-EVDP} extended the VDP equation by introducing three parameters to enhance the model applicability, and then a PINN was applied to infer these parameters, where the data were generated for a horizontal Rijke tube with acoustically compact heat sources. The optimized PINN-EVDP model was shown to be capable of replicating the thermoacoustic system instability behaviour. Further investigations were conducted to compare the Hopf bifurcation and amplitude death characteristics of the coupled extended VDP system with the coupled theoretical Rijke tube systems, showing the physical significance of the extended VDP system in the field of thermoacoustics.

\subsection{Combustion diagnostics}

Combustion diagnostics encompasses the techniques and methods for quantitatively measuring key parameters in combustion processes and fields. Acquiring high-fidelity experimental data is indispensable for understanding fundamental combustion physics, validating numerical models, and guiding the design of clean and efficient combustion systems. However, these measurements are challenged by the harsh conditions of practical combustors, limited optical access, and the ill-posed nature of inferring quantitative data from indirect signals like flame emissions. Since PIML has been widely used to aid experimental fluid mechanics (see Section \ref{sec: PIML_application}), combustion researchers have also been exploring its potential for combustion diagnostics.

Wang \textit{et al.} \cite{Wang202410_LB-PINN} used PINNs to reconstruct the 3D soot volume fraction in flames, where the NN took the 2D images of the accumulated projection value of the extinction coefficients as input, and output the local 3D extinction coefficients, as shown in Figure \ref{fig: other-combus}(c). The losses included the physical loss of the Lambert-Beer law, the boundary loss that constrained the positivity of the extinction coefficients, and the smoothness loss as a regularization term. Tested on two numerical cases and an experimental case, the model showed superior performance compared to conventional reconstruction methods. Later, Wang \textit{et al.} \cite{Wang202412_LoS} used PINNs to simultaneously reconstruct the soot temperature and volume fraction fields in experimental laminar ethylene flames. The inputs of the NN were the 2D matrix of the experimental soot radiation field and flame thickness in the line-of-sight (LoS), and the outputs were soot temperature, volume fraction fields, and an auxiliary variable. The physical loss was a first-order PDE for the auxiliary variable, which was derived from the LoS soot radiation integral equation. It was demonstrated that the physical constraint improved not only the predictive capability but also the adaptability and reliability across various combustion conditions.

Wang \textit{et al.} \cite{Wang202507_PI-shadowgraph} proposed a physics-informed shadowgraph network to reconstruct both density and temperature fields directly from shadowgraph images without labeled data, which first adopted a self-supervised AE to learn the light intensity fields and then employed a discrete PINN to reconstruct the refractive index fields under the constraint of the shadowgraph refractive equation. The temperature and density fields were computed based on the refractive index fields. This method effectively mitigates the sensitivity to BCs inherent in traditional numerical approaches and demonstrates high accuracy across various flow scenarios, including hot air jets, thermal plumes, and alcohol burner flames, as shown in Figure \ref{fig: other-combus}(d).

\subsection{Others}

Chi \textit{et al.} \cite{Chi202208_HRR} applied PIML to discover the heat release rate (HRR) marker for premixed NH$_3$/H$_2$/air flames. Specifically, a physical loss term was added to infer the free parameters in the approximation equation for HRR, as illustrated in Figure \ref{fig: other-combus}(e). Several markers were identified and then tested on various operating conditions to find the most accurate combinations. Hosseini \textit{et al.} \cite{Hosseini202210_PINN-Bratu} used the PINNs to solve the steady-state Bratu equation arising from solid biofuel combustion theory, which was a 1D nonlinear boundary value problem. Results showed that the trained PINNs were more accurate than the decomposition method and Laplace method. Yadov \textit{et al.} \cite{Yadav202210_PINN-flame-dyn} developed the PI-LSTM to predict the linear and nonlinear response of a premixed laminar flame to incoming velocity perturbations. The low-frequency limit for perfectly-premixed flames was incorporated as a physical loss term, which effectively improved the prediction accuracy of the flame transfer functions in the minimal data limit. Kim \textit{et al.} \cite{Kim202211_cetane} developed the PI-GNN to predict the cetane number (CN) with systematic data quality analysis. Based on the physical insights, the CN-related physical properties were embedded in the global state input, thus improving the model accuracy.

\section{Public resources}
\label{sec: resource}

\begin{figure}[ht]
    \centering
    \includegraphics[width=0.85\textwidth]{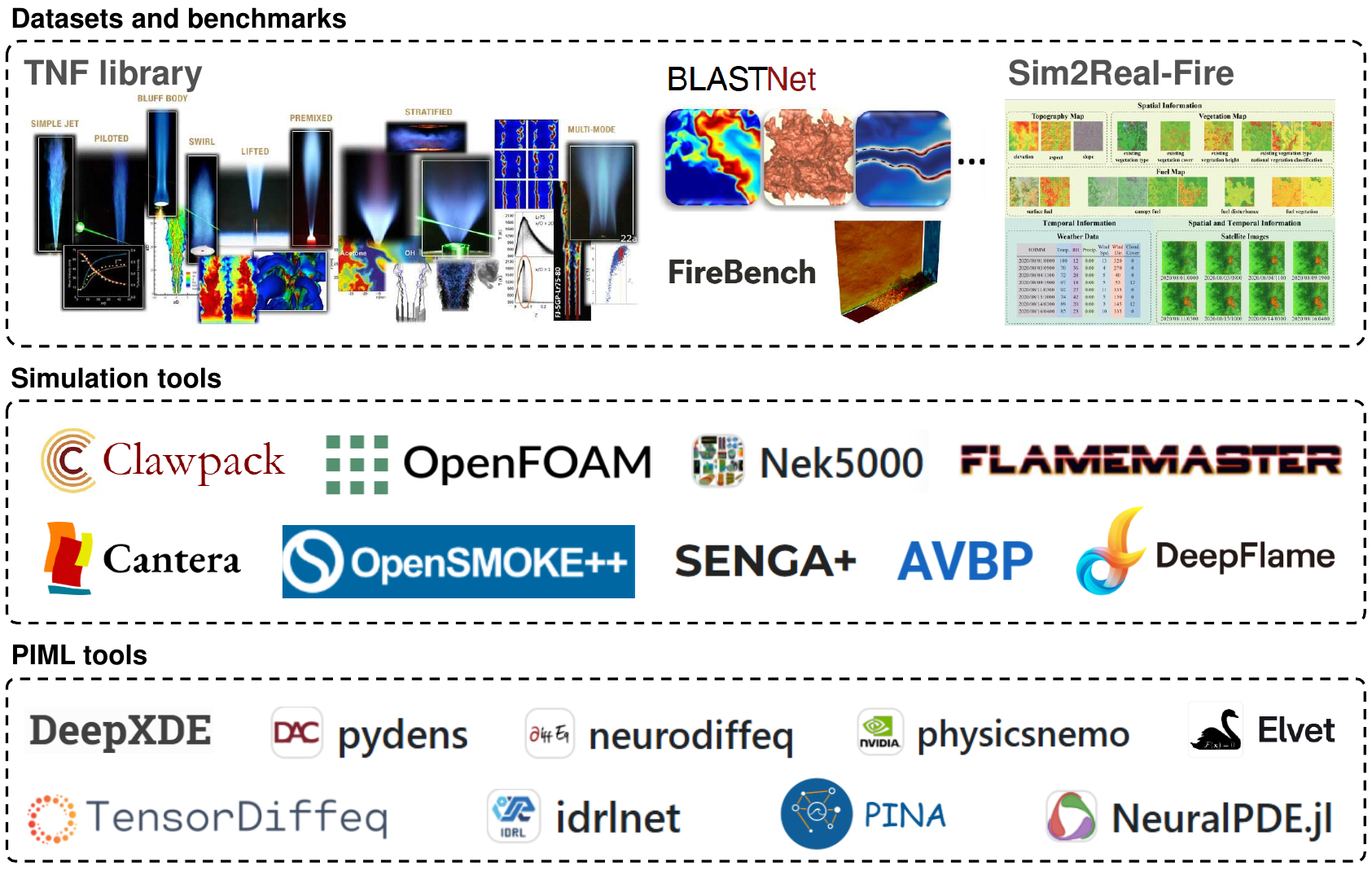}
    \caption{Some public resources for PIML for combustion, including the dataset/benchmarks, simulation tools, and PIML tools.}
    \label{fig: resources}
\end{figure}

This section will introduce some public resources that can promote the development of PIML for combustion, including the dataset/benchmarks, simulation tools, and PIML tools, some of which are illustrated in Figure \ref{fig: resources}. We list these resources as references not only for combustion researchers but also for AI researchers.

\subsection{Datasets and benchmarks}
\label{sec: data-bench}

It is known that data, algorithms, and computing power are the three driving forces behind AI development, reflecting the important role of data in AI. Specifically, data are the foundation for training models and their quality and quantity can directly influence the model performance. Therefore, it is important to establish public datasets and related benchmarks which can reduce redundant work, facilitate reproducibility, and standardize model evaluation and comparison. By definition, a benchmark is composed of the dataset(s) on which the benchmark is trained or evaluated, and a reference implementation in any language \cite{Thiyagalingam202204_SciMLBench}. Note that although PIML for forward problems is purely physics-driven and data-free, ground-truth data is also necessary for model evaluation, especially in academia. A large number of datasets and benchmarks have been established for various AI tasks, with the most famous of which is the ImageNet \cite{Deng200906_ImageNet} for CV. In contrast, the datasets and benchmarks for SciML and fluid, and combustion are fewer, but have been growing rapidly in recent years. Compared to traditional AI datasets, datasets for SciML, fluid, and combustion usually exhibit unique properties due to their physics-driven nature, such as high-dimensional, multiscale, costly-generated, and interdisciplinary, which pose challenges for the generation, storage, sharing, and use of the datasets.

Table \ref{table: data-bench} lists some representative public datasets and benchmarks for SciML, fluid, and combustion. Since fluid mechanics is an important topic of SciML, fluid datasets are included in many SciML dataset/benchmarks, such as those listed in Table \ref{table: data-bench}.
Takamoto \textit{et al.} \cite{Takamoto202210_PDEBench} developed the PDEBench, providing 35 datasets based on 11 well-known time-dependent and time-independent PDEs and evaluating popular SciML models, such as the FNO, U-Net, and PINNs, in an AR way (see Section \ref{sec: PINN_representation}). Luo \textit{et al.} \cite{Luo202309_CFDBench} constructed the CFDBench for evaluating the NOs on four CFD cases in both AR and NAR ways. Koehler \textit{et al.} \cite{Koehler202410_APEBench} established the APEBench not only for evaluating AR neural emulators, bust also providing an integrated differentiable simulation framework. As a PINN-specific benchmark, PINNacle developed by Hao \textit{et al.} \cite{Hao202306_PINNacle} provides a diverse dataset covering over 20 distinct PDEs from various domains. PINNacle also serves as a user-friendly toolbox that incorporate more than ten PINN methods, some of which are introduced in Section \ref{sec: PIML_develop}. A detailed comparison of fluid-contained SciML dataset/benchmarks can be found in the paper of FlowBench \cite{Tali202409_FlowBench}. Note that although some SciML benchmarks evaluate the models on AR tasks, their datasets can also be used for evaluating the classical NAR PIML models. In addition to the fluid data in SciML dataset/benchmarks, there are also some classical fluid datasets in the fluid community, such as the well-known Johns Hopkins turbulence database (JHTDB) \cite{Li200810_JHTDB} and the ERCOFTAC (European Research Community On Flow, Turbulence, And Combustion) classic collection database \cite{web_ERCOFTAC-data}. They were originally established for use in fluid tasks, such as turbulence modeling, but now they can also be used for SciML. A list of fluid databases and collections from the fluid community can be found in Reference \cite{web_ERCOFTAC-other-data}. In addition to datasets and benchmarks, knowledge bases also serve as important resources for both academia and industry. An example is the ERCOFTAC knowledge base wiki \cite{web_ERCOFTAC-wiki}, which provides structured knowledge and advice for CFD, covering non-reacting and reacting flows.

\begin{table}[ht]
	\small
	\centering
	\caption{Some public datasets and benchmarks for SciML, fluid, and combustion.}
	\label{table: data-bench}
	\begin{tabular}{>{\raggedright\arraybackslash}m{3cm}>{\raggedright\arraybackslash}m{11.5cm}>{\raggedright\arraybackslash}m{1cm}}
		\toprule[1.5pt]
		Name & Description & Link \\
		\midrule
		ScalarFlow \cite{Eckert201911_ScalarFlow} & A large-scale volumetric dataset of real-world scalar transport flows & \href{https://ge.in.tum.de/publications/2019-scalarflow-eckert}{\underline{URL}} \\
		PDEBench \cite{Takamoto202210_PDEBench} & A benchmark suite for evaluating SciML methods on time-dependent PDE-based tasks & \href{https://github.com/pdebench/PDEBench}{\underline{URL}} \\
		AirfRANS \cite{Bonnet202212_AirfRANS} & A dataset of solutions of 2D incompressible RANS equations over airfoils & \href{https://github.com/Extrality/AirfRANS}{\underline{URL}} \\
		PINNacle \cite{Hao202306_PINNacle} & A comprehensive benchmark of PINNs for solving PDEs & \href{https://github.com/i207M/PINNacle}{\underline{URL}} \\
		CFDBench \cite{Luo202309_CFDBench} & A benchmark for evaluating the generalization ability of NOs in CFD problems & \href{https://www.github.com/luo-yining/CFDBench}{\underline{URL}} \\
		FlowBench \cite{Tali202409_FlowBench} & A large-scale benchmark for flow simulation over complex geometries & \href{https://baskargroup.bitbucket.io}{\underline{URL}} \\
		APEBench \cite{Koehler202410_APEBench} & A benchmark suite for evaluating AR neural emulators for solving
		PDEs & \href{https://github.com/tum-pbs/apebench}{\underline{URL}} \\
		The Well \cite{Ohana202411_theWell} & A	large-scale simulation dataset of diverse spatiotemporal physical systems & \href{https://github.com/PolymathicAI/the_well}{\underline{URL}} \\
		JHTDB \cite{Li200810_JHTDB} & Johns Hopkins turbulence database & \href{https://turbulence.idies.jhu.edu}{\underline{URL}} \\
		ERCOFTAC \cite{web_ERCOFTAC-data} & ERCOFTAC classic collection database  & \href{http://cfd.mace.manchester.ac.uk}{\underline{URL}} \\
		NASA TMR \cite{web_NASA-turb} & Turbulence modeling resource of NASA Langley Research Center & \href{https://turbmodels.larc.nasa.gov}{\underline{URL}} \\
		Towne \textit{et al.} \cite{Towne202305_data-reduce} & A database for reduced-complexity modeling of fluid flows & \href{https://deepblue.lib.umich.edu/data/collections/kk91fk98z}{\underline{URL}} \\
		\midrule
		TNF library \cite{web_TNF} & A library of well-documented flames established by the TNF workshop & \href{https://tnfworkshop.org}{\underline{URL}} \\
		BLASTNet \cite{Chung202207_BLASTNet, Chung202210_BLASNet, Chung202309_BLASTNet2.0} & A large-scale community-involved network-of-datasets of reacting and non-reacting flows obtained from DNS for benchmarking ML methods & \href{https://blastnet.github.io}{\underline{URL}} \\
		GlobFire \cite{Artes201911_GlobFire} & A global wildfire dataset for the analysis of fire regimes and fire behaviour & \href{https://doi.org/10.1594/PANGAEA.895835}{\underline{URL}} \\
		Sim2Real-Fire \cite{Li202412_Sim2Real-Fire} & A multi-modal simulation dataset for forecast and backtracking of real-world forest fire & \href{https://github.com/TJU-IDVLab/Sim2Real-Fire}{\underline{URL}} \\
		FireBench \cite{Wang202411_FireBench} & An ensemble of LES dataset of 3D wildfire spread & \href{https://blastnet.github.io/firebench_wildfire_les}{\underline{URL}} \\
		\bottomrule[1.5pt]
	\end{tabular}
\end{table}

For combustion, the datasets and benchmarks are fewer than those for SciML and fluid. The TNF workshop is an international collaboration on turbulent combustion and established a library of well-documented flames that includes multiscalar and velocity data \cite{web_TNF}. The flame types cover from simple jet flames to piloted jet flames, bluff body flames, and swirl flames. Some classic flames such as the Sandia flames \cite{Barlow199808_Sandia-flame} are also included and have already been studied using PIML \cite{Peng202404_recons-flameD, Taassob202310_PINN-TC-closure} (see Section \ref{sec: turb-exp}). Chung \textit{et al.} \cite{Chung202207_BLASTNet, Chung202210_BLASNet} established the BLASTNet (Bearable Large Accessible Scientific Training Network-of-datasets), a network-of-datasets of reacting and non-reacting flows which combines community involvement, public data repositories, and lossy compression algorithms. Using the DNS data of a H$_2$-air lifted flame, the authors showcased the usage of BLASTNet by benchmarking the CNNs on both the tasks of combustion regime classification and filtered reaction rate regression \cite{Chung202210_BLASNet}. Later, the BLASTNet 2.0 \cite{Chung202309_BLASTNet2.0} was established with extended samples, configurations, and size, and has been used for benchmarking the scaling behaviour of five DL models on the 3D volumetric SR task. One of the five models adopted the physics-informed loss in a broad sense, as mentioned in Section \ref{sec: combus_SR}. BLASTNet has also been used for other PIML tasks, such as the PiMAPNet \cite{Sharma202505_PiMAPNet} introduced in Section \ref{sec: turb-exp}. Table \ref{table: data-bench} also provides some datasets of fire, including the GlobFire \cite{Artes201911_GlobFire}, Sim2Real-Fire \cite{Li202412_Sim2Real-Fire}, and FireBench \cite{Wang202411_FireBench}. The FireBench has also been uploaded to the BLASTNet. For a detailed comparison of datasets for wildfire analysis, readers can refer to the paper of Sim2Real-Fire \cite{Li202412_Sim2Real-Fire}.

\subsection{Simulation tools (data generators)}

In SciML, even with the growing number of public datasets and benchmarks, researchers still need to generate their own datasets using simulation tools, which are essential for customizing domain-specific cases, controlling data quality, and filling the inherent sparsity of public datasets. Table \ref{table: tools-simulation} gives a non-exhaustive list of public tools for ODE/PDE solution, fluid simulation, and combustion simulation. There are no clear boundaries between these three types of tools: Fluid simulation often falls under PDE solution, while combustion simulation often falls under either ODE solution or fluid simulation. Some ODE/PDE solution tools can be used for fluid/combustion simulation, while others cannot. For example, The CVODE \cite{Cohen199603_CVODE} in SUNDIALS \cite{Hindmarsh202509_SUNDIALS, Gardner202209_SUNDIALS2} has been used in many aforementioned combustion researches \cite{Kim202109_stiffNODE, Vijayarangan202311_dyn-informed-NODE, Zhang202208_Mscale-sample, Mao202307_DeepFlame}. The PyClaw \cite{Ketcheson201208_PyClaw}, Python version of Clawpack \cite{software_Clawpack, Mandli201608_Clawpack}, has been used for simulating supersonic combustion \cite{Wang202307_PINN-RDC}. For fluid simulation, the OpenFOAM \cite{Weller199811_FOAM, Jasak200709_OpenFOAM, Jasak200912_OpenFOAM, Moukalled2015_FVMinCFD} is a widely-used open-source CFD software and has many solvers for combustion, such as the chemFoam, reactingFoam, fireFoam, flameletFoam \cite{Mueller201306_flameletFoam}, and FGMFoam \cite{Bertsch201911_FGMFoam}. Recently, researchers have developed tools that utilize the large language models (LLMs) to automate OpenFOAM programming, including the MetaOpenFOAM \cite{Chen202407_MetaOpenFOAM, Chen202502_MetaOpenFOAM2.0} and OpenFOAMGPT \cite{Pandey202503_OpenFOAMGPT, Feng202504_OpenFOAMGPT2.0}. Cantera \cite{Goodwin2022_Cantera} is an open-source suite of tools for problems involving chemical kinetics, thermodynamics, and transport processes, and has been used in PIML for 1D laminar flames \cite{Wu202501_FlamePINN-1D}. Some AI-integrated simulators are also provided in Table \ref{table: tools-simulation}, including (1) the DeepFlame \cite{Mao202307_DeepFlame}, a deep learning empowered open-source platform for reacting flow simulations, which coupled the three tools of OpenFOAM, Torch \cite{Collobert201112_Torch7}, and Cantera, and (2) differentiable solvers, including the torchdiffeq \cite{Chen2018_torchdiffeq}, $\bf{\Phi} _\textrm{Flow}$ \cite{Holl202012_phiflow, Holl202405_PhiFlow}, JAX-Fluids \cite{Bezgin202209_JAX-Fluids, Bezgin202411_JAX-Fluids2.0}, Arrhenius.jl \cite{Ji202107_Arrhenius.jl}, and JANC \cite{Wen202504_JANC}. Table \ref{table: tools-simulation} also lists some tools that can be used for or are specifically designed for the DNS of fluid and combustion, including the Dedalus \cite{Burn202004_Dedalus}, Nek5000 \cite{web_Nek5000}, SENGA+ \cite{web_SENGA+}, and AVBP \cite{web_AVBP}. Note that most DNS tools for combustion are in-house codes, such as the NGA \cite{Desjardins200803_NGA}, S3D \cite{Chen200901_teraDNS}, and Athena-RFX \cite{Poludnenko201001_Athena-RFX}, which have been used for establishing the BLASTNet \cite{Chung202309_BLASTNet2.0}. See more examples in this review \cite{Domingo202208_recent-DNS-TC} and the paper of BLASTNet 2.0 \cite{Chung202309_BLASTNet2.0}. However, considering the extremely high cost of DNS for combustion, it is recommended to directly use the public datasets mentioned in Section \ref{sec: data-bench} rather than generating them, especially for AI-community researchers.

The tools in Table \ref{table: tools-simulation} have been used for generating datasets and benchmarks in Table \ref{table: data-bench}. For example, the Clawpack and $\bf{\Phi} _\textrm{Flow}$ were used to establish the PDEBench \cite{Takamoto202210_PDEBench}, while the Clawpack and Dedalus were used to establish the Well \cite{Ohana202411_theWell}. In addition to free tools, researchers can also use some widely-used commercial software for fluid/combustion simulation and dataset generation, such as the ANSYS Fluent \cite{software_Fluent}, Chemkin \cite{software_Chemkin}, and COMSOL Multiphysics \cite{software_COMSOL}. For example, COMSOL Multiphysics has been used for generating data in PIML for combustion \cite{Kim202301_Li-runaway} and generating the datasets of PINNacle \cite{Hao202306_PINNacle}. An overview of both free and commercial CFD software can be found in Reference \cite{web_overviewCFD}.

\begin{table}[ht]
	\small
	\centering
	\caption{Some public tools for ODE/PDE solution, fluid simulation, and combustion simulation.}
	\label{table: tools-simulation}
	\begin{tabular}{>{\raggedright\arraybackslash}m{3cm}>{\raggedright\arraybackslash}m{11.5cm}>{\raggedright\arraybackslash}m{1cm}}
		\toprule[1.5pt]
		Name & Description & Link \\
		\midrule
		SciPy \cite{Virtanen202002_SciPy} & An open-source scientific computing library in Python language & \href{https://scipy.org}{\underline{URL}} \\
		SUNDIALS \cite{Hindmarsh202509_SUNDIALS, Gardner202209_SUNDIALS2} & An open-source SUite of Nonlinear and DIfferential/ALgebraic Equation Solvers & \href{https://sundials.readthedocs.io}{\underline{URL}} \\
		GEKKO \cite{Beal201807_GEKKO} & A Python package for ML and optimization of mixed-integer and DAEs & \href{https://gekko.readthedocs.io}{\underline{URL}} \\
		torchdiffeq \cite{Chen2018_torchdiffeq} & Differentiable ODE solvers implemented in PyTorch & \href{https://github.com/rtqichen/torchdiffeq}{\underline{URL}} \\
		FEniCS \cite{Logg2012_FEniCS, Alnashs201507_FEniCS1.5, Baratta202312_DOLFINx} & An open-source computing platform for solving PDEs based on the FEM & \href{https://fenicsproject.org}{\underline{URL}} \\
		FreeFEM++ \cite{Hecht201212_FreeFEM++} & A free software for solving PDEs based on the FEM & \href{https://freefem.org}{\underline{URL}} \\
		FiPy \cite{Guyer200904_FiPy} & A PDE solver using Python based on the FVM & \href{https://www.ctcms.nist.gov/fipy}{\underline{URL}} \\
		Clawpack \cite{software_Clawpack, Mandli201608_Clawpack, Ketcheson201208_PyClaw} & A package that collects FVMs for hyperbolic systems of conservation laws & \href{https://www.clawpack.org}{\underline{URL}} \\
		Dedalus \cite{Burn202004_Dedalus} & An open-source framework for solving PDEs using spectral methods & \href{https://dedalus-project.org}{\underline{URL}} \\
		OpenFOAM \cite{Weller199811_FOAM, Jasak200709_OpenFOAM, Jasak200912_OpenFOAM} & An open-source CFD software & \href{https://openfoam.org}{\underline{URL}} \\
		Nek5000 \cite{web_Nek5000} & A fast and scalable open-source CFD solver using the spectral element method & \href{https://nek5000.mcs.anl.gov}{\underline{URL}} \\
		$\bf{\Phi} _\textrm{Flow}$ \cite{Holl202012_phiflow, Holl202405_PhiFlow} & A differentiable PDE solving framework for optimization and ML applications & \href{https://github.com/tum-pbs/PhiFlow}{\underline{URL}} \\
		JAX-Fluids \cite{Bezgin202209_JAX-Fluids, Bezgin202411_JAX-Fluids2.0} & A differentiable CFD solver for 3D compressible single-phase and two-phase flows & \href{https://github.com/tumaer/JAXFLUIDS}{\underline{URL}} \\
		CFD Python \cite{Barba201811_CFDPython} & An online learning module for CFD with Python & \href{https://lorenabarba.com/blog/cfd-python-12-steps-to-navier-stokes}{\underline{URL}} \\
		\midrule
		Cantera \cite{Goodwin2022_Cantera} & An open-source suite of tools for problems involving chemical kinetics, thermodynamics, and transport processes & \href{https://www.cantera.org}{\underline{URL}} \\
		FlameMaster \cite{Pitsch_FlameMaster} & An open source C++ package for 0D and 1D laminar combustion calculations & \href{https://www.itv.rwth-aachen.de/en/downloads/flamemaster}{\underline{URL}} \\
		OpenSMOKE++ \cite{Cuoci201502_OpenSMOKE++} & A framework for simulating reacting systems with detailed kinetic mechanisms & \href{https://www.opensmokepp.polimi.it}{\underline{URL}} \\
		SENGA+ \cite{web_SENGA+} & A DNS code for 3D compressible turbulent reacting flows &  \href{https://www.ukctrf.com/index.php/senga}{\underline{URL}} \\
		AVBP \cite{web_AVBP} & A suite of CFD tools for performing LES/DNS of compressible reacting flows & \href{https://avbp-wiki.cerfacs.fr}{\underline{URL}} \\
		DeepFlame \cite{Mao202307_DeepFlame} & A deep learning empowered CFD package for reacting flows & \href{https://deepflame.deepmodeling.com}{\underline{URL}} \\
		Arrhenius.jl \cite{Ji202107_Arrhenius.jl} & A differentiable combustion simulation package & \href{https://github.com/DENG-MIT/Arrhenius.jl}{\underline{URL}} \\
		JANC \cite{Wen202504_JANC} & A differentiable compressible reacting flow solver & \href{https://github.com/JA4S/JANC}{\underline{URL}} \\
		FDS \cite{McGrattan2025_FDS-guide} & Fire dynamics simulator, a LES code for fire-driven fluid flows & \href{https://pages.nist.gov/fds-smv}{\underline{URL}} \\
		\bottomrule[1.5pt]
	\end{tabular}
\end{table}

\subsection{PIML tools}

There are many tools for implementing the classical PIML, i.e., PINNs, some of which are presented in Table \ref{table: tools-PIML}. Most tools are written in Python, except for the ADCME \cite{Xu202011_ADCME} and NeuralPDE \cite{Zubov202107_NeuralPDE}, which are written in Julia. Most tools can serve as solvers in which users only need to define the problem, while other tools, such as the SciANN \cite{Haghighat202011_SciANN} and ADCME \cite{Xu202011_ADCME}, work as wrappers in which users still need to deal with most of the implementation details. Since AD has been implemented in modern AI frameworks, such as the PyTorch \cite{Paszke201912_PyTorch, Ansel202404_Pytorch2}, TensorFlow \cite{Abadi201611_TensorFlow}, and Keras \cite{Gulli2017_DL-Keras}, they are usually used as the backend for these PIML tools. Most PIML tools support one or two backends, except for the DeepXDE \cite{Lu202102_DeepXDE}, which supports the PyTorch, TensorFlow, JAX \cite{software_JAX}, and PaddlePaddle \cite{Ma201910_PaddlePaddle}. For more details of these PIML tools, readers can refer to these reviews \cite{Karniadakis202105_PIML, Cuomo202207_SciML-PINN, Wu202301_PIML-chem-eng, Zhao202410_PINN-complex-fluid}, where Reference \cite{Zhao202410_PINN-complex-fluid} also listed the merits and limitations of some tools.

\begin{table}[ht]
	\small
	\centering
	\caption{Some public tools of PIML.}
	\label{table: tools-PIML}
	\begin{tabular}{>{\raggedright\arraybackslash}m{3cm}>{\raggedright\arraybackslash}m{11.5cm}>{\raggedright\arraybackslash}m{1cm}}
		\toprule[1.5pt]
		Name & Description & Link \\
		\midrule
		DeepXDE \cite{Lu202102_DeepXDE} & A library for SciML and PIML & \href{https://deepxde.readthedocs.io}{\underline{URL}} \\
		PyDEns \cite{Koryagin201909_PyDEns} & A Python framework for solving ODE/PDEs using NNs & \href{https://github.com/analysiscenter/pydens}{\underline{URL}} \\
		NeuroDiffEq \cite{Chen202002_NeuroDiffEq, Liu202502_NeuroDiffEq2} & A Python package for solving DEs with NNs & \href{https://neurodiffeq.readthedocs.io}{\underline{URL}} \\
		NVIDIA PhysicsNeMo \cite{PhysicsNeMo} & An open-source framework for building, training, and fine-tuning SciML models & \href{https://github.com/NVIDIA/physicsnemo}{\underline{URL}} \\
		Elvet \cite{Araz202103_Elvet} & An NN-based DE and variational problem solver & \href{https://gitlab.com/elvet/elvet}{\underline{URL}} \\
		TensorDiffEq \cite{McClenny202103_TensorDiffEq} & A Python package on top of Tensorflow for scalable multi-GPU PINN solvers & \href{https://docs.tensordiffeq.io}{\underline{URL}} \\
		IDRLnet \cite{Peng202107_IDRLnet} & A PINN library on top of PyTorch & \href{https://idrlnet.readthedocs.io}{\underline{URL}} \\
		PINA \cite{Coscia202403_PINA} & A Python library designed to simplify and accelerate the development of SciML & \href{https://mathlab.github.io/PINA}{\underline{URL}} \\
		SciANN \cite{Haghighat202011_SciANN} & A Keras/TensorFlow wrapper for scientific computations and PIML using NNs & \href{https://www.sciann.com}{\underline{URL}} \\
		ADCME \cite{Xu202011_ADCME} & A Julia framework to solve inverse problems involving	physical simulations and NNs & \href{https://github.com/kailaix/ADCME.jl}{\underline{URL}} \\
		NeuralPDE \cite{Zubov202107_NeuralPDE} & A Julia package for solving PDEs using PINNs & \href{https://docs.sciml.ai/NeuralPDE}{\underline{URL}} \\
		\bottomrule[1.5pt]
	\end{tabular}
\end{table}

Some PIML tools have been used in PIML for combustion. For example, the NVIDIA PhysicsNeMo \cite{Chen202002_NeuroDiffEq}, previously known as NVIDIA Modulus and SimNet \cite{Hennigh202106_SimNet}, was used in Reference \cite{Liu202309_surr-flame}. The DeepXDE was used for implementing the FlamePINN-1D \cite{Wu202501_FlamePINN-1D} and optimizing the extended VDP system \cite{Xie202410_PINN-EVDP}. Some studies also implement the PIML for combustion directly using the modern AI frameworks, including the PyTorch \cite{Liu202312_recons-turb, Song202410_PINN-FPV, Cao202411_PTS-PINN-flame, Wang202307_PINN-RDC} and TensorFlow \cite{Taassob202310_PINN-TC-closure}, which is more flexible but cumbersome than using PIML tools. To show how popular PIML tools can be used for combustion research, a well-commented code demo is provided in Listing \ref{listing: demo}, which is based on the DeepXDE and adapted from the codes of FlamePINN-1D. Taking the 2D combustion reacting flow as example, the demo shows how to code the Equations \eqref{eq: mass} to \eqref{eq: T} using a PIML tool, in which the convenience of AD to compute derivatives is clearly demonstrated. To illustrate how to infer unknown parameters in inverse problems, the demo also defines an unknown correction coefficient for viscosity. It is straightforward to modify the demo for 3D combustion. For more details about the implementation of IC/BC/OCs, network, training, and evaluation, and how to code other equations in Section \ref{sec: combus_laws} in a differentiable way, readers can refer to the codes of FlamePINN-1D \cite{Wu202501_FlamePINN-1D}.

\section{Discussions and outlooks}
\label{sec: discuss-outlook}

After introducing the main existing work, it is necessary to analyze the limitations, provide useful suggestions, and point out potential future directions. This section will first introduce the current challenges and corresponding solutions, then discuss some of the choices that researchers and engineers may face, and finally point out promising opportunities for both combustion and AI.

\subsection{Challenges and solutions}
\label{sec: challenge}

\subsubsection{Limitations reinforced by combustion}
\label{sec: challenge_limitation}

PIML itself has many limitations in terms of convergence, accuracy, and generalization, while combustion, as an extremely complex multi-physical process, will pose even greater challenges for PIML. As mentioned in Section \ref{sec: PIML_develop}, PIML has many failure modes, such as multiscale representation \cite{Wang202106_eigen-bias-Fourier}, loss imbalance/conflict \cite{Wang202109_grad-flow}, mode differences in multiple subdomains \cite{Jagtap202011_XPINN}, unreasonable sampling \cite{Wu202210_comprehensive-sample}, and ill-conditioned PDEs \cite{Rathore202402_loss-landscape}. The inherent multiscale and multiphysics characteristics of combustion will certainly exacerbate these difficulties. For example, the wide range of scales involved in physical processes such as flow, chemical reaction, and diffusion in combustion reacting flows places higher demands on the multiscale representation of the ML models, while the large number of PDEs and corresponding IC/BCs poses challenges for loss balancing. Moreover, as mentioned in Section \ref{sec: PINN_other-topic}, most combustion PDE systems can be considered ill-conditioned. Even reaction problems that do not involve flow and diffusion have the inherent stiffness issue, as detailed in Section \ref{sec: solveODE}. In fact, the ill-conditioning issue also challenges the traditional numerical methods for combustion.

To deal with these difficulties, the methods introduced in Section \ref{sec: PIML_develop} should be appropriately used and improved. For example, the Fourier multiscale embedding can be used to enhance the multiscale representation \cite{Wang202106_eigen-bias-Fourier}, which has already been adopted in PIML for combustion \cite{Liu202309_surr-flame}. However, the introduction of hyperparameters is another minor issue, which can be resolved if the scale information is known. Moreover, there is no guarantee that this method will be effective on problems with a wider range of scales. In such cases, it is necessary to combine the domain decomposition methods, which are particularly important for large-domain and long-time problems. Domain decomposition can also facilitate parallel computation, thus accelerating the simulation. It is recommended to use knowledge-based domain decomposition if the approximate distribution characteristics of the solution are known in advance. Examples include combustors with reactant, mixing, reaction, and product regions. Adaptive domain decomposition can also be proposed for more complex problems. Unlike common domain decomposition methods, for combustion problems, each subdomain can use not only different networks but also different physical laws, which is similar to the adaptive combustion models \cite{Chung202101_dyn-submodel}. To deal with the multiphysics issue, imposing hard constraints on certain physical laws can always enhance the physical consistency and generalization ability. As for the ill-conditioning issue, some preconditioning strategies should be adopted, such as the normalization. For a deeper understanding, the ill-conditioning analysis should also be conducted for PIML solving combustion PDEs, which can be enlightened by the papers mentioned in Section \ref{sec: PINN_other-topic}. Although the reaction stiffness has been successfully solved in reaction ODE problems (see Section \ref{sec: solveODE}), it has not been thoroughly solved in reacting flow problems. As summarized in Table \ref{table: reacting-flow}, when using the direct finite-rate chemistry in PIML, the most detailed reaction mechanism is only a two-step mechanism. Other studies either discard, simplify, tabulate, pre-learn, or online-learn the reaction part. Among these, only tabulating and pre-learning can represent the known detailed mechanism and may serve as feasible ways to tackle the stiffness issue in reacting flows, as long as they are differentiable. In fact, tabulating and pre-learning with NNs can be viewed as preconditioning strategies and can be categorized as building surrogate models of chemical kinetics, as mentioned in Section \ref{sec: surr}, not for CFD, but for PIML.

\subsubsection{Discrepancies between physics, (physical) model, and data}
\label{sec: challenge_discrepancy}

There are many physical mechanisms involved in combustion that are not yet fully understood, such as the interaction between turbulence and chemical reactions, as well as the reaction mechanisms of specific problems. Therefore, there are no corresponding precise physical models. For PIML, which requires the coupling of physical laws, this presents both a challenge and an opportunity. The challenge lies in the need to carefully determine the physical model, while the opportunity lies in the feasibility to use PIML to model these unknown physical phenomena. Moreover, when real-world data are available, data-physics and data-model discrepancies also pose challenges. In these inverse problems, careful consideration must be given to the relative importance/weights of fitting the data and minimizing the physical model residuals.

As exemplified in Sections \ref{sec: PIML_0Dcombus} and \ref{sec: PIML_reacting-flow}, to deal with these discrepancies between physics, model, and data, three strategies can be adopted: (1) Use reduced models, i.e., keep only the certain models and discard the uncertain parts. For example, the mass/continuity equation was adopted in most studies, as shown in Table \ref{table: reacting-flow}, because it is certain, simple, and do not involve source terms that may need to be modeled for other equations. For the same reason, the energy and species equations are discarded in many studies. (2) Use simplified models. Examples include using RANS or LES models for turbulence and using simplified reaction mechanisms for complex reactions. Filtered PDEs have been proven to be effective for PIML to handle real-world data \cite{Zhang202411_FPDE}. (3) Learn the models, especially the turbulence and reaction closures, which have been illustrated in References \cite{Taassob202310_PINN-TC-closure, Taassob202406_PI-DeepONet-TC-closure, Liu202406_recons-soot, Yadav202412_RF-PINN}. These works demonstrate that the potential of PIML in inverse problems is not limited to field reconstruction and parameter inference, but can also assist in physical modeling. In addition to designing strategies to address these discrepancies, it is also necessary to have methods to measure these discrepancies, i.e., uncertainty quantification (UQ) \cite{Psaro202301_UQinSciML, Deng202508_SciML-combus}. The methods for quantifying data and model uncertainties mentioned in Section \ref{sec: PINN_other-topic} can be applied to PIML for combustion.

\subsubsection{Reproducibility}

Reproducibility is the foundation of academic research, ensuring the reliability, transparency, and implementability of the adopted methods in the papers, thereby guaranteeing the credibility of their conclusions. Many fields face reproducibility issues, or the so-called reproducibility crisis \cite{Baker201605_repro-crisis, Huston201802_AI-repro-crisis}, and we believe that PIML for combustion is no exception. According to Reference \cite{McGreivy202409_overoptimism}, overoptimism exists in ML for fluid-related PDEs, which is caused by weak baselines and reporting biases. The former does not apply to most studies on PIML for combustion because their goal is not to exceed a baseline model. However, reporting bias is highly likely to exist, which includes selective reporting, outcome switching, and publication bias \cite{McGreivy202409_overoptimism}. In addition to the reporting bias, the reproducibility issues can also be caused by other pitfalls, including the sensitivity to hyperparameters, ambiguity in implementation details, and inaccessibility of data and codes. As demonstrated in Tables \ref{table: chemical-kinetics} and \ref{table: reacting-flow}, most studies on PIML for combustion do not share codes, which is very different from the open source ecosystem of the AI community.

To deal with the reproducibility issues, McGreivy \textit{et al.} \cite{McGreivy202409_overoptimism} call for bottom-up cultural changes to minimize biased reporting and top-down structural reforms to reduce perverse incentives. We briefly outline some solutions at the researcher level: (1) Provide detailed model configurations, parameter settings, and evaluation metrics. (2) Improve data accessibility by either making data available or using publicly available data, such as those in Table \ref{table: data-bench}. When establishing public datasets, it is preferable to follow the principles of FAIR (Findability, Accessibility, Interoperability, and Reusability) \cite{Wilkinson201603_FAIR}. When establishing public benchmarks, some suggestions can be found in Reference \cite{Thiyagalingam202204_SciMLBench}. (3) It is preferable to provide the codes, as we have done in References \cite{Wu202501_FlamePINN-1D, Wang202408_PI-RGAN-RDC} and Listing \ref{listing: demo}. When developing public software, some suggestions can be found in Reference \cite{Niemeyer201903_software-practice}. (4) Report not only positive results but also negative results. (5) Conduct replicate experiments and report the mean and variance of the results to demonstrate that the impact of randomness is negligible. (6) Conduct hyperparameter sensitivity analysis. For hyperparameters that are highly sensitive, sufficient explanation should be provided. Finally, we emphasize that the reproducibility issues can also be alleviated by the establishment of an open, active, and rigorous community for PIML for combustion, which involves the standardization of processes, tools, and definitions, as well as the open sourcing of data, codes, and tools. There is still a long way to go in this regard.

\subsection{Discussions on implementation choices}
\label{sec: choices}

\textbf{Forward vs inverse problems}. As shown in Tables \ref{table: chemical-kinetics} and \ref{table: reacting-flow}, PIML is currently used more for inverse problems than for forward problems, and this is not limited to the field of combustion. There are many reasons for this: (1) Inverse problems are usually simpler than forward problems when both types of problems can be studied because of the involvement of data. When solving forward problems fails, inverse problems are the only option. (2) Inverse problems are more common than forward problems in practice. Forward problems have strict requirements and need complete equations and IC/BCs, while inverse problems have more relaxed requirements. Therefore, it is not that researchers do not study forward problems, but that the specific problems themselves are inverse problems. (3) Inverse problems have richer application significance than forward problems in terms of field reconstruction, parameter inference, physical modeling, and kinetics discovery. Forward problems often only involve simulation, which can also be achieved by traditional numerical methods. (4) Inverse problems are better suited to leveraging the advantages of PIML than forward problems. It is well known that the overall performance of PIML in forward problems is far inferior to that of traditional numerical methods \cite{McGreivy202409_overoptimism}. However, one advantage of PIML is that it can easily couple data and incomplete equations to solve inverse problems, which is not straightforward for traditional numerical methods. In fact, the data-driven characteristic is an inherent advantage of ML, which can be leveraged in PIML. Nevertheless, the forward problems still have two important aspects: (1) In terms of academic research, the forward problem is the basis of the inverse problem. Research on the forward problems helps deepen our understanding of PIML, which in turn helps us understand the inverse problems. Furthermore, many proposed methods on the forward problems in terms of representation models, optimization processes, and PDE systems can be seamlessly transferred to the inverse problems. (2) In terms of practical applications, although PIML is inferior to traditional solvers in solving single PDE, it has demonstrated advantages in parametric/surrogate solving. Examples can be found in Section \ref{sec: PINN_other-topic}, Section \ref{sec: PIML_application}, and Reference \cite{Liu202309_surr-flame}. Compared with traditional solvers, it is more efficient, provides a continuous representation in the parameter space, and can easily perform parameter sensitivity analysis. Note that parametric/surrogate modeling can also be implemented in the way of inverse problems, where the data can be utilized. To sum up, both types of problems are important. For researchers and engineers, the choice between forward problems and inverse problems should depend on the specific circumstances and objectives.

\textbf{Continuous vs discrete PINNs}. As defined in Section \ref{sec: PINN_representation}, continuous PINNs follow the coordinate-to-variable mapping way, i.e., $\widehat{\mathbf{u}} = \mathrm{NN} (\mathbf{z})$, while discrete PINNs follow the field-to-field mapping way, i.e., $\widehat{\mathbf{U}}_{\mathrm{out}} = \mathrm{NN} (\mathbf{U}_{\mathrm{in}})$. Both have their own advantages and disadvantages \cite{Luo202309_CFDBench}. Continuous PINNs provide a continuous representation of the solution, making them mesh-independent and allowing for easy query of values at any points. However, this greatly increases the difficulty of learning, as it requires learning the entire solution in the spatiotemporal domain. In contrast, discrete PINNs only need to learn the field-to-field mapping. The AD used in continuous PINNs eliminates the error in differentiation, but also increases the computational cost. Additionally, continuity makes it difficult for continuous PINNs to learn large gradients and discontinuous features, which are easier for discrete PINNs. Moreover, discrete PINNs are easier to implement hard constraints on PDEs, which will be elaborated in the part of "soft vs hard constraints" in this section. The choice between continuous PINNs and discrete PINNs should depend on the specific task type. For example, continuous PINNs are more suitable for learning continuous spatiotemporal solutions, reconstructing non-grid-based distributed data, and performing zero-error differentiation. Discrete PINNs are more suitable for tasks such as surrogate modeling of chemical kinetics (Section \ref{sec: surr}), SR (Section \ref{sec: combus_SR}), and AR simulations \cite{Sharma202505_PiMAPNet}.

\textbf{Complete/detailed vs reduced/simplified physics}. As mentioned in Sections \ref{sec: challenge_limitation} and \ref{sec: challenge_discrepancy}, not all studies on PIML for combustion adopt complete/detailed physical models, such as detailed reaction mechanisms, complete PDEs, non-averaged NS equations, and closed turbulence terms. The choice between complete/detailed or reduced/simplified physical models should depend on the specific circumstances. For forward problems, the physical model can be simplified but must be complete, otherwise the problem will be ill-posed. For inverse problems with data, the choice is more flexible, and either a complete/detailed or a reduced/simplified physical model can be used. In fact, sometimes using only a few physical constraints can greatly enhance the overall performance. For example, some studies in Sections \ref{sec: surr} and \ref{sec: NODE_surr} show that using only species and element conservations can achieve a significant improvement. When there are discrepancies between the physics, model, and data, a reduced/simplified physical model can be used. See Section \ref{sec: challenge_discrepancy} for details. When both complete models and data are available, comparative studies can be conducted to identify which models have minimal impact on the results and can be omitted, thereby reducing computational costs, as demonstrated in Reference \cite{Yadav202412_RF-PINN}.

\textbf{Soft vs hard constraints}. Classical PIML integrates physical laws through loss functions, which is a soft constraint, so there is still a risk of violating the laws. This is why hard constraints are necessary. As mentioned in the Section \ref{sec: PINN_other-topic}, there are many ways to implement hard constraints for IC/BCs, PDEs, and other physical laws. For example, IC/BCs for continuous PINNs can be enforced through input and output transformations, while for discrete PINNs, they can be enforced through padding operations. Discrete PINNs can also easily implement hard constraints on PDEs using convolution kernels or traditional numerical methods/solvers. However, it is difficult to implement hard constraints on PDEs for continuous PINNs. Even for the hard constraint projection (HCP) \cite{Chen202108_TgHCP}, it is still necessary to minimize the PDE losses after projection. This is one of the advantages of discrete PINNs over continuous PINNs. For physical laws in the form of algebraic equations, i.e., most of the equations in Section \ref{sec: combus_laws}, hard constraints can be implemented simply by directly specifying them, for both continuous and discrete PINNs. For example, the law of mass action and the Arrhenius law are hard constraints in all the literature that considers them introduced in this paper, because they are explicitly embedded in the intermediate steps of the method rather than constrained by the loss. Hard constraints have both advantages and disadvantages. Imposing hard constraints on certain physical laws will alleviate the burden of loss optimization. For example, enforcing species conservation can reduce one equation of $Y$ that need to be solved. However, hard constraints can sometimes increase the difficulty of optimization, especially for IC/BCs, as they often involve hyperparameters and "hyperfunctions". In some cases, such as PDEs, complexity is also a disadvantage of hard constraints, which are less straightforward than soft constraints. The choice between soft constraints and hard constraints depends on many factors. By default, hard constraints should be considered first. Soft constraints can be used in the following situations: (1) when the constraint requirements are not strict; (2) when hard constraints are difficult to implement, such as for PDEs in continuous PINNs; (3) when hard constraints are extremely cumbersome to implement; (4) when hard constraints impair or greatly slow down the optimization and convergence, which requires \textit{a posteriori} testing.

\subsection{Opportunities for combustion and AI}

\textbf{Combustion-side opportunities}.
\begin{itemize}
	\item \textbf{\textit{Academic research}}. First, there are still many limitations that need to be further solved, as mentioned in Section \ref{sec: challenge}, including large-domain long-time problems, stiffness caused by detailed reaction mechanisms, model-data-physics discrepancies, and reproducibility issues. Second, there are still many combustion scenarios that can be studied using PIML, such as multiphase combustion, pollutant control, ammonia combustion, microgravity combustion, explosions, and fires. Finally, there are still many tasks that can be achieved using PIML, such as improving combustion models, gaining insights from experimental data, and developing hybrid solvers.

    \item \textbf{\textit{Practical applications}}. Combustion has important applications in various fields, including power generation, transportation, fire safety, manufacturing, chemical synthesis, waste disposal, and space exploration. Therefore, the application of PIML for combustion in these fields remains to be explored. One promising application scenario is the construction of digital twins for combustion equipment, with functions of real-time monitoring, detection, prediction, and control. Good digital twins often require both big data and physical knowledge, which is the main advantage of PIML.

	\item \textbf{\textit{Integration of advanced AI methods}}. As introduced in this paper, various AI methods haven been used in PIML for combustion, including the basic MLP, KAN \cite{Koenig202507_ChemKAN}, NO, NODE, GAN, LSTM, AE, self-supervised learning, and transfer learning \cite{Liu202509_3Dfrom2D}. However, there are still many advanced AI methods that have potential for application, such as DM, reinforcement learning, foundation models, and PDE discovery. Among these, DM has already been applied to fluid mechanics \cite{Shu202302_PI-DM, Li202404_Lag-turb-DM, Guastoni202506_FM-DM} and combustion \cite{Liu202509_3Dfrom2D, Huang202409_PCDM-SR-flame}. Foundation models are a hot topic due to their notable applications in the NLP field, including LLMs such as GPT \cite{OpenAI202304_GPT-4}, Llama \cite{Touvron202307_Llama2}, and DeepSeek \cite{Liu202412_DeepSeek-V3}. In the SciML field, foundation models are a current key direction \cite{Meng202505_phy-meet-ML, Subramanian202306_FM4SciML}. Some foundation models for PDEs have been developed, including the PDEformer \cite{Ye202402_PDEformer-1, Ye202507_PDEformer-2}, UPS \cite{Shen202403_UPS}, Poseidon \cite{Herde202405_Poseidon}, PROSE-PDE \cite{Sun202503_PROSE-PDE}, and PROSE-FD for fluid PDEs \cite{Liu202409_PROSE-FD}. Therefore, establishing foundation models for combustion is a feasible and promising direction \cite{Ihme202408_AI-catalyst-combus}, which also needs to be informed by combustion physics.
\end{itemize}

\textbf{AI-side opportunities}.
\begin{itemize}
    \item \textbf{\textit{Extension of strengths}}. Combustion is an field where AI and PIML can further demonstrate and extend their interdisciplinary capabilities, broad applicability, and powerful versatility. This can be achieved in many of the aforementioned scenarios, such as computation acceleration, data assimilation, combustion diagnostics, fire safety, and digital twins.

	\item \textbf{\textit{Mitigation of limitations}}. Combustion is a challenging testing ground for AI and PIML, where the shortcomings and limitations of existing model/algorithms may be exposed in terms of multiscale modeling, multiphysics modeling, generalization, and physical consistency. This is an opportunity to improve existing model/algorithms and develop new model/algorithms.

	\item \textbf{\textit{Combustion-inspired AI}}. Physics-inspired AI paradigms are an important topic, in which AI methods are often proposed or improved by drawing inspiration from physical laws, phenomena, or principles. The inspiration can be achieved in many ways, including embedding physical laws, mimicking physical processes, and borrowing physical principles. Therefore, physics-inspired AI includes physics-informed AI, or it can be considered as physics-informed AI in a broad sense. Jiao \textit{et al.} \cite{Jiao202408_AImeetPhy} have reviewed the AI methods that are inspired by classical mechanics, electromagnetism, statistical physics, and quantum mechanics. Additionally, two currently popular generative models are also inspired by physics: DM \cite{Sohl201503_DPM, Ho202006_DDPM} is inspired by nonequilibrium thermodynamics, while the continuity equation in the flow matching \cite{Lipma202210_FM} originates from fluid mechanics. Therefore, it is natural to ask: can we develop combustion-inspired AI? In fact, CRNN can be regarded as one such effort \cite{Ji202101_CRNN}, with its network architecture designed based on the law of mass action and Arrhenius law. However, given the abundance of combustion laws, phenomena, and principles, there are still many possibilities to explore, which may give rise to some revolutionary AI methods.
\end{itemize}

\section{Conclusions}
\label{sec: conclusion}

This paper provides a comprehensive review on PIML for combustion, systematically summarizing the fundamental principles, major studies, key advances, and available resources, while also providing constructive discussions and recommendations for researchers and engineers. The main conclusions are as follows: (1) PIML has been widely and successfully applied to combustion chemical kinetics, combustion reacting flows, and other combustion problems, addressing both forward and inverse problems. These applications demonstrate the growing attention, proven effectiveness, and significant potential of PIML in combustion research. (2) Numerous approaches exist for integrating combustion laws into ML. A primary focus of this review is to explain how such laws can be integrated into ML frameworks--either through soft or hard constraints, via loss functions or representation models, within coordinate-to-variable or field-to-field paradigms, and using either complete/detailed or reduced/simplified physics--highlighting the flexibility of PIML. (3) By incorporating combustion knowledge, the physical consistency and accuracy of ML models are always enhanced compared to purely data-driven ML. (4) PIML serves as a unified bridge between physics, models, and data, enabling both model-to-data simulation/reconstruction tasks and data-to-model modeling tasks under the guidance of physics This integration produces three key outcomes that represent or contain physics: enhanced data, improved physical models, and more reliable ML models. When discrepancies arise among physics, physical models, and data, PIML can address them through a range of strategies. (5) Challenges remain in terms of the inherent limitations of PIML, physics-model-data discrepancies, and reproducibility. Furthermore, many potential applications and research topics remain unexplored. Both of these points indicate promising directions for the future. In conclusion, PIML for combustion is a rapidly evolving field that calls for increased collaboration between combustion and AI researchers to address significant problems, develop improved strategies, and explore emerging opportunities.

\section*{Acknowledgments}
\addcontentsline{toc}{section}{Acknowledgments}
This work is supported by the National Key R\&D Program of China (No. 2021YFA0716202), the National Natural Science Foundation of China (No. 62273149), the Science and Technology Plan Project of Yunnan Province (No. 202302AQ370003-2), and the Creative Seed Fund of Shanxi Research Institute for Clean Energy, Tsinghua University.

\appendix

\section{A code demo of PIML for combustion}

\definecolor{codegreen}{RGB}{0,135,0}
\definecolor{codepurple}{RGB}{178,51,255}
\definecolor{codered}{RGB}{186,69,69}

\lstdefinestyle{mystyle}{
    lineskip=-0.5pt,
    basicstyle=\ttfamily\footnotesize,
    keywordstyle=\color{codepurple},
    commentstyle=\color{codegreen}\itshape,
    stringstyle=\color{codered},
    breaklines=true,
    breakatwhitespace=false,
    captionpos=b,
    keepspaces=true,
    columns=flexible,
    showspaces=false,
    showstringspaces=false,
    showtabs=false,
    tabsize=2,
    frame=single,
}
\lstset{style=mystyle}

\begin{center}
\begin{lstlisting}[
    language=Python,
    caption={A code demo for the PDEs of 2D unsteady combustion reacting flows using the DeepXDE, with PyTorch as the backend. The Dufour and Soret effects are neglected.}]
import torch
import deepxde as dde
import deepxde.grad.jacobian as jacobian
import deepxde.grad.hessian as hessian

def pde(self, xyt, uvpTYs):
    """
    shape of xyt: (n_pnt, 3)
    shape of uvpTYs: (n_pnt, 4 + n_spe)
    n_pnt: the number of sampling points on the domain for the current batch
    n_spe: the number of species
    """
    gas2d = args.gas2d  # a class containing gas information and some differentiable functions
    n_spe = gas2d.n_spe
    # scaling/normalizing factors:
    scale_x, scale_y, scale_t = args.scales["x"], args.scales["y"], args.scales["t"]
    scale_u, scale_v, scale_p, scale_T = args.scales["u"], args.scales["v"], args.scales["p"], args.scales["T"]
    scale_Ys = torch.tensor(args.scales["Ys"], dtype=dtype)  # shape: (n_spe, )

    u, v, p, T, Ys = uvpTYs[:, 0:1], uvpTYs[:, 1:2], uvpTYs[:, 2:3], uvpTYs[:, 3:4], uvpTYs[:, 4:]

    # compute the derivatives
    u_x = jacobian(uvpTYs, xyt, i=0, j=0)
    u_y = jacobian(uvpTYs, xyt, i=0, j=1)
    u_t = jacobian(uvpTYs, xyt, i=0, j=2)
    # u_xx = hessian(uvpTYs, xyt, component=0, i=0, j=0)
    # u_yy = hessian(uvpTYs, xyt, component=0, i=1, j=1)
    v_x = jacobian(uvpTYs, xyt, i=1, j=0)
    v_y = jacobian(uvpTYs, xyt, i=1, j=1)
    v_t = jacobian(uvpTYs, xyt, i=1, j=2)
    # v_xx = hessian(uvpTYs, xyt, component=1, i=0, j=0)
    # v_yy = hessian(uvpTYs, xyt, component=1, i=1, j=1)
    p_x = jacobian(uvpTYs, xyt, i=2, j=0)
    p_y = jacobian(uvpTYs, xyt, i=2, j=1)
    p_t = jacobian(uvpTYs, xyt, i=2, j=2)
    T_x = jacobian(uvpTYs, xyt, i=3, j=0)
    T_y = jacobian(uvpTYs, xyt, i=3, j=1)
    T_t = jacobian(uvpTYs, xyt, i=3, j=2)
    T_xx = hessian(uvpTYs, xyt, component=3, i=0, j=0)
    T_yy = hessian(uvpTYs, xyt, component=3, i=1, j=1)
    Ys_x = torch.hstack([jacobian(uvpTYs, xyt, i=4+k, j=0) for k in range(n_spe)])
    Ys_y = torch.hstack([jacobian(uvpTYs, xyt, i=4+k, j=1) for k in range(n_spe)])
    Ys_t = torch.hstack([jacobian(uvpTYs, xyt, i=4+k, j=2) for k in range(n_spe)])
    # Ys_xx = torch.hstack([hessian(uvpTYs, xyt, component=4+k, i=0, j=0) for k in range(n_spe)])
    # Ys_yy = torch.hstack([hessian(uvpTYs, xyt, component=4+k, i=1, j=1) for k in range(n_spe)])

    # compute some necessary properties using T and Ys
    rho = gas2d.cal_rho(T, Ys)  # density, shape: (n_pnt, 1)
    cps = gas2d.cal_cps(T)  # heat capacities, shape: (n_pnt, n_spe)
    cp = gas2d.cal_cp(T, Ys)  # mean heat capacity, shape: (n_pnt, 1)
    mu = gas2d.cal_mu(T, Ys)  # mean viscosity, shape: (n_pnt, 1)
    lam = gas2d.cal_lam(T, Ys)  # mean thermal conductivity, shape: (n_pnt, 1)
    Dkm_apo = gas2d.cal_Dkms_apo(T, Ys)  # mean diffusion coefficients, shape: (n_pnt, n_spe)
    W = gas2d.cal_W(Ys)  # mean molecular weight, shape: (n_pnt, 1)
    Xs = gas2d.cal_Xs(Ys)  # mole fractions, shape: (n_pnt, n_spe)
    wdot = gas2d.cal_omega_dot_mass(T, Ys)  # species production rates, shape: (n_pnt, n_spe)
    wTdot = gas2d.cal_omegaT_dot(T, Ys)  # reaction heat, shape: (n_pnt, 1)

    # infer the correction coefficient for viscosity (cmu) in inverse problems, if users want
    cmu = self.cmu_infe_s / args.scales["cmu"] if "cmu" in args.infer_paras else 1.0
    mu = mu * cmu

    # compute some terms in the continuity and momentum equations
    rho_x = jacobian(rho, xyt, i=0, j=0)
    rho_y = jacobian(rho, xyt, i=0, j=1)
    rho_t = jacobian(rho, xyt, i=0, j=2)
    div_U = u_x + v_y
    tau_11 = mu * (4 / 3) * u_x
    tau_12 = mu * div_U
    tau_21 = mu * div_U
    tau_22 = mu * (4 / 3) * v_y
    tau_11_x = jacobian(tau_11, xyt, i=0, j=0)
    tau_12_y = jacobian(tau_12, xyt, i=0, j=1)
    tau_21_x = jacobian(tau_21, xyt, i=0, j=0)
    tau_22_y = jacobian(tau_22, xyt, i=0, j=1)

    # compute some terms in the species equations
    Xs_x = torch.hstack([jacobian(Xs, xyt, i=k, j=0) for k in range(n_spe)])  # (n_pnt, n_spe)
    Xs_y = torch.hstack([jacobian(Xs, xyt, i=k, j=1) for k in range(n_spe)])  # (n_pnt, n_spe)
    js1 = -rho * (gas2d.Ws / W) * Dkm_apo * Xs_x  # species mass flux on x, (n_pnt, n_spe)
    js2 = -rho * (gas2d.Ws / W) * Dkm_apo * Xs_y  # species mass flux on y, (n_pnt, n_spe)
    js1 = js1 - Ys * torch.sum(js1, dim=-1, keepdim=True)  # correct the mixture-averaged model
    js2 = js2 - Ys * torch.sum(js2, dim=-1, keepdim=True)
    js1_x = torch.hstack([jacobian(js1, xyt, i=k, j=0) for k in range(n_spe)])  # (n_pnt, n_spe)
    js2_y = torch.hstack([jacobian(js2, xyt, i=k, j=1) for k in range(n_spe)])  # (n_pnt, n_spe)

    # compute some terms in the energy equation
    lam_x = jacobian(lam, xyt, i=0, j=0)
    lam_y = jacobian(lam, xyt, i=0, j=1)
    heat_cond = lam *(T_xx + T_yy) + lam_x * T_x + lam_y * T_y
    js1_cp_sum = torch.sum(js1 * cps, dim=-1, keepdim=True)  # (n_pnt, 1)
    js2_cp_sum = torch.sum(js2 * cps, dim=-1, keepdim=True)  # (n_pnt, 1)
    heat_spediff = -js1_cp_sum * T_x - js2_cp_sum * T_y
    heat_pressure = p_t + u * p_x + v * p_y
    heat_viscous = tau_11 * u_x + tau_12 * u_y + tau_21 * v_x + tau_22 * v_y

    # assemble the equations/residuals
    continuity = rho_t + rho * div_U + (u * rho_x + v * rho_y)
    momentum_x = rho * (u_t + u * u_x + v * u_y) - (tau_11_x + tau_12_y) + p_x
    momentum_y = rho * (v_t + u * v_x + v * v_y) - (tau_21_x + tau_22_y) + p_y
    energy = rho * (T_t + u * T_x + v * T_y) - (heat_cond + heat_spediff + wTdot + heat_pressure + heat_viscous)
    species = rho * (Ys_t + u * Ys_x + v * Ys_y) + js1_x + js2_y - wdot  # shape: (n_pnt, n_spe)

    # scale/normalize the equations
    continuity *= scale_u / scale_x
    momentum_x *= scale_u * scale_u / scale_x
    momentum_y *= scale_v * scale_u / scale_x
    energy *= scale_T * scale_u / scale_x / 1000  # 1000 for scaling cp
    species *= scale_Ys * scale_u / scale_x

    return [continuity, momentum_x, momentum_y, energy] + [species[:, k:k+1] for k in range(n_spe)]
\end{lstlisting}
\label{listing: demo}
\end{center}

\scriptsize
\bibliographystyle{unsrturl}
\clearpage  
\phantomsection  
\addcontentsline{toc}{section}{Reference}  

\bibliography{references}

\end{document}